\documentclass{elsart}
\usepackage[dvips]{epsfig}
\usepackage[dvips]{graphicx}
\usepackage{amsfonts}
\usepackage{epsf}

\def\undemi{\hbox{$\frac{1}{2}$}}
\def\R{\mathbb{R}}

\newcommand{\ka}{\kappa}

\begin{document}

\begin{frontmatter}

\title{A Boussinesq system for two-way propagation of interfacial waves}

\author[a]{Hai Yen Nguyen, Fr\'ed\'eric Dias}

\address[a]{CMLA, ENS Cachan, CNRS, PRES UniverSud,
61, avenue du Pr\'esident Wilson, 94230 Cachan cedex, France}

\begin{abstract}
The theory of internal waves between two layers of immiscible fluids is important both for its applications
in oceanography and engineering, and as a source of interesting mathematical model equations that exhibit nonlinearity
and dispersion. A Boussinesq system for two-way propagation of interfacial waves in a rigid lid configuration is derived.
In most cases, the nonlinearity is quadratic. However, when the square of the depth ratio is close to the density ratio,
the coefficients of the quadratic nonlinearities become small and cubic nonlinearities must be considered. The
propagation as well as the collision of solitary waves and/or fronts is studied numerically.
\end{abstract}

\end{frontmatter}

\section{Introduction}

As emphasized by Helfrich \& Melville \cite{HM} in their recent survey article
on long nonlinear internal waves, observations over the past four decades have demonstrated that
internal solitary-like waves are ubiquitous features of coastal oceans and marginal seas.
Solitary waves are long nonlinear waves consisting
of a localized central core and a decaying tail. They arise whenever there is a balance
between dispersion and nonlinearity. They have been
proved to exist in specific parameter regimes, and are often
conveniently modelled by Korteweg--de Vries (KdV)
equations or Boussinesq systems. As explained by Evans \& Ford \cite{EF}, the differences
between ``free-surface'' and ``rigid lid'' internal waves are small for internal waves of interest. Therefore
the ``rigid lid'' configuration remains popular for investigating internal waves even if it
does not allow for generalized solitary waves, which are long nonlinear waves consisting
of a localized central core and periodic non-decaying oscillations
extending to infinity. Such waves arise whenever there is a resonance
between a linear long wave speed of one wave mode in the system
and a linear short wave speed of another mode \cite{FDG}.

When dealing with interfacial waves with rigid boundaries in the framework of the full Euler equations, the
amplitude of the central core is bounded by the configuration. In
the case of solitary waves, it is known that when the wave speed
approaches a critical value the solution reaches a maximum
amplitude while becoming indefinitely wider; these waves are often
called `table-top' waves. In the limit as the width of the central
core becomes infinite, the wave becomes a front \cite{DV03}. Such
behavior is conveniently modelled by an extended Korteweg--de
Vries (eKdV) equation, i.e. a KdV equation with a cubic nonlinear
term \cite{Funakoshi}. Sometimes the terminology `modified KdV equation' or
`Gardner equation' is also used. KdV-type equations only describe one-way wave propagation.
The natural extension toward two-way wave propagation is the class of
Boussinesq systems. We will derive two sets of Boussinesq systems, one with quadratic nonlinearities
and another one with quadratic and cubic nonlinearities. We will use the terminology `extended'
for a Boussinesq system with both quadratic and cubic terms. Some
questions arise when dealing with `table-top' solitary waves. What are their properties? How do they interact?
The main goal of this work is to learn more about these waves by studying and integrating numerically an extended Boussinesq
system which allows a comparison between fronts and the more standard solitary waves. More general models have also
been derived by Choi \& Camassa \cite{CC}. They considered shallow water as well as deep water configurations. In the
shallow water case, their set of equations is the two-layer version of the Green--Naghdi equations. The equations
derived in \cite{CC} were recently extended to the free-surface configuration \cite{Barros}.
Solitary waves for two-layer flows have also been computed numerically as
solutions to the full incompressible Euler equations in the
presence of an interface by various authors -- see for example
\cite{LD}. Similarly fronts have been computed for example in \cite{DV03,DV04}. 

The paper is organized as follows. In \S\,2, we present the governing equations and the corresponding boundary conditions. 
A first Boussinesq system of three equations is derived in \S\,3. Then it is shown in \S\,4 how to reduce this system
to a system of two equations, one for the evolution of the interface shape and the other one for the evolution 
of a combination of the horizontal velocities in each layer. The 
numerical scheme and the numerical solutions are described in \S\,5. Results are shown for the propagation of a single wave,
for the co-propagation of two waves and for the collision of two waves of equal as well as unequal sizes.
When the square of the depth ratio is close to the density ratio,
the coefficients of the quadratic nonlinearities become small and cubic nonlinearities must be considered. An extended
Boussinesq system is derived in \S\,6. Numerical solutions of the extended Boussinesq system are described in \S\,7. In
particular, the collision of `table-top' waves is considered. A short conclusion is given in \S\,8. In the Appendices,
we provide very accurate results for wave run-up and phase shift, as well as some intermediate steps in the derivation
of the extended Boussinesq system.

\section{Governing equations}

The origin of the systems of partial differential equations that will be derived below is explained in this section. The 
methods are standard, but to our knowledge some of these equations are derived for the first time. 

\begin{figure} 
\begin{center}
\begin{tabular}{c c}
(a) in physical space & (b) in dimensionless variables \\
\includegraphics[width=6cm]{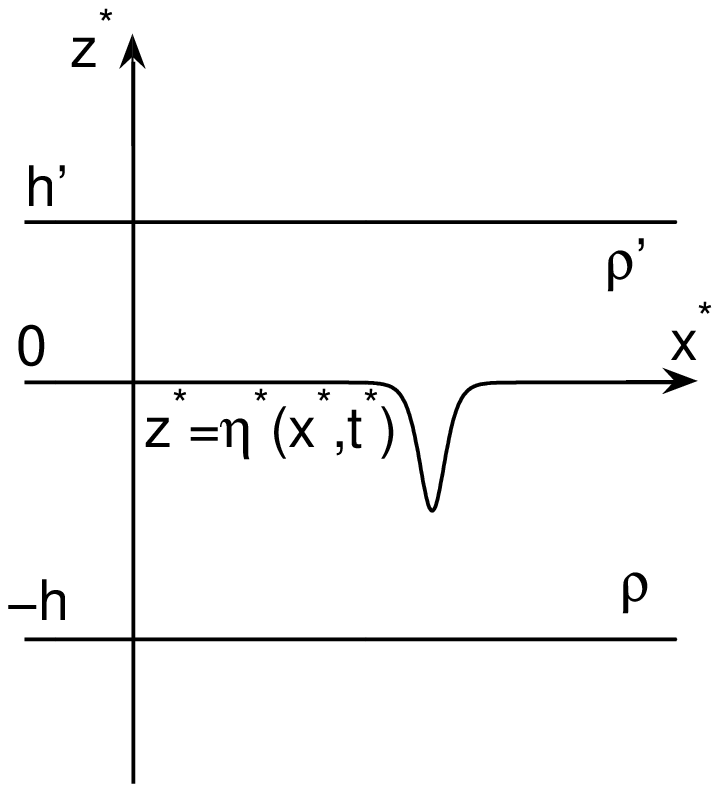}&\includegraphics[width=6cm]{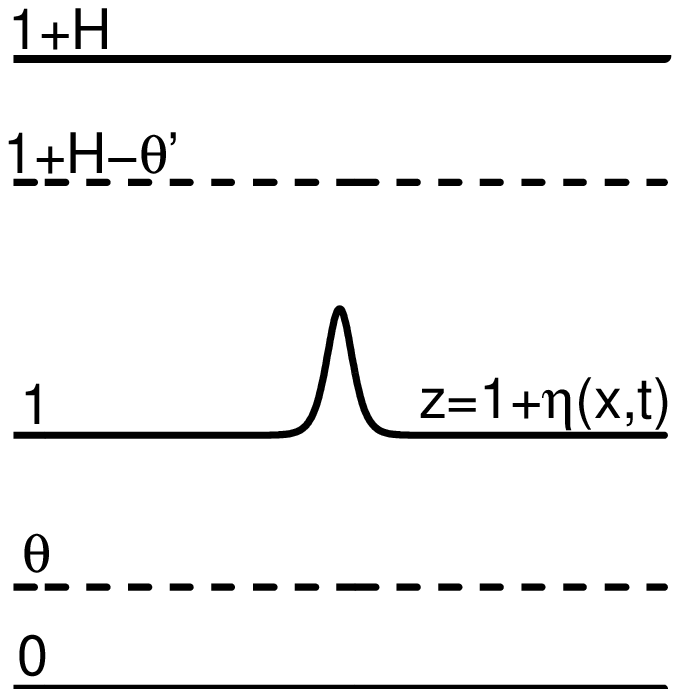}
\end{tabular}
\end{center}
\caption{Sketch of solitary waves propagating at the interface between two fluid layers with different densities $\rho'$ and $\rho$. The 
top and the bottom of the fluid domain are flat and rigid boundaries, located respectively at $z^*=h'$ and $z^*=-h$. (a)
Sketch of a solitary wave of depression in physical space; (b) Sketch of a solitary wave of elevation in dimensionless coordinates, 
with the thickness $h$ of the bottom layer taken as unit length
and the long wave speed $c$ as unit velocity. The dashed lines represent arbitrary fluid levels $\theta$ and $1+H-\theta'$ in each 
layer. The dimensionless number $H$ is equal to $h'/h$.}
\label{sketch_wave}
\end{figure}

Waves at the interface between two fluids are considered. The bottom as well as the upper boundary are assumed to be
flat and rigid. A sketch is given in Figure \ref{sketch_wave}. The analysis is restricted to two-dimensional flows. In other words, 
there is only one horizontal direction, $x^*$, 
in addition to the vertical direction, $z^*$. The interface is described by $z^*=\eta^*(x^*,t^*)$. The bottom layer 
$\Omega_{t^*}=\{(x^*,z^*): x^* \in \R, -h<z^*<\eta^*(x^*,t^*)\}$ and the upper layer
$\Omega'_{t^*}=\{(x^*,z^*): x^* \in \R, \eta^*(x^*,t^*)<z^*<h'\}$ are filled with inviscid, incompressible fluids, with densities 
$\rho$ and $\rho'$
respectively. All quantities related to the upper layer are denoted with a prime. All physical variables are denoted with a star.

In addition the flows are assumed to be irrotational. Therefore we are dealing with potential flows
and only stable configurations with $\rho>\rho'$ are considered. Velocity potentials $\phi^*=\phi^*((x^*,z^*),t^*)$ in
$\Omega_{t^*}$ and $\phi^{*'}=\phi^{*'}((x^*,z^*),t^*)$ in $\Omega'_{t^*}$ are introduced, so that the velocity vectors 
${\bf v}^*$ and ${\bf v}^{*'}$ are given by
\begin{eqnarray}
{\bf v}^*=\nabla{\phi^*}, \label{poten1}\\
{\bf v}^{*'}=\nabla{\phi^{*'}}. \label{poten2}
\end{eqnarray}
Writing the continuity equations in each layer leads to 
\begin{eqnarray}
\phi^*_{x^*x^*}+\phi^*_{z^*z^*} &= & 0 \quad \mbox{for} \;\;  -h<z^*<\eta^*(x^*,t^*), \label{la1}\\
\phi^{*'}_{x^*x^*}+\phi^{*'}_{z^*z^*} & = & 0  \quad \mbox{for} \;\; \eta^*(x^*,t^*)<z^*<h'. \label{la2}
\end{eqnarray}
The boundary of the system $\{\Omega_{t^*},\Omega'_{t^*}\}$ has two parts: the flat bottom $z^*=-h$ and the flat roof $z^*=h'$.
The impermeability conditions along these rigid boundaries give
\begin{eqnarray}
\phi^*_{z^*} & = & 0 \quad \mbox{at} \;\; z^*=-h, \label{fo1}\\
\phi^{*'}_{z^*} & = & 0 \quad \mbox{at} \;\; z^*=h'. \label{fo2}
\end{eqnarray}
The kinematic conditions along the interface, namely $D(\eta^*-z^*)/Dt^*=0$, give  
\begin{eqnarray}
\eta^*_{t^*}=\phi^*_{z^*}-\phi^*_x\eta^*_x \quad \mbox{at} \;\; z^*=\eta^* (x^*,t^*),  \label{ci1}\\
\eta^*_{t^*}=\phi^{*'}_{z^*}-\phi^{*'}_x\eta^*_x \quad \mbox{at} \;\; z^*=\eta^* (x^*,t^*). \label{ci2}
\end{eqnarray}
The dynamic boundary condition imposed on the interface, namely the continuity of pressure since surface tension effects
are neglected, gives
\begin{equation}\label{in}
\rho \left(\frac{\partial\phi^*}{\partial t^*}+\frac{1}{2}|\nabla\phi^*|^2+gz^*\right)
=\rho'\left(\frac{\partial\phi^{*'}}{\partial t^*}+\frac{1}{2}|\nabla\phi^{*'}|^2+gz^*\right) \;\; \mbox{at} \;\; z^*=\eta^* (x^*,t^*),
\end{equation}
where $g$ is the acceleration due to gravity.
The system of seven equations (\ref{la1})--(\ref{in}) represents the starting model for the study of
wave propagation at the interface between two fluids. Combined with initial conditions or periodicity conditions, it 
is the classical interfacial wave problem, which has been
studied for more than a century. A nice feature of this formulation is that the pressures in both layers have been removed. 
In some cases, it is advantageous to keep the pressures in the equations. For example, Bridges \& Donaldson \cite{BD}
in their study of the criticality of two-layer flows provide an appendix on the inclusion of the lid pressure in the
calculation of uniform flows. In the next sections, we will derive simplified models based on certain additional
assumptions on wave amplitude, wavelength and fluid depth.

\section{System of three equations in the limit of long, weakly dispersive waves}

The derivation follows closely that of \cite{BCS} for a single layer.
Let us now consider waves whose typical amplitude, $A$, is small compared to the depth of the bottom layer $h$,
and whose typical wavelength, $\ell$, is large compared to the depth of the bottom layer\footnote{There is some arbitrariness
in this choice since there are two fluid depths in the problem. We could have also chosen the depth of the top layer
as reference depth. In fact, we implicitly make the assumption that the ratio of liquid depths is neither too small nor too large,
without going into mathematical details. Models valid for arbitrary depth ratio have been derived for example by
Choi \& Camassa \cite{CC}.}. Let us define the three following
dimensionless numbers, with their characteristic magnitude:
$$\alpha=\frac{A}{h} \ll 1, \quad \beta=\frac{h^2}{\ell^2} \ll 1, \quad 
S=\frac{\alpha}{\beta} = \frac{A\ell^2}{h^3} \approx 1. $$
Here $S$ is the Stokes number. Let us also introduce the dimensionless density ratio $r$ as well as the depth ratio $H$:
$$ r = \frac{\rho'}{\rho}, \quad H = \frac{h'}{h}. $$
Obviously $r$ takes values between 0 and 1, the case $r=0$ corresponding to water waves\footnote{In a recent paper, Kataoka
\cite{TK} showed that when $H$ is near unity, the stability of solitary waves changes drastically for small density ratios
$r$. Therefore one must be careful in evaluating the stability of air-water solitary waves. In other words, there may be
differences between $r=0$ and the true value $r=0.0013$.} while the case $r \approx 1$
corresponds to two fluids with almost the same density such as an upper, warmer layer extending down to the interface
with a colder, more saline layer. The depth ratio takes theoretical values between 0 and $\infty$ but as said above
values $H \ll 1$ or $H \gg 1$ should be avoided in the framework of our weakly nonlinear analysis.

The procedure is most transparent when working with the variables scaled in such a way that the dependent quantities appearing
in the problem are all of order one, while the assumptions about small amplitude and long wavelength appear explicitly
connected with small parameters in the equations of motion. Such consideration leads to the scaled, dimensionless variables
$$x^*=\ell {x}, \quad z^*=h({z}-1), \quad \eta^*=A{\eta}, \quad t^*=\ell {t}/c_0, \quad 
\phi^*=gA\ell {\phi}/c_0, \quad \phi^{*'}=gA\ell {\phi'}/c_0, $$
where $c_0=\sqrt{gh}$. The speed $c_0$, which represents
the long wave speed in the limit $r \to 0$, is not necessarily the most natural choice for interfacial waves. 
The natural choice would be to take
$$ c_0 = \sqrt{gh} \sqrt{\frac{1-r}{1+r/H}}, $$
which is the speed of long waves in the configuration shown in Figure \ref{sketch_wave}. It does not matter for the
asymptotic expansions to be performed later.
%\begin{figure}
%\begin{center}
% \includegraphics[width=6cm,height=5cm]{eq3_1.eps}
% \includegraphics[width=6cm,height=5cm]{eq3_2.eps}
%\end{center}
%\caption{Dimensionless variables vs physical variables}
%\end{figure}

In these new variables, the set of equations (\ref{la1})--(\ref{in}) becomes after reordering
\begin{eqnarray}
 \beta\phi_{xx}+\phi_{zz} &=& 0 \quad \mbox{in} \;\; 0<z<1+\alpha\eta, \label{7.1}\\ 
 \phi_z &=& 0 \quad \mbox{on} \;\; z=0, \label{7.2}\\ 
 \eta_t +\alpha\phi_x\eta_x -\frac{1}{\beta}\phi_z &=& 0 \quad \mbox{on} \;\;  z=1+\alpha\eta, \label{7.3}\\
 \beta\phi'_{xx} +\phi'_{zz} &=& 0 \quad \mbox{in} \;\; 1+\alpha\eta<z<1+H,  \label{7.4}\\
 \phi'_z &=& 0 \quad \mbox{on} \;\;  z=1+H,   \label{7.5}\\
 \eta_t +\alpha\phi'_x\eta_x -\frac{1}{\beta}\phi'_z &=& 0 \quad \mbox{on} \;\; z=1+\alpha\eta,  \label{7.6}
\end{eqnarray}
\begin{equation}
\left(\eta +\phi_t +\frac{1}{2} {\alpha}\phi_x^2 +\frac{1}{2}\frac{\alpha}{\beta}\phi_z^2\right)
= r\left(\eta+\phi'_t +\frac{1}{2} {\alpha}\phi_x^{'2}+\frac{1}{2}\frac{\alpha}{\beta}\phi_z^{'2}\right)
\;\; \mbox{on} \;\; z=1+\alpha\eta.  \label{7.7}
\end{equation}

We represent the potential $\phi$ as a formal expansion,
$$\phi((x,z),t)=\sum_{m=0}^\infty f_m(x,t)z^m.$$
Demanding that $\phi$ formally satisfy Laplace's equation (\ref{7.1}) leads to the recurrence relation
\begin{equation}  
(m+2)(m+1)f_{m+2}(x,t)=-\beta (f_m(x,t))_{xx}, \;\; \forall m=0,1,2,\ldots. \label{recu}
\end{equation}
Let $F(x,t)=f_0(x,t)$ denote the velocity potential at the bottom $z=0$ and use (\ref{recu}) repeatedly to obtain
\begin{eqnarray*}
f_{2k}(x,t) & = & \frac{(-1)^k\beta^k}{(2k)!}\frac{\partial^{2k} F(x,t)}{\partial x^{2k}}, 
\;\; \forall k=0,1,2,\ldots, \\
f_{2k+1}(x,t) & = & \frac{(-1)^k\beta^k}{(2k+1)!}\frac{\partial^{2k} f_1(x,t)}{\partial x^{2k}}, 
\;\; \forall k=0,1,2,\ldots .
\end{eqnarray*}
Equation (\ref{7.2}) implies that $f_1(x,t)=0$, so
\begin{equation}
f_{2k+1}(x,t)=0, \;\; \forall k=0,1,2,\ldots,
\end{equation}
and therefore
$$\phi((x,z),t)=\sum_{k=0}^\infty \frac{(-1)^k\beta^k}{(2k)!} \frac{\partial^{2k} F(x,t)}{\partial x^{2k}}z^{2k}.$$
Let $\partial F(x,t)/\partial x=u(x,t)$. Substitute the latter representation into (\ref{7.3}) to obtain
\begin{equation}\label{eq3_1}
\eta_t +u_x+\alpha(u\eta)_x-\frac{1}{6}\beta u_{xxx}-\frac{1}{2}\alpha\beta(\eta u_{xx})_x
      +\frac{1}{120}\beta^2u_{xxxxx}+O(\beta^3)=0.
\end{equation}

Similarly we represent the potential $\phi'$ as a formal expansion,
$$\phi'((x,z),t)=\sum_{m=0}^\infty f'_m(x,t)(1+H-z)^m. $$
Demanding that $\phi'$ formally satisfy Laplace's equation (\ref{7.4}) leads to the recurrence relation
\begin{equation} 
(m+2)(m+1)f'_{m+2}(x,t)=-\beta (f'_m(x,t))_{xx}, \;\; \forall m=0,1,2,\ldots. \label{recup}
\end{equation}
Let $F'(x,t)=f'_0(x,t)$ denote the velocity potential on the roof $z=1+H$ and use (\ref{recup}) repeatedly to obtain
\begin{eqnarray*}
f'_{2k}(x,t) & = & \frac{(-1)^k\beta^k}{(2k)!}\frac{\partial^{2k} F'(x,t)}{\partial x^{2k}},
\;\; \forall k=0,1,2,\ldots, \\
f'_{2k+1}(x,t) & = & \frac{(-1)^k\beta^k}{(2k+1)!}\frac{\partial^{2k} f'_1(x,t)}{\partial x^{2k}},
\;\; \forall k=0,1,2,\ldots .
\end{eqnarray*}
Equation (\ref{7.5}) implies that $f'_1(x,t)=0$, so
\begin{equation}
f'_{2k+1}(x,t)=0, \;\; \forall k=0,1,2,\ldots,
\end{equation}
and therefore
$$\phi'((x,z),t)=\sum_{k=0}^\infty \frac{(-1)^k\beta^k}{(2k)!} \frac{\partial^{2k} F'(x,t)}{\partial x^{2k}}(1+H-z)^{2k}.$$
Let $\partial F'(x,t)/\partial x=u'(x,t)$. Substitute the latter representation into (\ref{7.6}) to obtain
\begin{eqnarray}
\eta_t -Hu'_x+\alpha (u'\eta)_x+\frac{1}{6}\beta H^3u'_{xxx}-\frac{1}{2}\alpha\beta H^2(\eta u'_{xx})_x
 & & \nonumber \\
-\frac{1}{120}\beta^2 H^5u'_{xxxxx}+O(\beta^3) & = & 0. \label{eq3_2}
\end{eqnarray}
It is important at this stage that $H=O(1)$.

Substitute the representations for $\phi$ and $\phi'$ into the dynamic condition $(\ref{7.7})$ to obtain the third equation
\begin{eqnarray*}
(1-r)\eta+F_t-rF'_t-\frac{1}{2}\beta \left(u_{xt}- rH^2u'_{xt}\right) \hspace{2cm} & & \\
-\alpha\beta\eta (u_{xt}+rHu'_{xt})+\frac{1}{24}\beta^2 \left(u_{xxxt} 
-rH^4u'_{xxxt}\right)  \hspace{2cm} & & \\
+\frac{1}{2}\alpha(u^2-\beta uu_{xx})
-\frac{1}{2}\alpha r(u'^2-\beta H^2u'u'_{xx})+\frac{1}{2}\alpha\beta(u^2_x-rH^2u'^2_x)
+O(\beta^3) & = & 0.
\end{eqnarray*}
Differentiating with respect to $x$ yields
\begin{eqnarray}
(1-r)\eta_x+u_t-ru'_t-\frac{1}{2}\beta (u_{xxt}-rH^2u'_{xxt})+\alpha(uu_x-ru'u'_x) & & \nonumber \\
 -\alpha\beta(\eta u_{xt})_x-\alpha\beta rH(\eta u'_{xt})_x 
+\frac{1}{24}\beta^2 (u_{xxxxt}-rH^4u'_{xxxxt}) & & \nonumber \\
-\frac{1}{2}\alpha\beta(uu_{xx}-rH^2u'u'_{xx})_x
+\alpha\beta(u_xu_{xx}-rH^2u'_xu'_{xx})+O(\beta^3)& = & 0. \label{eq3_3}
\end{eqnarray}

The three equations (\ref{eq3_1}),(\ref{eq3_2}) and (\ref{eq3_3}) provide a 
Boussinesq system of equations describing waves at the interface $\eta(x,t)$ between two fluid layers based on the horizontal
velocities $u$ and $u'$ along the bottom and the roof, respectively. It is correct up to second order in $\alpha$, $\beta$. 

One can derive a class of systems which are formally equivalent to the system we just derived. This will be accomplished
by considering changes in the dependent variables and by making use of lower-order relations in higher-order terms. 
Toward this goal, begin by letting $w(x,t)$ be the scaled horizontal velocity corresponding to the physical depth $(1-\theta)h$ below
the unperturbed interface, and $w'(x,t)$ be the scaled horizontal velocity corresponding to the physical depth $(H-\theta')h$
above the unperturbed interface.
The ranges for the parameters $\theta$ and $\theta'$ are $0\leq\theta\leq 1$ and $0\leq\theta'\leq H$. Note that 
$(\theta,\theta')=(0,0)$ leads to $w=u$ and $w'=u'$,
while $(\theta,\theta')=(1,H)$ leads to both velocities evaluated along the interface. A
formal use of Taylor's formula with remainder shows that
\begin{eqnarray*}
w=\phi_x|_{z=\theta}&=&\left(F_x-\frac{1}{2}\beta F_{xxx}\theta^2+\frac{1}{24}\beta^2\theta^4F_{xxxxx}\right)+O(\beta^3)\\
     &=& u-\frac{1}{2}\beta\theta^2 u_{xx}+\frac{1}{24}\beta^2\theta^4u_{xxxx}+O(\beta^3 )
\end{eqnarray*}
as $\beta \to 0$. In Fourier space, the latter relationship may be written as
$$\hat{w}=\left(1+\frac{1}{2}\beta\theta^2k^2+\frac{1}{24}\beta^2\theta^4k^4\right)\hat{u}+O(\beta^3).$$
Inverting the positive Fourier multiplier yields
\begin{eqnarray*}
\hat{u}&=&\left(1+\frac{1}{2}\beta\theta^2k^2+\frac{1}{24}\beta^2\theta^4k^4\right)^{-1}\hat{w}+O(\beta^3)\\
       &=&\left(1-\frac{1}{2}\beta\theta^2k^2+\frac{5}{24}\beta^2\theta^4k^4\right)\hat{w}+O(\beta^3)
\end{eqnarray*} 
as $\beta \to 0$. Thus there appears the relationship
\begin{equation}\label{u}
u=w+\frac{1}{2}\beta\theta^2w_{xx}+\frac{5}{24}\beta^2\theta^4w_{xxxx}+O(\beta^3).
\end{equation}
Similarly
\begin{eqnarray*}
w'=\phi'_x|_{z=1+H-\theta'}&=&\left(F'_x-\frac{1}{2}\beta F'_{xxx}\theta'^2+\frac{1}{24}\beta^2F'_{xxxxx}\theta'^4\right)+O(\beta^3)\\
&=&u'-\frac{1}{2}\beta \theta'^2 u'_{xx}+\frac{1}{24}\beta^2\theta'^4u'_{xxxx}+O(\beta^3)
\end{eqnarray*}
and
$$\hat{w'}=\left(1+\frac{1}{2}\beta\theta'^2 k^2+\frac{1}{24}\beta^2\theta'^4 k^4 \right)\hat{u'}+O(\beta^3).$$
Inverting the positive Fourier multiplier yields
$$
\hat{u'}=\left(1-\frac{1}{2}\beta\theta'^2k^2+\frac{5}{24}\beta^2\theta'^4 k^4\right)\hat{w'}+O(\beta^3)
$$
and thus the relationship
\begin{equation}\label{uu}
u'=w'+\frac{1}{2}\beta\theta'^2w'_{xx}+\frac{5}{24}\beta^2\theta'^4w'_{xxxx}+O(\beta^3).
\end{equation}
Substitute the expressions (\ref{u}) and (\ref{uu}) for $u$ and $u'$ into (\ref{eq3_1}) and (\ref{eq3_2}), respectively, to obtain
\begin{eqnarray}
\eta_t + w_x+\alpha (w\eta)_x+\frac{1}{2}\beta\left(\theta^2-\frac{1}{3}\right) w_{xxx} & & \nonumber \\
+\frac{1}{2}\alpha\beta(\theta^2-1)(\eta w_{xx})_x 
+\frac{5}{24}\beta^2\left(\theta^2-\frac{1}{5}\right)^2 w_{xxxxx}+O(\beta^3) & = & 0 \nonumber \\ & & \label{eq1u'} \\
\eta_t-Hw'_x+\alpha (w'\eta)_x -\frac{1}{2}\beta H\left(\theta'^2-\frac{1}{3}H^2\right)w'_{xxx} & & \nonumber \\
+\frac{1}{2}\alpha\beta \left(\theta'^2-H^2\right)(\eta w'_{xx})_x 
-\frac{5}{24}\beta^2H\left(\theta'^2
-\frac{1}{5}H^2\right)^2w'_{xxxxx}+O(\beta^3) & = & 0. \nonumber \\ & & \label{eq2u'}
\end{eqnarray}
Substitute the expressions (\ref{u}) and (\ref{uu}) for $u$ and $u'$ into (\ref{eq3_3}) to obtain
\begin{eqnarray}\label{eq3u'}
(1-r)\eta_x+ w_t-rw'_t 
+\frac{1}{2}\beta\left[(\theta^2-1)w-r(\theta'^2-H^2)w'\right]_{xxt} +\alpha(ww_x-rw'w'_x) & & \nonumber \\ 
+\frac{1}{24}\beta^2\left[(\theta^2-1)(5\theta^2-1)w_{xxxxt}-r(\theta'^2-H^2)(5\theta'^2-H^2)w'_{xxxxt}\right]
& & \nonumber \\
-\alpha\beta\left[(\eta w_{xt})_x + rH(\eta w'_{xt})_x\right]  
+\frac{1}{2}\alpha\beta\left[(\theta^2-1)ww_{xxx}-r(\theta'^2-H^2)w'w'_{xxx}\right] & & \nonumber \\
+\frac{1}{2}\alpha\beta\left[(\theta^2+1)w_xw_{xx} - r(\theta'^2+H^2)w'_xw'_{xx}\right] +O(\beta^3)&=&0. \nonumber \\ & & 
\end{eqnarray}

The system of three equations (\ref{eq1u'})--(\ref{eq3u'}) is formally
equivalent to the previous system but it allows one to choose the fluid levels $\theta$ and $\theta'$ as reference for
the horizontal velocities. Among all these systems that model the same physical problem one can select those with the
best dispersion relations.
Neglecting terms of $O(\alpha^2,\beta^2,\alpha\beta)$, the system (\ref{eq1u'})--(\ref{eq3u'}) reduces to
\begin{equation}\label{bo}
%    \left\{
    \begin{array}{rcc}
\eta_t + w_x+\alpha (w\eta)_x+\frac{1}{2}\beta(\theta^2-\frac{1}{3}) w_{xxx} & = & 0 \\
\eta_t-Hw'_x+\alpha (w'\eta)_x -\frac{1}{2}\beta H(\theta'^2-\frac{1}{3}H^2)w'_{xxx} & = & 0 \\
(1-r)\eta_x+ w_t-rw'_t+\frac{1}{2}\beta[(\theta^2-1)w-r(\theta'^2-H^2)w']_{xxt}+\alpha(ww_x -rw'w'_x)
& = & 0
     \end{array}
%    \right.
\end{equation}

\section{System of two equations}

The systems obtained in the previous section are not appropriate for numerical computations. One would like 
to obtain a system of two evolution equations for the
variables $\eta$ and $W=w-rw'$. In fact, Benjamin and Bridges \cite{BB} (see also \cite{DB94,CGK,ADK} ) formulated 
the interfacial wave problem using Hamiltonian formalism and showed that 
the canonical variables for interfacial waves are $\eta^*(x^*,t^*)$ and $\rho\phi^*(x^*,\eta^*,t^*)-\rho'\phi^{*'}(x^*,\eta^*,t^*)$.

At leading order, the first two equations of system (\ref{bo}) give
$$    \left\{
    \begin{array}{ll}
\eta_t + w_x=0,\\
\eta_t -Hw'_x=0.
     \end{array}
    \right. 
$$
Assuming the fluids to be at rest as $x\rightarrow\infty $, one has $w= -Hw'$. Therefore
\begin{equation} \label{wWw}
    w=\frac{H}{r+H}W+O(\beta), \quad w'=\frac{-1}{r+H}W+O(\beta).
\end{equation}

Adding $H$ times the first equation to $r$ times the second equation of system (\ref{bo}) yields
\begin{equation}\label{somme}
\begin{array}{rcl}
(r+H)\eta_t +H(w-rw')_x+\alpha [(Hw+rw')\eta]_x & & \\
+\frac{H}{2}\beta\Big[(\theta^2-\frac{1}{3})w_{xxx}
-r\Big(\theta'^2-\frac{1}{3}H^2\Big)w'_{xxx}\Big] & = & 0.
\end{array}
\end{equation}
Using (\ref{wWw}) and neglecting higher-order terms, one obtains
$$
\eta_t=-\frac{H}{r+H}W_x-\alpha\frac{H^2-r}{(r+H)^2}(W\eta)_x-\beta\left(\frac{1}{2}\frac{H^2S}{(r+H)^2}
+\frac{1}{3}\frac{H^2(1+rH)}{(r+H)^2}\right)W_{xxx},$$
where
$$ S=(\theta^2-1)+\frac{r}{H}\left(\theta'^2-H^2\right). $$
In the third equation of system (\ref{bo}), the term with the $xxt-$derivatives can be written as
$$
\frac{1}{2}\beta\left[ (\theta^2-1)w_{xxt}- r(\theta'^2-H^2)w'_{xxt} \right]
=\frac{1}{2}\beta\frac{HS}{r+H}W_{xxt}.
$$
The quadratic terms of the third equation of system (\ref{bo}) can be written as
$$
\alpha (ww_x - rw'w'_x) = \alpha\frac{H^2-r}{(r+H)^2}WW_x.
$$
Then the third equation of system (\ref{bo}) becomes
$$W_t=-(1-r)\eta_x-\frac{1}{2}\beta\frac{HS}{r+H}W_{xxt}-\alpha\frac{H^2-r}{(r+H)^2} WW_x.$$
The final system of two equations for interfacial waves in the limit of long, weakly dispersive waves, can be written 
in terms of the horizontal velocities at arbitrary fluid levels as (in dimensionless form)
\begin{equation}
\left\{
\begin{array}{ll}
\eta_t=-\frac{H}{r+H}W_x-\alpha\frac{H^2-r}{(r+H)^2}(W\eta)_x-\beta\left(\frac{1}{2}\frac{H^2S}{(r+H)^2}
+\frac{1}{3}\frac{H^2(1+rH)}{(r+H)^2}\right)W_{xxx}\\
 W_t=-(1-r)\eta_x-\alpha\frac{H^2-r}{(r+H)^2} WW_x-\frac{1}{2}\beta\frac{HS}{r+H}W_{xxt},
\end{array}
\right.
\end{equation}
or as (in physical variables)
\begin{equation}\label{2eqphy}
   \left\{
          \begin{array}{ll}
            \eta^*_{t^*}=-h d_1 W^*_{x^*}-d_4(W^*\eta^*)_{x^*}-h^3 d_2 W^*_{x^*x^*x^*}, \\ 
               W^*_{t^*}=-g(1-r)\eta^*_{x^*}-d_4W^*W^*_{x^*}-h^2 d_3 W^*_{x^*x^*t^*},
          \end{array}
    \right.
\end{equation}
where
\begin{equation} \label{dcoef} 
d_1=\frac{H}{r+H}, \quad d_2=\frac{H^2}{2(r+H)^2}\left(S+\frac{2}{3}(1+rH)\right), 
\quad d_3=\frac{1}{2}Sd_1, \quad d_4=\frac{H^2-r}{(r+H)^2}. 
\end{equation}
Notice that Choi \& Camassa \cite{CC} also derived a system of two equations (see their equations (3.33) and (3.34)),
but it is different from ours. In particular, their coefficient $d_2$ is equal to $0$, and their equation for $W_t$
possesses an extra quadratic term $\eta\eta_x$. The reason is that their `$W$' is the mean horizontal velocity through the
upper layer. The value of $S$ which best approximates the Choi \& Camassa equations is $S=-\frac{2}{3}(1+rH)$. Indeed the
coefficient $d_2$ then vanishes. This particular value for $S$ can be explained as follows. The leading order correction
to the horizontal velocity is given by
$$ w(z) = u - \undemi \beta z^2 u_{xx}. $$
The value of $z$, say $z=\theta$, for which the mean velocity
$$ \overline{w} = \int_0^1 w(z) \, dz $$ 
is equal to $w(\theta)$ is given by $\theta = 1/\sqrt{3}$. Similarly, one finds $\theta'= (1/\sqrt{3})H$ for the upper layer.
Therefore $S=-\frac{2}{3}(1+rH)$.

Recall that the scaling that led to our Boussinesq system is given by
$$\frac{x^*}{h}=\frac{x}{\sqrt\beta}, \quad \frac{t^*}{h/c_0} = \frac{t}{\sqrt\beta}, \quad \frac{\eta^*}{h} 
= \alpha \eta, \quad \frac{W^*}{gh/c_0} = \alpha W, $$
with $c_0=\sqrt{gh}$, $\alpha \ll 1$, $\beta \ll 1$ and $\alpha = O(\beta)$.
Linearizing system (\ref{2eqphy}) and looking for solutions $(\eta^*,W^*)$ proportional to $\exp(ikx^*-i\omega t^*)$ leads 
to the dispersion relation
$$\frac{\omega^2}{k^2}=\frac{gh(1-r)(d_1-d_2 k^2h^2)}{1- d_3 k^2h^2}.$$ Plots of the dispersion relation are given 
in the next section. Since $0\leq \theta\leq 1$ and $0\leq\theta'\leq H$, the definition of $S$ implies that 
$$-1-rH\leq S\leq 0.$$
It follows that $d_3\leq 0$ and therefore the denominator $1-d_3h^2k^2$ is positive.
In order to have well-posedness (that is $\omega^2/k^2$ positive for all values of $k$), 
$d_2$ must be negative, which is the case if $S\leq -\frac{2}{3}(1+rH)$.
Finally the condition we want to impose on $S$ is that
\begin{equation}\label{contrainS}
-(1+rH)\leq S\leq -\frac{2}{3}(1+rH).
\end{equation}
It is satisfied if one takes the horizontal velocities on the bottom and on the roof ($S=-(1+rH)$) or the mean
horizontal velocities in the bottom and upper layers ($S=-\frac{2}{3}(1+rH)$), but it is not if one takes the
horizontal velocities along the interface ($S=0$).

\section{The numerical scheme and numerical solutions}

\begin{figure} 
\begin{center}
\begin{tabular}{c c}
(a) & (b)\\
\includegraphics[width=6cm,height=5cm]{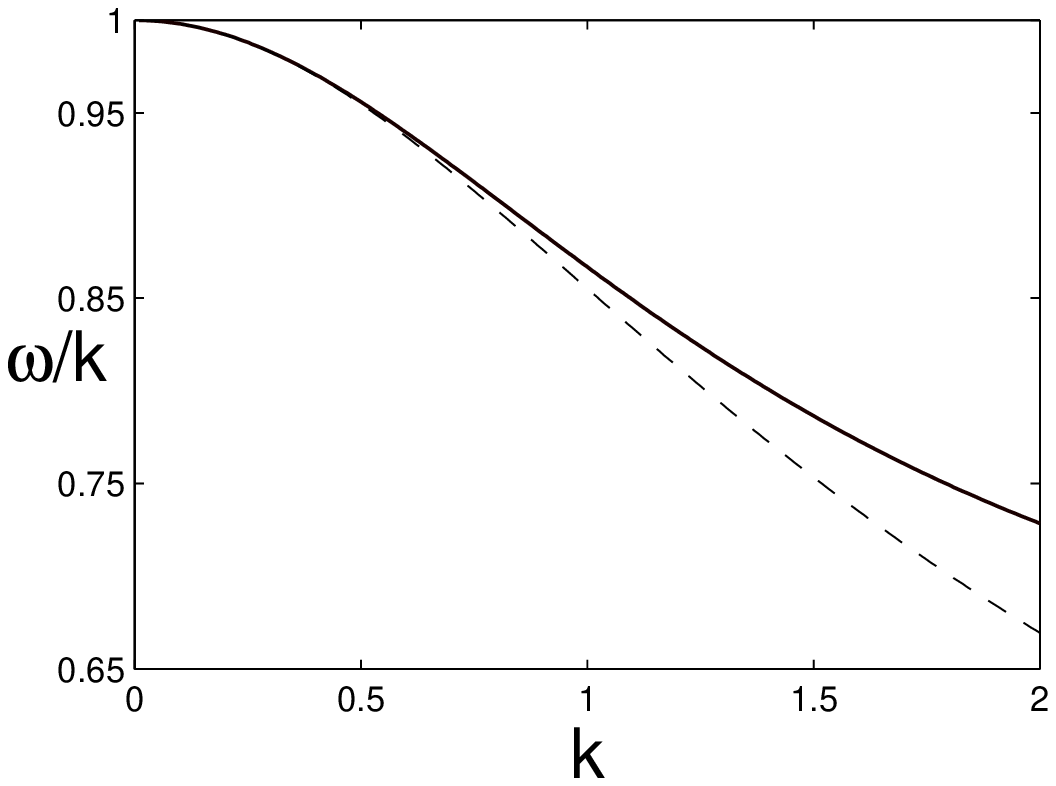} & \includegraphics[width=6cm,height=5cm]{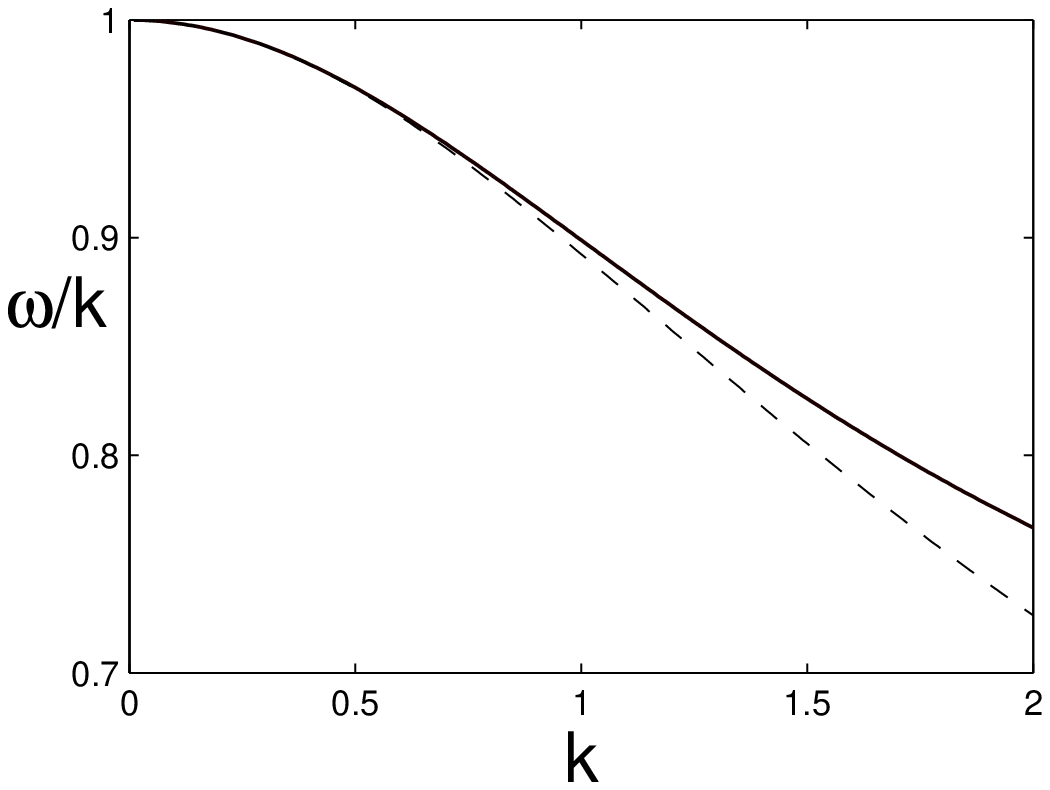}
\end{tabular}
\end{center}
\caption{Dispersion relation (\ref{rel_dispersion}) for the Boussinesq system (\ref{2eq_clean_AB}) with $S=-1-rH$,  $r=0.9$: 
(a) $H=1.2$, (b) $H=0.8$. The dashed curves represent the dispersion relation for the linearized interfacial wave equations,
without the long wave assumption (see for example \cite{LD}).}
\label{plot_dispersion}
\end{figure}

In order to integrate numerically the Boussinesq system (\ref{2eqphy}), we introduce a slightly different change of variables,
where the stars still denote the physical variables and no new notation is introduced for the dimensionless
variables: 
$$x=\frac{x^*}{h}, \;\; \eta=\frac{\eta^*}{h}, \;\; t=\frac{c}{h}t^*, \;\; W=\frac{W^*}{c}, \quad \mbox{with}  \;\;
c^2=gh\frac{H(1-r)}{r+H}.$$
The system (\ref{2eqphy}) becomes
\begin{equation}\label{2eq_clean_AB}
   \left\{
          \begin{array}{ll}
            \eta_t=-d_1W_x-d_4(W\eta)_x-d_2W_{xxx} \\ 
               W_t=-\displaystyle{\frac{1}{d_1}}\eta_x-d_4WW_x-d_3W_{xxt}
          \end{array}
    \right., 
\end{equation}
with dispersion relation
\begin{equation}\label{rel_dispersion}
\frac{\omega^2}{k^2}=\frac{d_1-d_2k^2}{d_1(1-d_3k^2)}.
\end{equation}
As $k \to 0$, $\omega/k \to 1$. As $k \to \infty$, 
$$ \frac{\omega^2}{k^2} \to \frac{d_2}{d_1 d_3} = 1 + \frac{2(1+rH)}{3S}. $$
Typical plots of the dispersion relation (\ref{rel_dispersion}) are given in Figure \ref{plot_dispersion}.
Comparisons between the approximate and the exact dispersion relations, given by
$$ \frac{\omega^2}{k^2}=\frac{\tanh k \tanh kH}{d_1 k (\tanh kH + r \tanh k)} $$
are also shown. A very good agreement is found for small $k$.
Taking the Fourier transform of the system (\ref{2eq_clean_AB}) gives
$$
   \left\{
          \begin{array}{ll}
            \hat{\eta_t}= (d_2k^2-d_1)ik\hat{W}-d_4 ik\widehat{W\eta} \\
             \hat{W_t}=-\displaystyle{\frac{1}{d_1(1-d_3k^2)}ik\hat{\eta}-\frac{d_4}{2(1-d_3k^2)}ik\widehat{W^2}}
          \end{array}
    \right. .
$$
The system of differential equations is solved by a pseudo-spectral method in space with a number $N$ of Fourier modes on a periodic 
domain of length $L$. 
For most applications, $N=1024$ was found to be sufficient. The time integration is performed using the classical fourth-order
explicit Runge--Kutta scheme. The time step $\Delta t$ was optimized through a trial and
error process and was found to have a dependence in $1/N$. 

Since the main goal is to study the propagation and the collision of solitary waves, we first look for solitary wave
solutions of the system (\ref{2eq_clean_AB}). As opposed to the KdV equation, there are no explicit solitary wave solutions of 
the Boussinesq system that are physically relevant. Therefore we look for an approximate solitary wave solution to 
(\ref{2eq_clean_AB}) as in \cite{BC} (see also \cite{DD} for the existence of solitary wave solutions). The leading-order terms give
$$ \eta_t=-d_1W_x, \quad W_t=-\frac{1}{d_1}\eta_x. $$
A solution representing a right-running wave is 
$$W(x-t)=\frac{1}{d_1}\eta(x-t).$$
Let us look for solutions of system (\ref{2eq_clean_AB}) in the form 
$$W(x,t)=\frac{1}{d_1}[\eta(x,t)+M(x,t)],$$ 
where $M$ is assumed to be small compared to $\eta$ and $W$. Substituting the expression for $W$ into (\ref{2eq_clean_AB}) and 
neglecting higher-order terms yields
\begin{equation}\label{cherM}
   \left\{
          \begin{array}{ll}
            \eta_t=\displaystyle{-\eta_x-M_x-\frac{d_4}{d_1}(\eta^2)_x-\frac{d_2}{d_1}\eta_{xxx}}\\ 
            \eta_t=\displaystyle{-\eta_x-M_t-\frac{1}{2}\frac{d_4}{d_1}(\eta^2)_x-d_3\eta_{xxt}} 
          \end{array}
    \right. .
\end{equation}
Assuming that the solitary wave goes to the right, one has $M_t \approx -M_x$. Therefore
$$M_x=-\frac{1}{4}\frac{d_4}{d_1}(\eta^2)_x-\frac{1}{2}\frac{d_2}{d_1}\eta_{xxx}+\frac{1}{2}d_3\eta_{xxt}. $$
Substituting the expression for $M_x$ into one of the equations of system $(\ref{cherM})$ yields
\begin{equation}\label{cherM1}
\eta_t+\eta_x+\frac{3d_4}{4d_1}(\eta^2)_x+\frac{d_2}{2d_1}\eta_{xxx}+\frac{d_3}{2}\eta_{xxt}=0.
\end{equation} 
This is essentially the model equation that was studied in \cite{BPS}.

Looking for solitary wave solutions of (\ref{cherM1}) in the form 
\begin{equation}\label{onde_sol}
\eta=\eta_0\,{\rm sech}^2[\ka(x+x_0-Vt)]
\end{equation}
leads to two equations for $\ka$ and $V$:
$$
\left\{
     \begin{array}{l}
          -V+1+2(d_2/d_1)\ka^2-2d_3\ka^2 V =0 \\
          d_4\eta_0-4d_2\ka^2+4d_1d_3\ka^2 V  =0 
      \end{array}
\right. .
$$
Solving for $\kappa^2$ and $V$ yields
$$ \ka^2=\frac{d_4\eta_0}{4\left(d_2-d_1d_3-\frac{1}{2}d_3d_4\eta_0\right)}, \quad
           V=1+\frac{d_4\eta_0}{2d_1},
$$
and, assuming $M(\pm\infty)=0$, one obtains explicitly the following expression for $M$:
$$
M = -\frac{d_4}{4d_1}\eta^2-\frac{d_2}{2d_1}\eta_{xx}+\frac{d_3}{2}\eta_{xt}.
$$
For a given pair $(r,H)$, one must only consider values of $\eta_0$ which are such that $\ka^2 > 0$.
%\begin{equation}\label{condi}
%\frac{I\eta_0}{6\Big(K-L(1+\frac{2I\eta_0}{3})\Big)}>0.
%\end{equation}
In addition one has the condition (\ref{contrainS}) on $S$.
The sign of $d_4$ depends on the relation between $H^2$ and $r$. Let us assume first that $H^2>r$ so that $d_4>0$. 
In order for the condition $\ka^2>0$ to be satisfied,
one needs $$\eta_0 \left(d_2-d_1d_3-\frac{1}{2}d_3d_4\eta_0\right) >0. $$
The values of $\eta_0$ for which the left-hand side of the inequality vanishes are
$$
\eta_{01}=0, \quad
\eta_{02}=\frac{4H(r+H)(1+rH)}{3(H^2-r)S}.
$$
Since $S<0$, $\eta_{02}<0$ and therefore $\eta_{02}<\eta_{01}$. The coefficient of $\eta_{0}^2$ in the inequality is positive.
Consequently one must have
$$
\eta_0>\eta_{01}=0 \quad \mbox{or} \quad
\eta_0<\eta_{02}=\frac{4H(r+H)(1+rH)}{3(H^2-r)S}.
$$
This second branch is not acceptable since
$$
\frac{4H(r+H)(1+rH)}{3(H^2-r)} > 1+rH > -S > 0.
$$
Therefore $$\frac{4H(r+H)(1+rH)}{3(H^2-r)S}<-1,$$ which gives an amplitude larger than the depth!

Similarly, when $H^2<r$ one finds a second branch which is not acceptable.
The summary of acceptable values for $\eta_0$ is given in the table 
\begin{center}
\begin{tabular}{|c|c|}		                                                  
\hline
$H^2-r>0$      & $0<\eta_0<H$ 	\\					          
\hline
$H^2-r<0$      & $-1<\eta_0<0$  \\                  
\hline
\end{tabular}
\end{center}
For a ``thick'' upper layer ($H^2>r$), the solitary waves are of elevation, while they are of depression for a ``thick'' bottom layer
($H^2<r$). The weakly nonlinear theory developed in the present section does not provide any bounds on the amplitude of the solitary 
waves. We have added a physical constraint based on the fact that both layers are bounded by flat solid boundaries. It
is well-known in the framework of the full interfacial wave equations (see for example \cite{LD}) that the rigid top and bottom 
provide natural bounds on the solitary wave amplitudes. As the speed increases, the wave amplitude reaches a limit. In
the next section, we extend our weakly nonlinear analysis to cubic terms so that this effect can be incorporated. 

Once the approximate solitary wave (\ref{onde_sol}) has been obtained, it is possible to make it cleaner by iterative
filtering. This technique has been used by several authors, including \cite{BC,BDM}, and is explained in Appendix A. In
order to study run-ups and phase shifts during collision of solitary waves, it is important to use clean solitary waves for
the initial conditions. On the other hand, in order to show only the qualitative behavior, it is not necessary. Therefore results 
in this Section are given for non-filtered solitary waves. Some results with filtered waves are described in Appendix A.

\begin{figure}
\begin{center}
\begin{tabular}{c c}
(a) $t=0$ & (b) $t=50$ \\
\includegraphics[width=6cm]{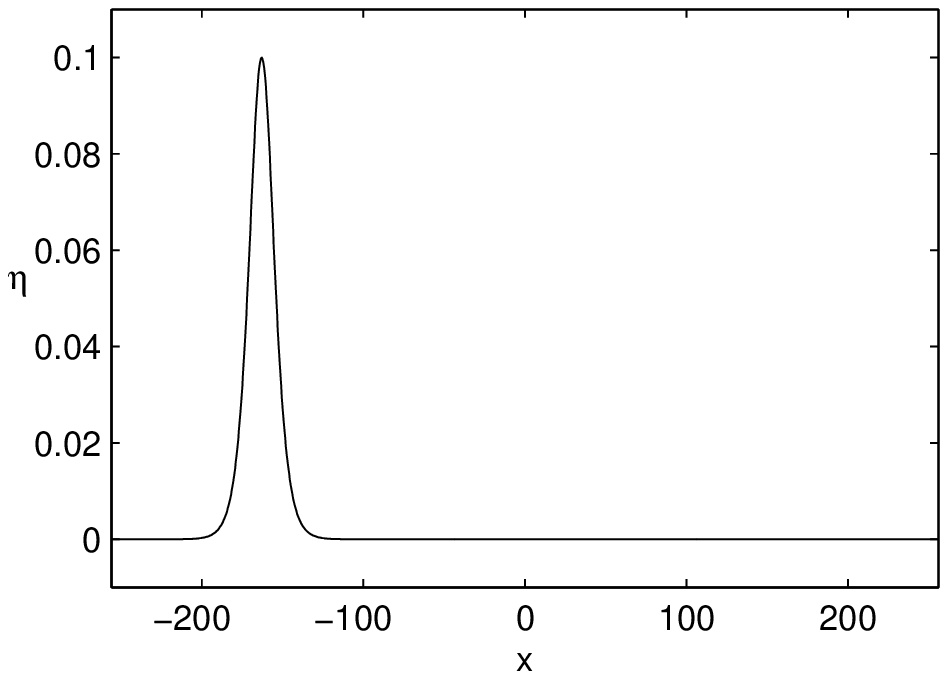}& \includegraphics[width=6cm]{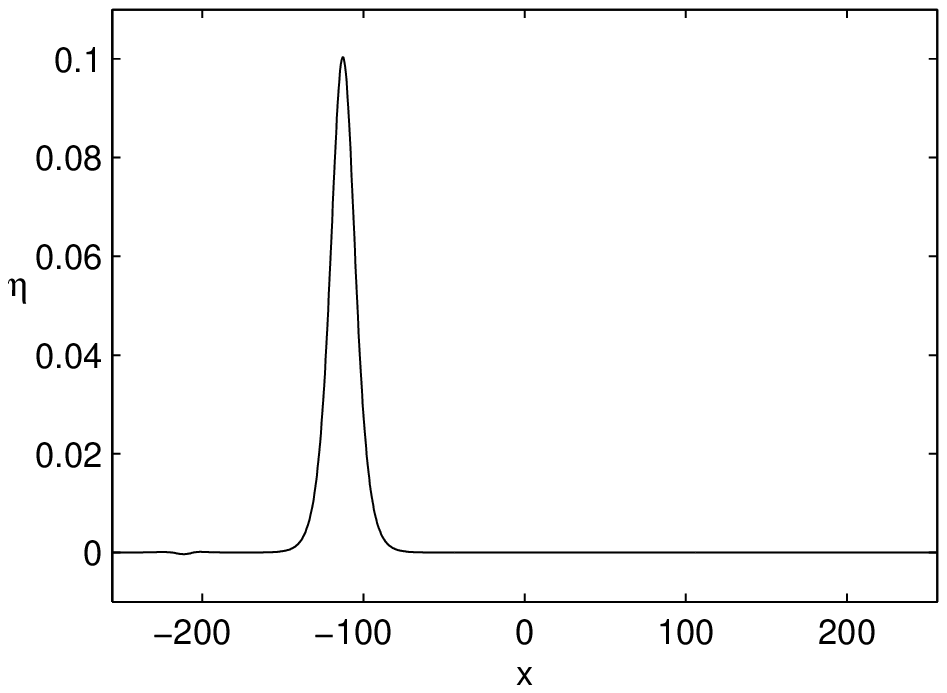} \\
(c) $t=160$ & (d) $t=220$ \\
\includegraphics[width=6cm]{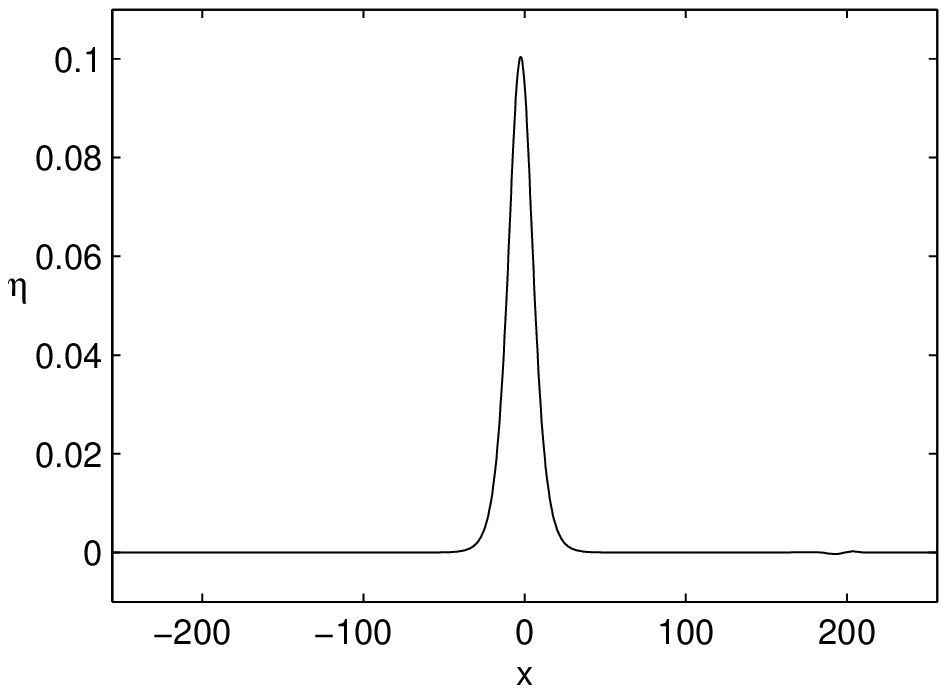} & \includegraphics[width=6cm]{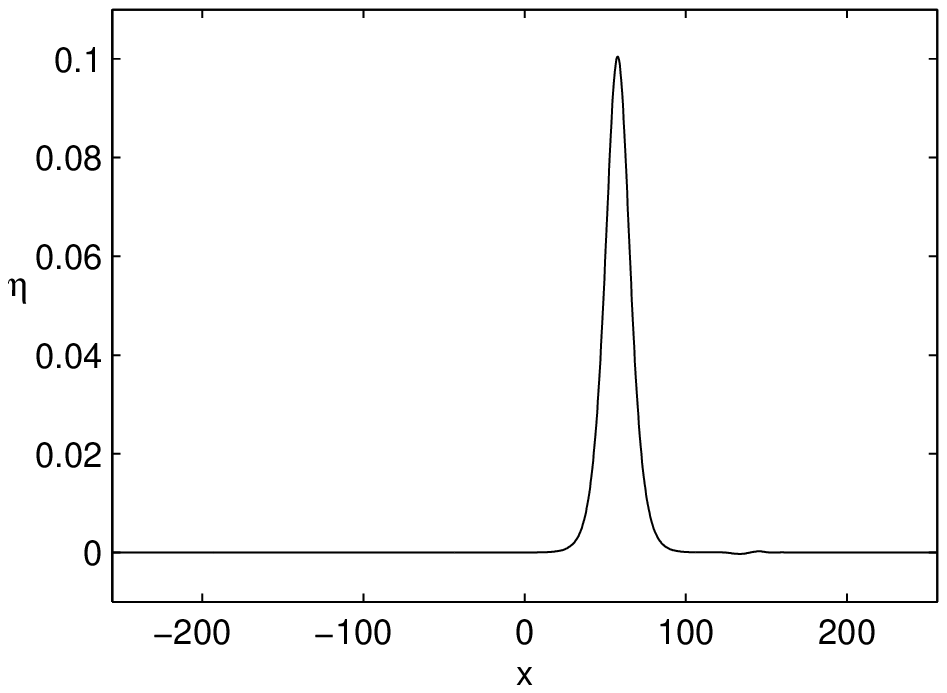}\\
(e) $t=380$ & (f) evolution in time\\
\includegraphics[width=6cm]{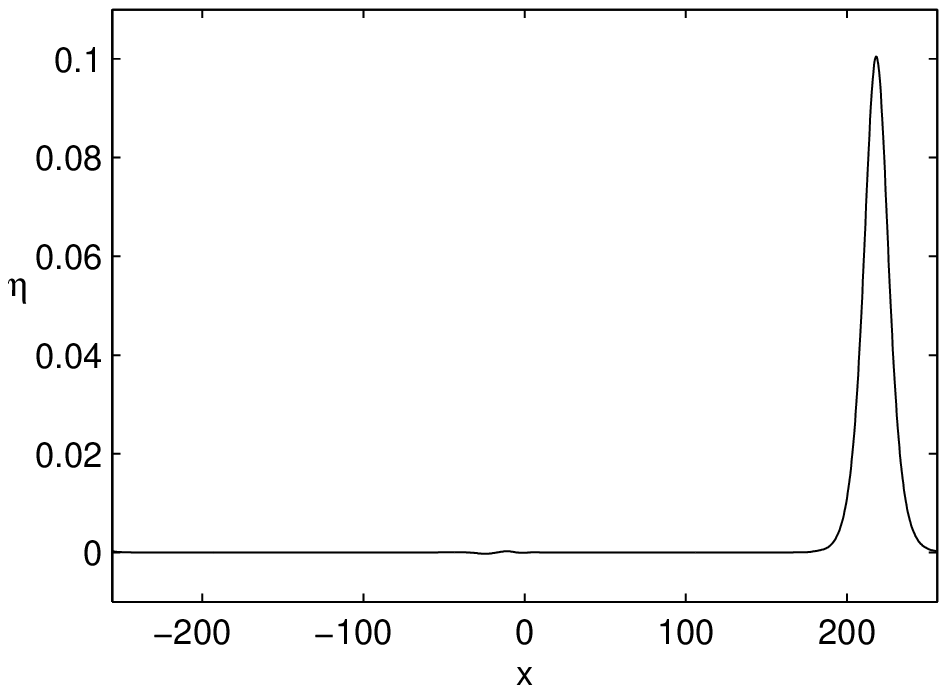} & \includegraphics[width=6cm]{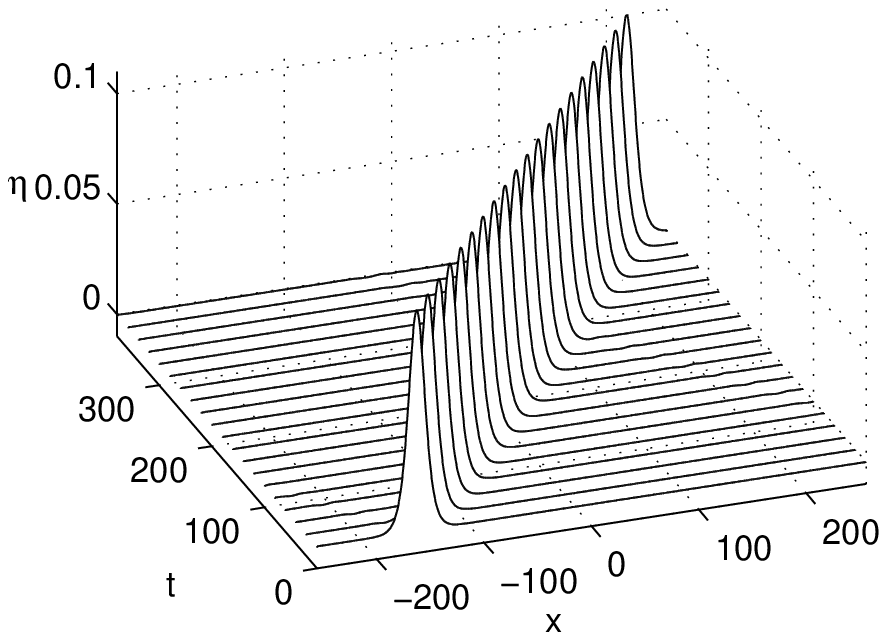}\\
\end{tabular}
\caption{An approximate solitary wave propagating to the right. This is a solution to the system of quadratic Boussinesq equations
(\ref{2eq_clean_AB}), with parameters $H=1.1$, $r=0.9$, $L=512$, $N=1024$, $S=-1-rH$, $\eta_0=0.1$.}
\label{quadratique_general}
\end{center}
\end{figure}
In Figure \ref{quadratique_general}, we show the propagation of an almost perfect right-running solitary wave of
elevation. Even though all computations are performed with dimensionless variables, it is interesting to provide numerical
applications for a configuration that could be realized in the laboratory \cite{MB}. Keeping $r=0.9$ as in the figure, one could take for 
example $h=10$ cm, $h'=11$ cm $(H=1.1)$. The solitary wave amplitude is $1$ cm, its speed $c \approx 23.2$ cm/s, the length of 
the domain $51.2$ m (a bit long!). The plots (b)--(e) would then correspond to snapshots at $t=21.5$ s, $t=68.9$ s, $t=94.8$ s 
and $t=163.7$ s. 

\begin{figure}
\begin{center}
\begin{tabular}{c c}
(a) $t=0$ & (b) $t=40$\\
\includegraphics[width=6cm]{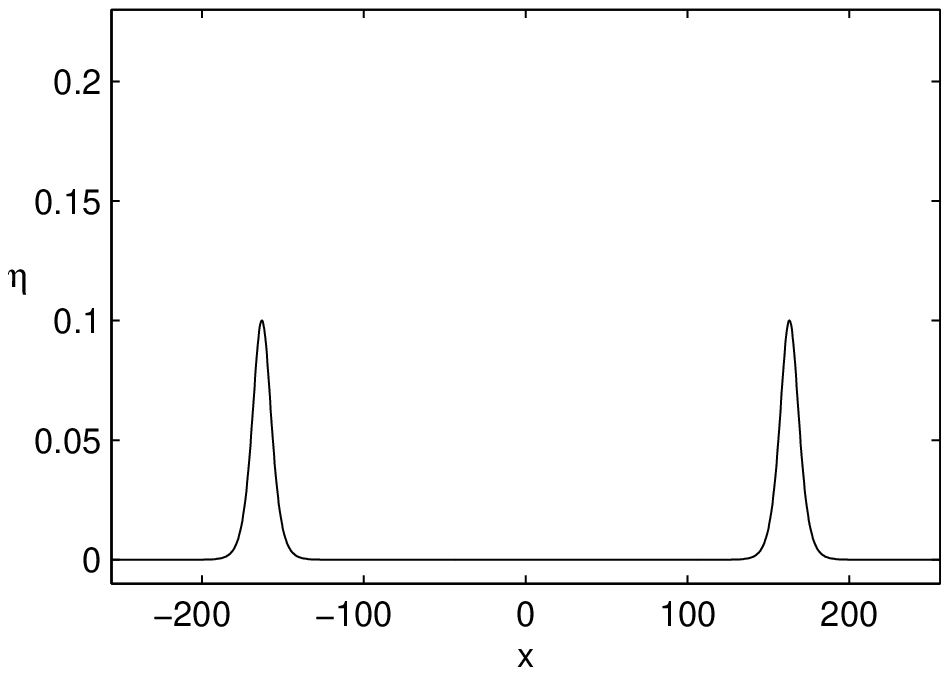}& \includegraphics[width=6cm]{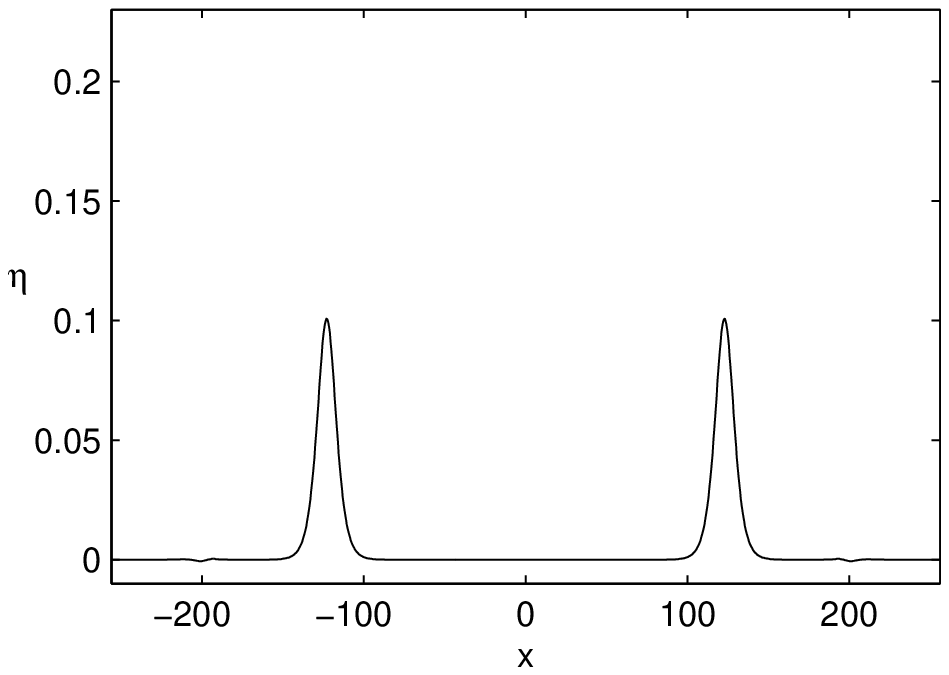} \\
(c) $t=161$ & (d) $t=170$\\
\includegraphics[width=6cm]{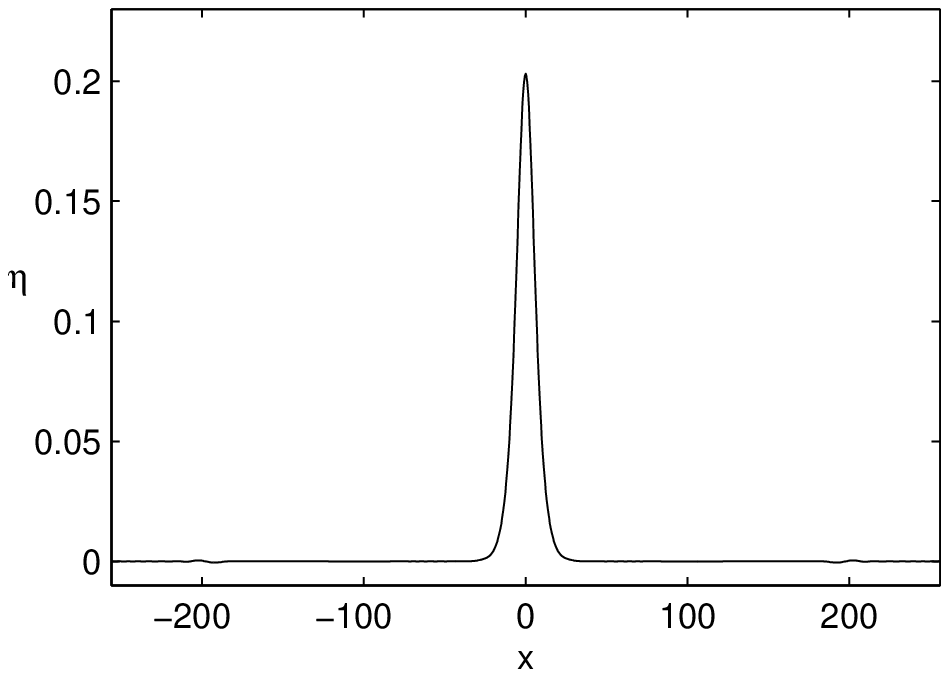} & \includegraphics[width=6cm]{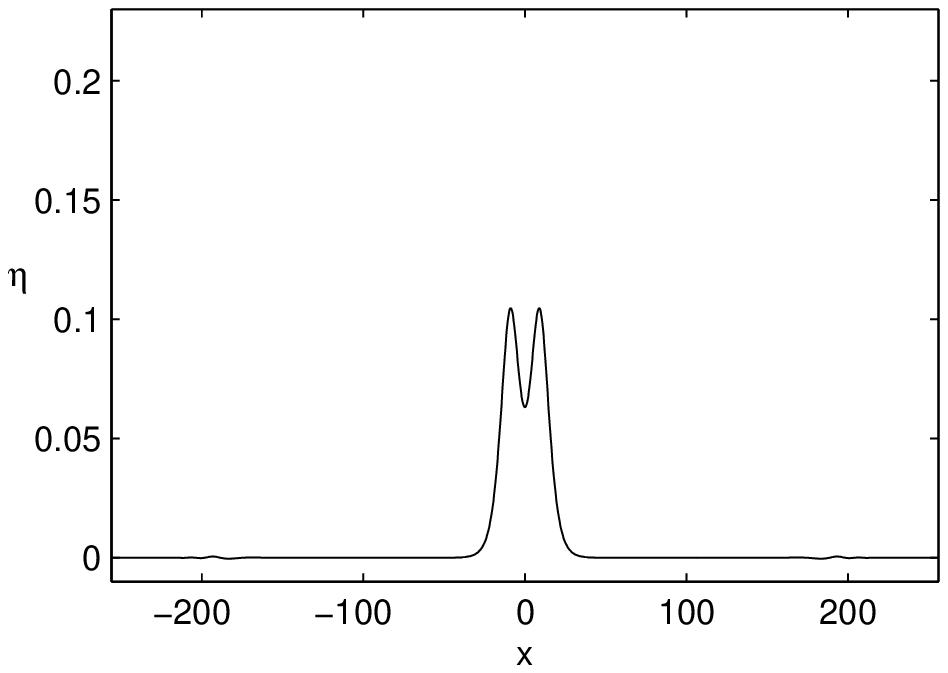}\\
(e) $t=250$ & (f) evolution in time\\
\includegraphics[width=6cm]{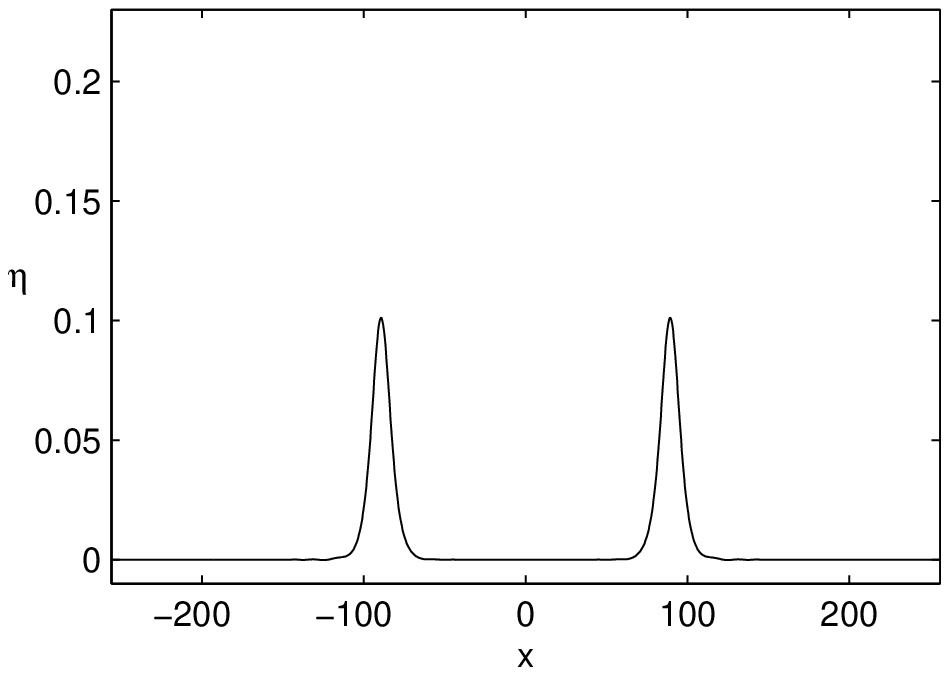} & \includegraphics[width=6cm]{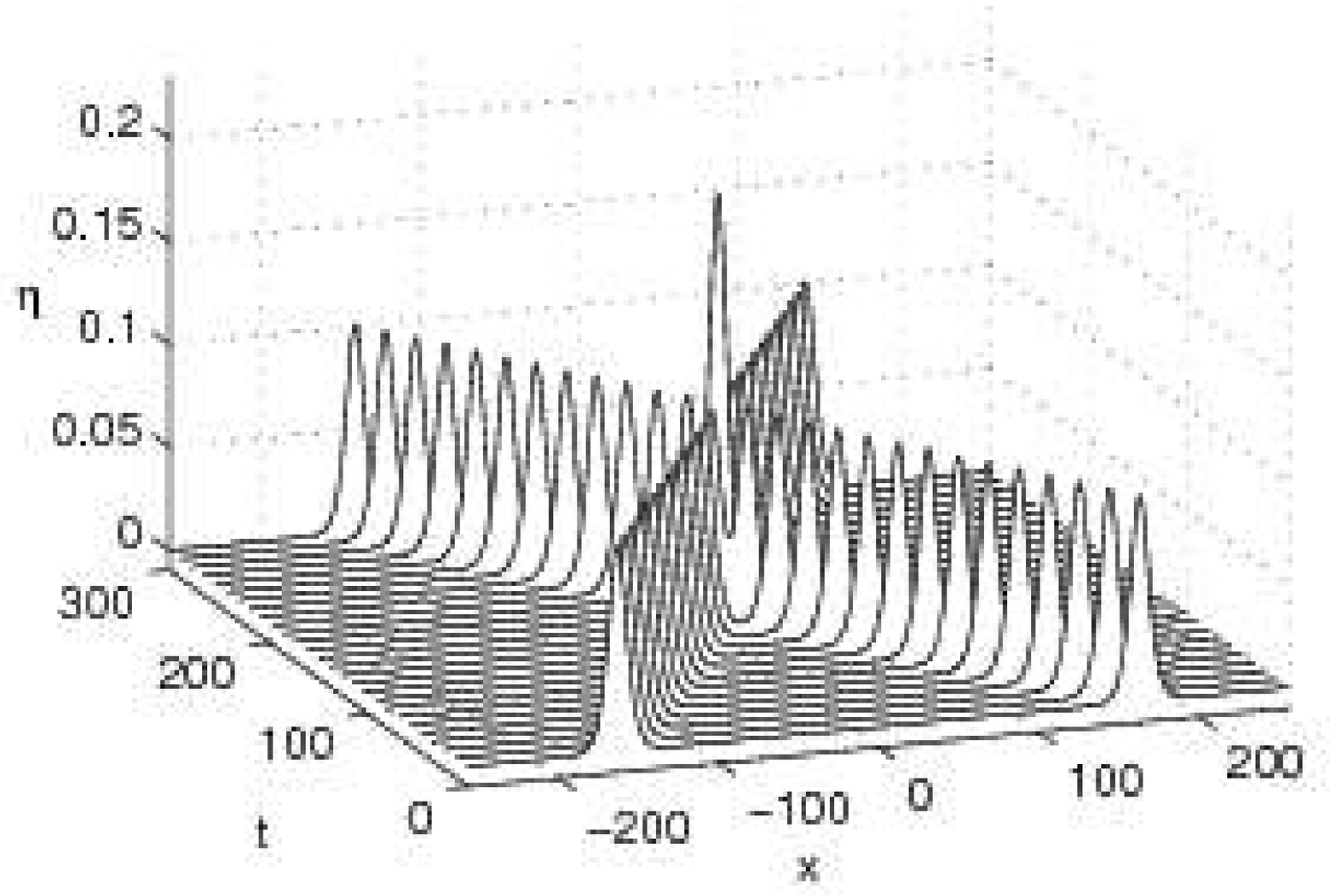}\\
\end{tabular}
\caption{Head-on collision of two approximate solitary waves of elevation of equal size. This is a solution to the system of quadratic 
Boussinesq equations (\ref{2eq_clean_AB}), with parameters $H=1.2$, $r=0.8$, $L=512$, $N=1024$, $S=-1-rH$, 
$\eta_0^{\ell}=\eta_0^r=0.1$, where the superscripts $\ell$ and $r$ stand for left and right respectively.}
\label{quadratique_collision}
\end{center}
\end{figure}
In Figure \ref{quadratique_collision}, we show the head-on collision of two almost perfect solitary waves of elevation of equal 
amplitude moving
in opposite directions. As in the one-layer case, the solution rises to an amplitude slightly larger than the sum of the amplitudes
of the two incident solitary waves (see Appendix A). After the collision, two similar waves emerge and return to the form of
two separated solitary waves. As a result of this collision, the amplitudes of the two resulting solitary waves are slightly
smaller than the incident amplitudes and their centers are slightly retarded from the trajectories of the incoming centers
(see again Appendix A). 
% (see for example \cite{CGHHS} for a description of these features).

\begin{figure}
\begin{center}
\begin{tabular}{c c}
(a) $t=0$ & (b) $t=40$\\
\includegraphics[width=6cm]{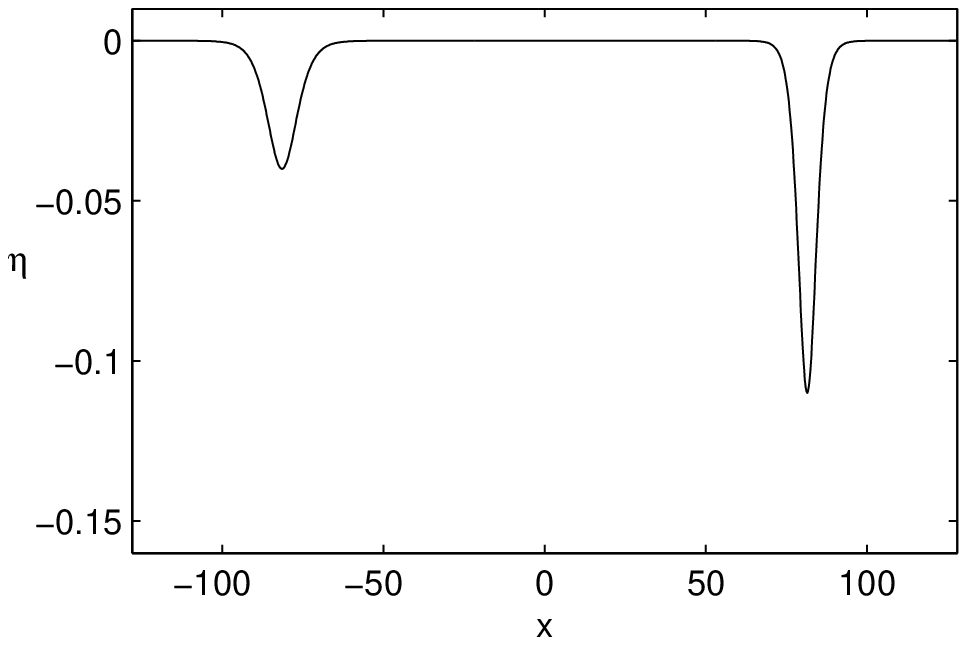}& 
\includegraphics[width=6cm]{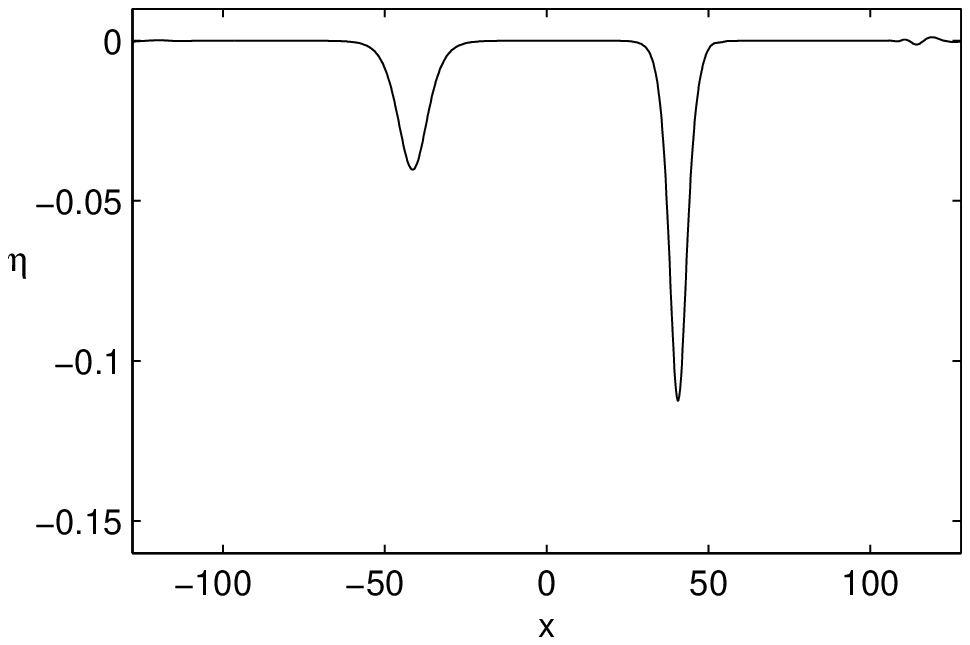} \\
(c) $t=70$ & (d) $t=80$\\
\includegraphics[width=6cm]{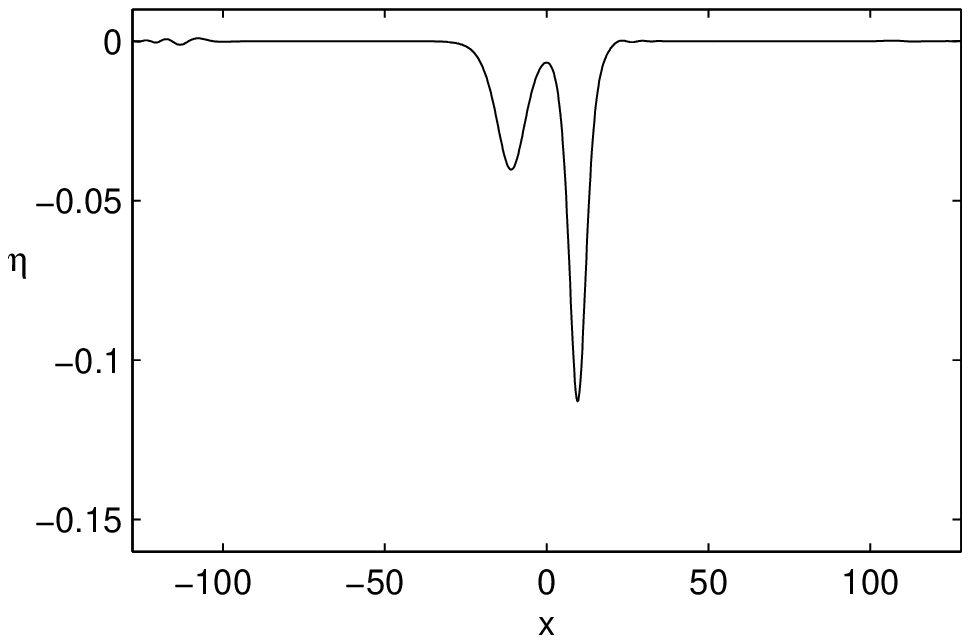} & 
\includegraphics[width=6cm]{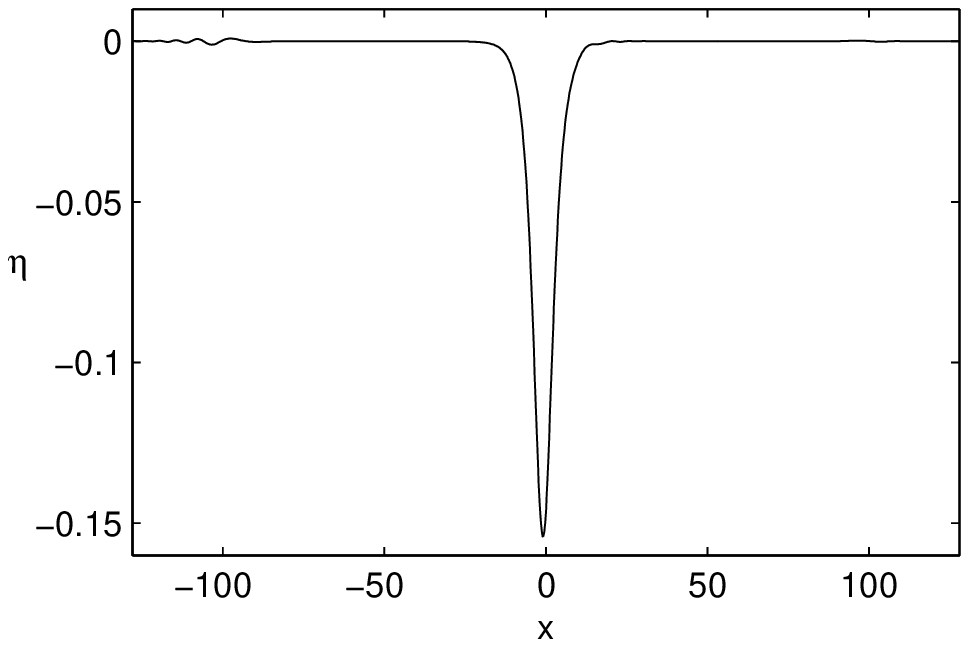}\\
(e) $t=110$ & (f) evolution in time\\
\includegraphics[width=6cm]{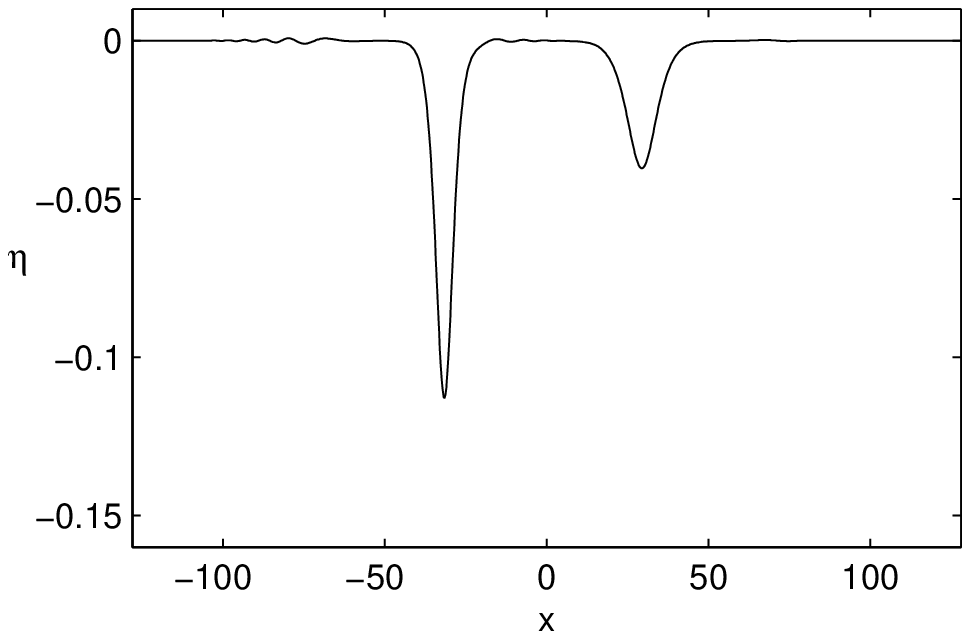} & 
\includegraphics[width=6cm]{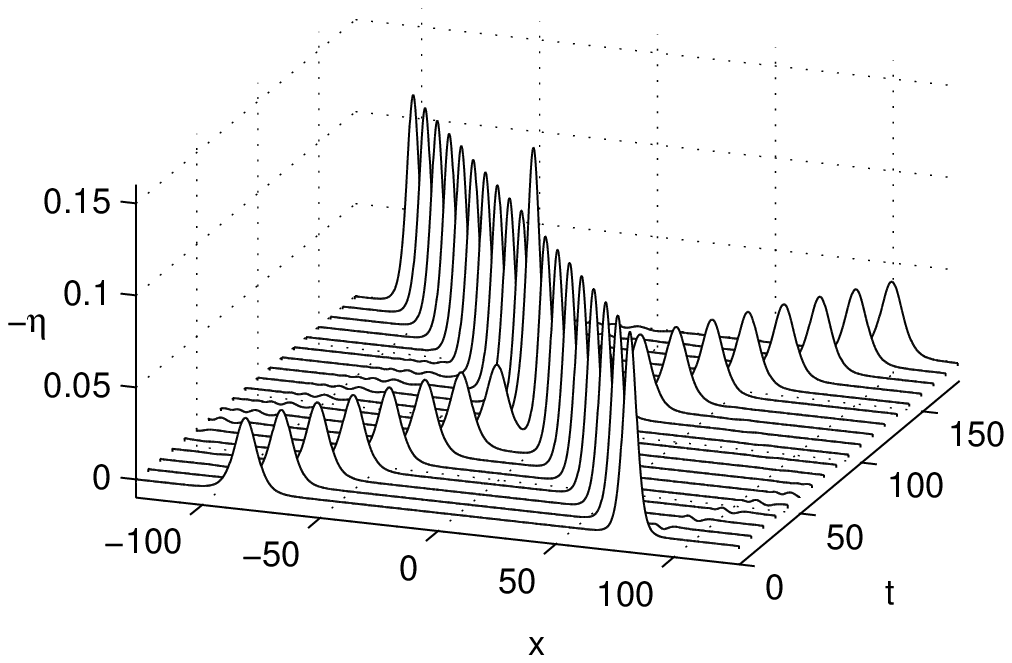}
\end{tabular}
\caption{Head-on collision of two almost perfect solitary waves of depression of different sizes. This is a solution to the system of 
quadratic Boussinesq equations (\ref{2eq_clean_AB}), with parameters $H=0.6$, $r=0.85$, $L=256$, $N=1024$, $S=-1-rH$, 
$\eta_0^{\ell}=-0.04$, $\eta_0^r=-0.11$, where the superscripts $\ell$ and $r$ stand for left and right respectively. In plot (f), 
note that $-\eta(x,t)$ has been plotted for the sake of clarity.}
\label{quadratique_count_diff_negative}
\end{center}
\end{figure}

In Figure \ref{quadratique_count_diff_negative}, we show the collision of two almost perfect solitary waves of depression of unequal 
amplitudes moving
in opposite directions. The numerical simulations exhibit a number of the same features that have been observed in the
symmetric case. 
%In particular we found that the phase lag is asymmetric, with the smaller solitary wave being delayed more 
%significantly than the larger (this cannot be seen in Figure \ref{quadratique_count_diff_negative}f).

\begin{figure}
\begin{center}
\begin{tabular}{c c}
(a) $t=0$ & (b) $t=4160$\\
\includegraphics[width=6cm]{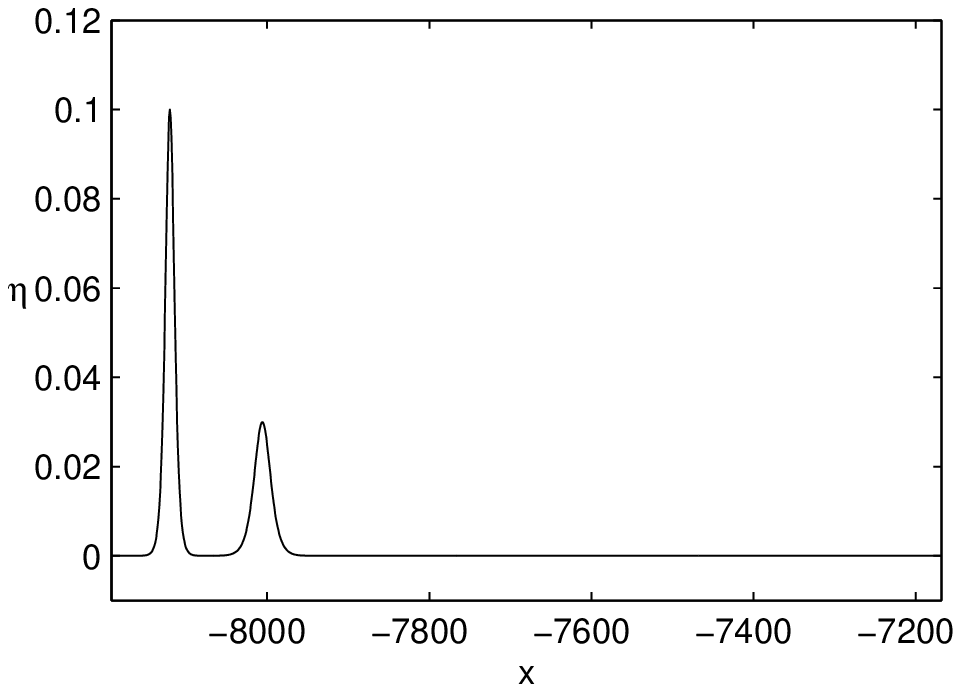}& \includegraphics[width=6cm]{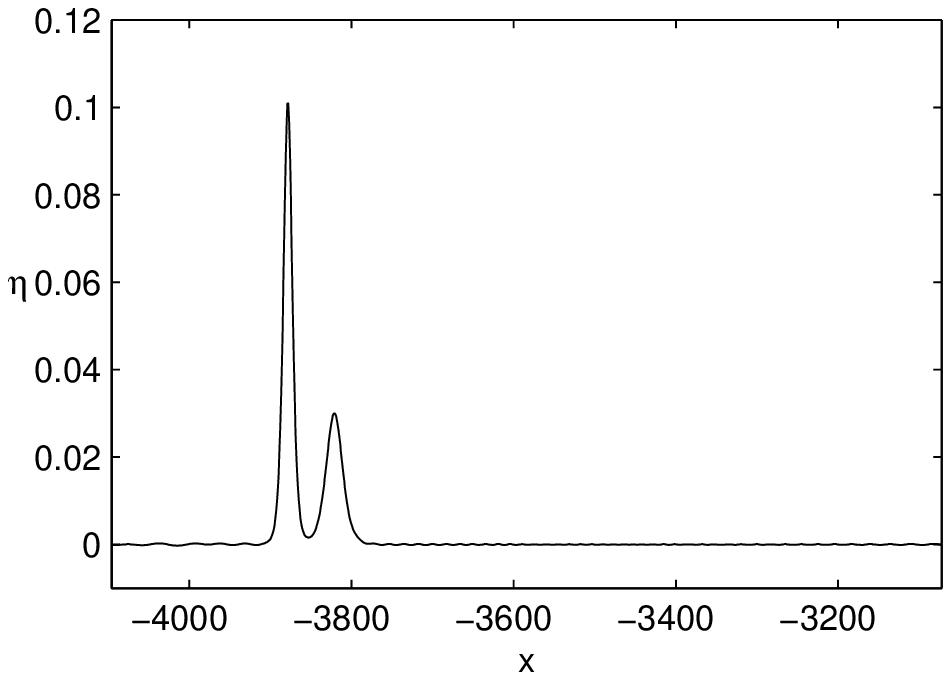} \\
(c) $t=6350$ & (d) $t=7450$\\
\includegraphics[width=6cm]{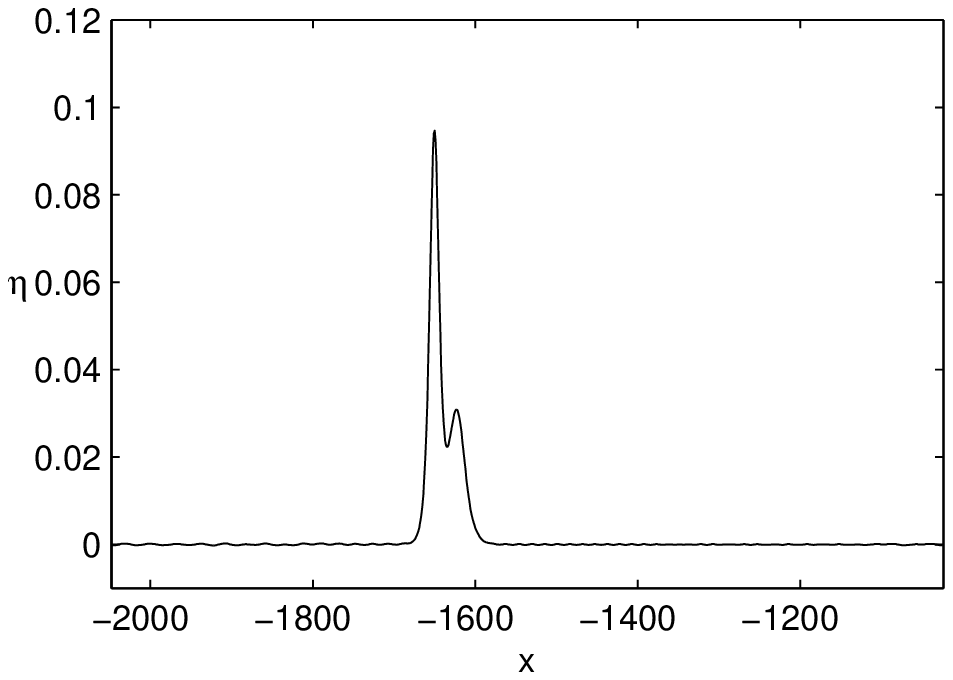} & \includegraphics[width=6.4cm]{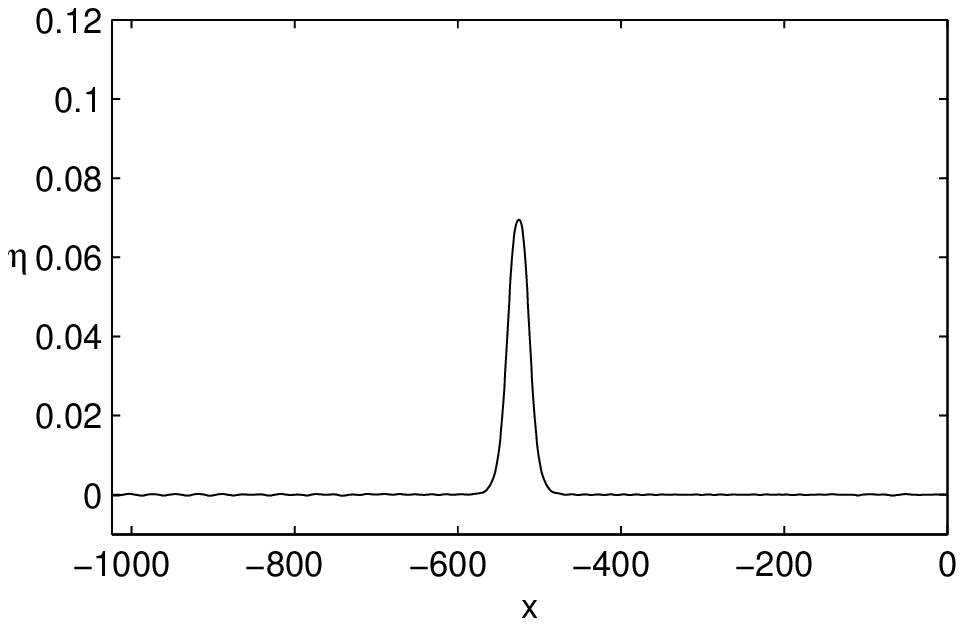}\\
(e) $t=9750$ & (f) $t=15900$\\
\includegraphics[width=6cm]{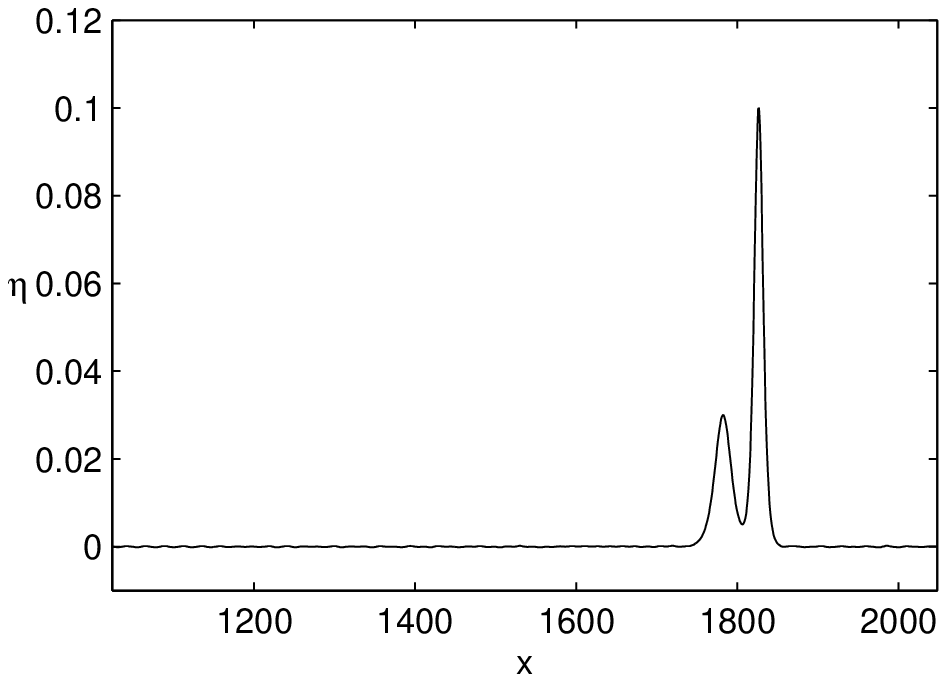} & \includegraphics[width=6cm]{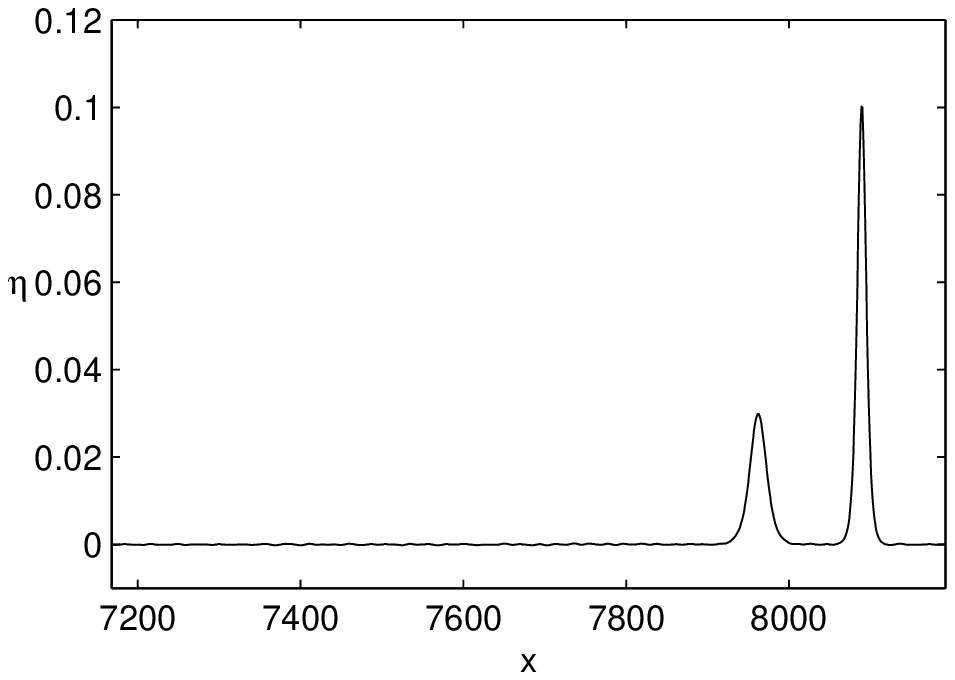}\\
\end{tabular}
\caption{Co-propagation of two almost perfect solitary waves of elevation of different sizes. This is a solution to the system of 
quadratic Boussinesq equations (\ref{2eq_clean_AB}), with parameters $H=1.6$, $r=0.95$, $L=2^{14}$, $N=2^{14}$, $S=-1-rH$, 
$\eta_0^{\ell}=0.1$, $\eta_0^r=0.03$, where the superscripts $\ell$ and $r$ stand for left and right respectively.}
\label{quadratique_copro}
\end{center}
\end{figure}

In Figure \ref{quadratique_copro}, we show the co-propagation of two solitary waves of elevation of different amplitudes. A
sequence of spatial profiles is shown. The
larger one, which is faster, eventually passes the smaller one, which is slower. Again there is a phase shift after
the interaction. The amplitude of the solution $\eta(x,t)$ never exceeds that of the larger solitary wave, nor does
it dip below the amplitude of the smaller. 
%The larger overtaking wave is shifted forward while the smaller is shifted backward. 

\section{Extended Boussinesq system of two equations with cubic terms}

When $|H^2-r|$ is small, one needs to go one step beyond and take into consideration the cubic terms.
Again one would like to obtain a system of two equations for the variables $\eta$ and $W=w-rw'$. We derive first
a general system of two equations with cubic terms. Then we introduce a specific scaling for the case where 
$|H^2-r|$ is small. A lot of terms in the system drop out because they are of higher order. 

The leading order terms lead to the same equation as before, namely $w= -Hw'$. And again
\begin{equation}\label{W}
    w=\frac{H}{r+H}W+O(\beta), \quad w'=\frac{-1}{r+H}W+O(\beta).              
\end{equation}
At the next order, the first two equations of (\ref{bo}) give
$$w_x+\alpha(w\eta)_x+\frac{1}{2}\beta\left(\theta^2-\frac{1}{3}\right)w_{xxx}
=-Hw'_x+\alpha(w'\eta)_x-\frac{1}{2}\beta H\left(\theta'^2-\frac{1}{3}H^2\right)w'_{xxx}.$$
Since the speeds $w$ and $w'$ vanish as $x\to \infty$ one has
$$
w= -Hw'+\alpha (w'-w)\eta -\frac{1}{2}\beta\left( H\left(\theta'^2-\frac{1}{3}H^2\right)w'_{xx}
+\left(\theta^2-\frac{1}{3}\right) w_{xx}\right).
$$
Using (\ref{W}) for the terms containing $\alpha$ or $\beta$ and neglecting terms of $O(\beta^2)$, one obtains
\begin{eqnarray}
    w&=&-Hw'-\alpha \frac{1+H}{r+H} W\eta+\frac{1}{2}\beta H
\frac{\left( \theta'^2-\frac{1}{3}H^2 \right)-\left(\theta^2-\frac{1}{3}\right)}{r+H} W_{xx}, \label{ww'}\\
    w'&=&-\frac{w}{H}-\alpha \frac{1+H}{H(r+H)}W\eta +\frac{1}{2}\beta 
\frac{\left( \theta'^2-\frac{1}{3}H^2 \right)-\left(\theta^2-\frac{1}{3}\right)}{r+H}W_{xx}.  \label{w'w}          
\end{eqnarray}

In Appendix B, after several substitutions, one obtains the system of two equations (\ref{sommedeux}) and (\ref{trois}). 
Switching back to the physical variables 
$$x^*=\ell {x}, \quad \eta^*=A{\eta}, \quad t^*=\ell {t}/c_0, \quad 
W^*=gAW/c_0, \quad \mbox{with} \;\; c_0=\sqrt{gh},$$ 
the system (\ref{sommedeux})-(\ref{trois}) becomes
\begin{equation}\label{sommedeux_phy}
   \begin{array}{ll}
(r+H)\eta^*_{t^*} + hHW^*_{x^*}+\frac{H^2-r}{r+H}(W^*\eta^*)_{x^*}
+\frac{1}{2}h^3H\frac{H(\theta^2-\frac{1}{3})+r(\theta'^2-\frac{1}{3}H^2)}{r+H}W^*_{x^*x^*x^*}\\ 
-\frac{1}{h} \frac{r(1+H)^2}{(r+H)^2} (W^*\eta^{*2})_{x^*}
+\frac{1}{2}h^2rH(1+H)\frac{(\theta'^2-\frac{1}{3}H^2)-(\theta^2-\frac{1}{3})}{(r+H)^2}(W^*\eta^*)_{x^*x^*x^*}\\ 
+\frac{1}{2}h^2\left(rH(1+H)\frac{(\theta'^2-\frac{1}{3}H^2)-(\theta^2-\frac{1}{3})}{(r+H)^2} 
+\frac{H^2(\theta^2-1)-r(\theta'^2-H^2)}{r+H}\right)(W^*_{x^*x^*}\eta^*)_{x^*}\\ 
-\frac{1}{4}h^5\left(\frac{rH^2\left((\theta'^2-\frac{1}{3}H^2)-(\theta^2-\frac{1}{3})\right)^2}{(r+H)^2} 
-\frac{5}{6}\frac{H^2(\theta^2-\frac{1}{5})^2+rH(\theta'^2-\frac{1}{5}H^2)^2}{r+H}\right) W^*_{x^*x^*x^*x^*x^*}=0,
  \end{array}
\end{equation}
\begin{equation}\label{trois_phy}
   \begin{array}{ll}
g(1-r)\eta^*_{x^*}+ W^*_{t^*} +\frac{H^2-r}{(r+H)^2} W^*W^*_{x^*}
+\frac{1}{2}h^2\frac{H(\theta^2-1)+r(\theta'^2-H^2)}{r+H}W^*_{x^*x^*t^*}\\
-\frac{1}{h}\frac{r(1+H)^2}{(r+H)^3}(W^{*2}\eta^*)_{x^*}
+\frac{1}{2}h^2rH(1+H)\frac{(\theta'^2-\frac{1}{3}H^2)-(\theta^2-\frac{1}{3})}{(r+H)^3}(W^*W^*_{x^*x^*})_{x^*}\\ 
+h\frac{H(1-r)}{r+H}(\eta^* W^*_{x^*t^*})_{x^*}
+\frac{1}{2}h^2\frac{H^2(\theta^2-1) -r(\theta'^2-H^2)}{(r+H)^2} W^*W^*_{x^*x^*x^*}\\ 
+\frac{1}{2}h^2\frac{H^2(\theta^2+1)-r(\theta'^2+H^2)}{(r+H)^2} W^*_{x^*}W^*_{x^*x^*}
-\frac{1}{2}hrH(1+H)\frac{(\theta^2-1)-(\theta'^2-H^2)}{(r+H)^2}(W^*\eta^*)_{x^*x^*t^*}\\ 
+h^4\left(\frac{rH\left((\theta^2-1)-
(\theta'^2-H^2)\right)\left((\theta'^2-\frac{1}{3}H^2)-(\theta^2-\frac{1}{3})\right)}{4(r+H)^2}
+\frac{H(\theta^2-1)(5\theta^2-1)+r(\theta'^2-H^2)(5\theta'^2-H^2)}{2(r+H)}\right)\\
 W^*_{x^*x^*x^*x^*t^*}=0.
  \end{array}
\end{equation}
The specific scaling for small values of $|H^2-r|$,
$$ \frac{x^*}{h}=\frac{x}{\beta}, \quad \frac{t^*}{h/c_0} = \frac{t}{\alpha}, \quad \frac{\eta^*}{h} = \alpha \eta, 
\quad \frac{W^*}{gh/c_0} = \alpha W,
\quad H^2-r = \alpha {\mathcal C}, $$
with $c_0=\sqrt{gh}$, $\alpha \ll 1$, $\beta \ll 1$, $\alpha=O(\beta)$,
will lead to a new Boussinesq system with cubic terms. 
A lot of terms in (\ref{sommedeux_phy})-(\ref{trois_phy}) drop out because they are of higher order. Keeping terms of order 
$\alpha^2$ and $\alpha^4$ and going back to physical variables, the system of two equations becomes
 \begin{equation}\label{ordre5_phy}
\begin{array}{rll}  \eta^*_{t^*}&=&-h\frac{H}{r+H}W^*_{x^*}-h^3\left(\frac{1}{2}\frac{H^2S}{(r+H)^2}
+\frac{1}{3}\frac{H^2(1+rH)}{(r+H)^2}\right)W^*_{x^*x^*x^*} \\ & & 
-\frac{H^2-r}{(r+H)^2}(W^*\eta^*)_{x^*}+ \frac{1}{h}\frac{r(1+H)^2}{(r+H)^3} (W^*\eta^{*2})_{x^*}\\
 W^*_{t^*}&=&-g(1-r)\eta^*_{x^*}
 -\frac{1}{2}h^2\frac{HS}{r+H}W^*_{x^*x^*t^*}
 -\frac{H^2-r}{(r+H)^2}W^*W^*_{x^*}+\frac{1}{h}\frac{r(1+H)^2}{(r+H)^3}(W^{*2}\eta^*)_{x^*}.
 \end{array}
 \end{equation}
 This is the same system as (\ref{2eqphy}) with two extra terms, the cubic terms. We will call it
 a system of extended Boussinesq equations (see also \cite{CGK}). Linearizing (\ref{ordre5_phy}) gives the same dispersion relation 
as before. 
%%%%%%%%%%%%%%%%%%%%%%%%%%%%%%%%%%%%%%%%%%%%%%%%%%%%%%%%
\section{Numerical solutions of the extended Boussinesq system}

In order to integrate numerically the extended Boussinesq system (\ref{ordre5_phy}), we introduce a slightly different change of 
variables, where the stars still denote the physical variables and no new notation is introduced for the dimensionless
variables: 
$$x=\frac{x^*}{h}, \;\; \eta=\frac{\eta^*}{h}, \;\; t=\frac{c}{h}t^*, \;\; W=\frac{W^*}{c}, \;\; \mbox{with} \;\;
c^2=\frac{ghh'(\rho-\rho')}{\rho' h+\rho h'}=\frac{ghH(1-r)}{r+H}.$$
Using the same coefficients as in (\ref{dcoef}), we rewrite system (\ref{ordre5_phy}) with the new variables as
\begin{equation}\label{ordre5S}
\begin{array}{rll}  \eta_t&=&-d_1 W_x- d_2 W_{xxx}
-d_4(W\eta)_x+ d_5(W\eta^2)_x \\ 
 W_t&=&-(1/d_1)\eta_x -d_3 W_{xxt}
 -d_4 WW_x+d_5(W^2\eta)_x
 \end{array}
 \end{equation}
 where the new coefficient $d_5$ is equal to
 $$ d_5 = \frac{r(1+H)^2}{(r+H)^3}. $$
When $(\theta,\theta')=(0,0)$, one recovers the system with horizontal velocities on the bottom and on the roof.

Taking the Fourier transform of the system (\ref{ordre5S}) gives
      \begin{eqnarray*}
       \hat{\eta_t} & = & (d_2k^2-d_1)ik\hat{W}-d_4 ik\widehat{(W\eta )}+d_5 ik\widehat{(W\eta^2)}, \\
       (1-d_3k^2)\hat{W_t} & = & -\frac{1}{d_1}ik\hat{\eta}-\frac{d_4}{2}ik\widehat{(W^2)}+d_5 ik\widehat{(W^2\eta)}.
\end{eqnarray*}
The system of differential equations is integrated numerically with the same method as in \S\,5. 

Again we look for approximate solitary wave solutions to (\ref{ordre5S}).
As before we look for solutions of the form
$$W(x,t)=\frac{1}{d_1}[\eta(x,t)+M(x,t)],$$ 
where $M$ is assumed to be small compared to $\eta$ and $W$. Substituting the expression for $W$ into (\ref{ordre5S}) and neglecting
higher-order terms yields
\begin{equation}\label{M}
M_x=-\frac{1}{4}\frac{d_4}{d_1}(\eta^2)_x-\frac{1}{2}\frac{d_2}{d_1}\eta_{xxx}+\frac{1}{2}d_3\eta_{xxt}. 
\end{equation}
Substituting the expression for $M_x$ into one of the equations of system (\ref{ordre5S}) yields
\begin{equation}\label{eKdV}
\eta_t+\eta_x+\frac{3d_4}{4d_1}(\eta^2)_x-\frac{d_5}{d_1}(\eta^3)_x+\frac{d_2}{2d_1}\eta_{xxx}+\frac{d_3}{2}\eta_{xxt}=0.
\end{equation} 
We have checked that the extended KdV equation (\ref{eKdV}) is in agreement with previously derived eKdV equations such as
in \cite{DV04}.

Let $V=1+c_1$ be the wave speed, with $c_1$ small. In the moving frame of reference $X=x-(1+c_1)t, T=t$, 
equation (\ref{eKdV}) becomes
$$ -c_1\eta_X+\eta_T+\frac{3d_4}{4d_1}(\eta^2)_X-\frac{d_5}{d_1}(\eta^3)_X+\frac{d_2}{2d_1}\eta_{XXX}
+\frac{d_3}{2}\left[-(1+c_1)\eta_{XXX}+\eta_{XXT}\right]=0.
$$
Looking for stationary solutions and integrating with respect to $X$ yields
\begin{equation}\label{7}
-c_1\eta+\frac{3d_4}{4d_1}\eta^2-\frac{d_5}{d_1}\eta^3+\frac{1}{2}\left(\frac{d_2}{d_1}-d_3-c_1d_3\right)\eta_{XX}=0.
\end{equation}
Letting
$$
\alpha_1=\frac{3}{2}\frac{H^2-r}{H(r+H)}, \;\; \beta_1=3\frac{r(1+H)^2}{H(r+H)^2}, \;\;
\lambda_1=\frac{1}{6}\frac{H(rH+1)}{r+H}-\frac{1}{4}\frac{HS}{r+H}c_1, $$
equation (\ref{7}) becomes
$$
-c_1\eta +\frac{1}{2}\alpha_1 \eta^2-\frac{1}{3}\beta_1 \eta^3 +\lambda_1 \eta_{XX}=0.
$$
It has solitary wave solutions
$$
\eta(X)=\left(\frac{\alpha_1}{\beta_1}\right)\frac{1-\epsilon^2}{1+\epsilon \cosh(\sqrt{\frac{c_1}{\lambda_1}}X)}, 
\quad \mbox{with} \;\; \epsilon = \frac{\sqrt{\alpha_1^2-6\beta_1 c_1}}{|\alpha_1|}.$$
\begin{figure}
\begin{center}
\begin{tabular}{c c}
 (a)  & (b) \\
\includegraphics[width=8cm,height=5cm]{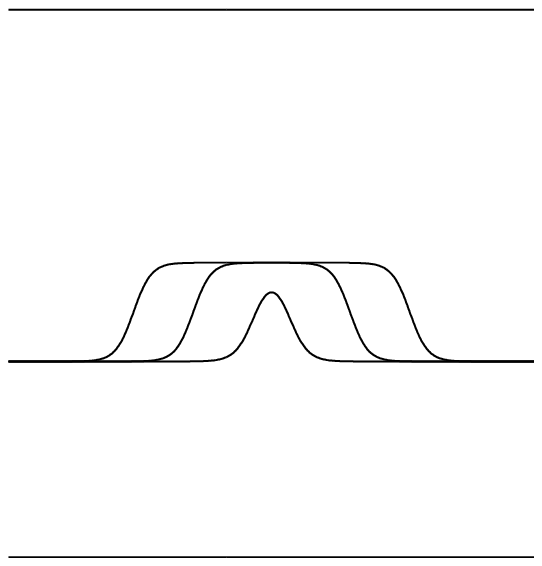}& \includegraphics[width=6cm,height=5cm]{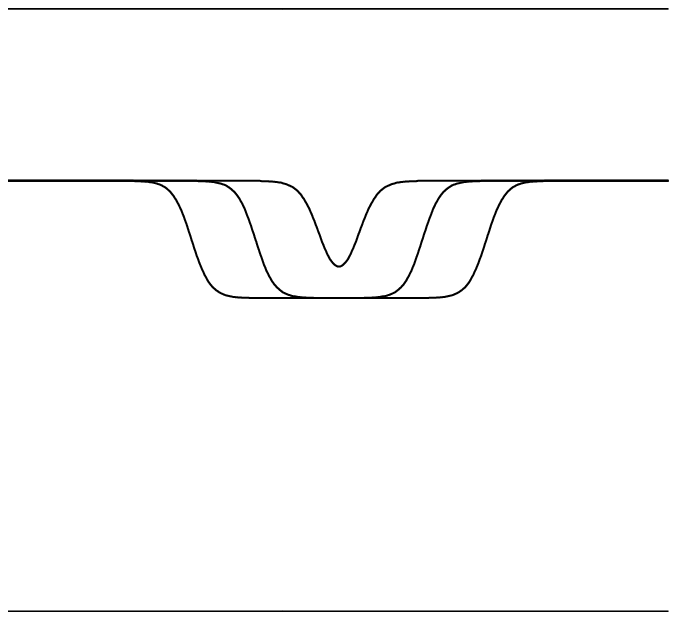}\\ 
\end{tabular}
\caption{`Table-top' solitary waves which are approximate solutions of the extended Boussinesq system
(\ref{ordre5S}). The horizontal velocities are taken on the top and the bottom so that $S=-(1+rH)$.
(a) $H=1.8$, $r=0.8$. The wave speeds $V$ are, going from the smallest to the widest solitary wave,
$V_{\rm max}-V \sim 10^{-3}, 10^{-9}, 10^{-15}$;
(b) $H=0.4$, $r=0.9$. The wave speeds $V$ are, going from the smallest to the widest solitary wave,
$V_{\rm max}-V \sim 10^{-3}, 10^{-9}, 10^{-14}$.}
\label{table_top}
\end{center}
\end{figure}
In the fixed frame of reference, the profile of the solitary waves is given by
\begin{equation}\label{onde_plate}
\eta(x,t)=\left(\frac{\alpha_1}{\beta_1}\right)\frac{1-\epsilon^2}{1+\epsilon \cosh\left(\sqrt{\frac{V-1}{\lambda_1}}(x-Vt)\right)}.
\end{equation}
When $H^2>r$ the solitary waves are of elevation. When $H^2<r$ they are of depression.
The parameter $\epsilon$ can take values ranging from $0$ (infinitely wide solution) to $1$ (solution of infinitesimal amplitude). 
Assuming $M(\pm\infty)=0$, one can compute $M$ explicitly by integrating equation (\ref{M}) with respect to $x$:
$$
M = -\frac{d_4}{4d_1}\eta^2-\frac{d_2}{2d_1}\eta_{xx}+\frac{d_3}{2}\eta_{xt}.
$$

Typical approximate solitary waves solutions are shown in Figure \ref{table_top}. Notice that the condition $|H^2-r|$ small
is not really satisfied for the selected values of $H$ and $r$. The reason is that otherwise the waves would have been too small to be 
clearly visible. Of course we still have the conditions on $S$ for well-posedness:
$$-(1+rH)\leq S \leq -\frac{2}{3}(1+rH). $$
The solitary waves are characterized by wave velocities
larger than $1$ $(c_1>0)$. The maximum wave velocity $V_{\rm max}$
is obtained when $\epsilon \to 0$. One finds $c_1 \to \alpha_1^2/6\beta_1$, so that 
$$V_{\rm max} = 1 + \frac{(H^2-r)^2}{8rH(1+H)^2}. $$

Once the approximate solitary wave (\ref{onde_plate}) has been obtained, it is again possible to make it cleaner by iterative
filtering. Qualitative results for non-filtered solitary waves are given in this Section. Some accurate results for
run-ups and phase shifts with filtered waves are described in Appendix A.

In Figure \ref{cubic_collision_table}, we show the head-on collision of two almost perfect `table-top' solitary waves of 
elevation of equal amplitude moving in opposite directions. As in the case with only quadratic nonlinearities, the solution 
rises to an amplitude larger than the sum of the amplitudes
of the two incident solitary waves. After the collision, two similar waves emerge and return to the form of
two separated `table-top' solitary waves. As a result of this collision, the amplitudes of the two resulting solitary waves are slightly
smaller than the incident amplitudes and their centers are slightly retarded from the trajectories of the incoming centers.
\begin{figure}
\begin{center}
\begin{tabular}{c c}
 (a) $t=0$ & (b) $t=500$\\
\includegraphics[width=6cm]{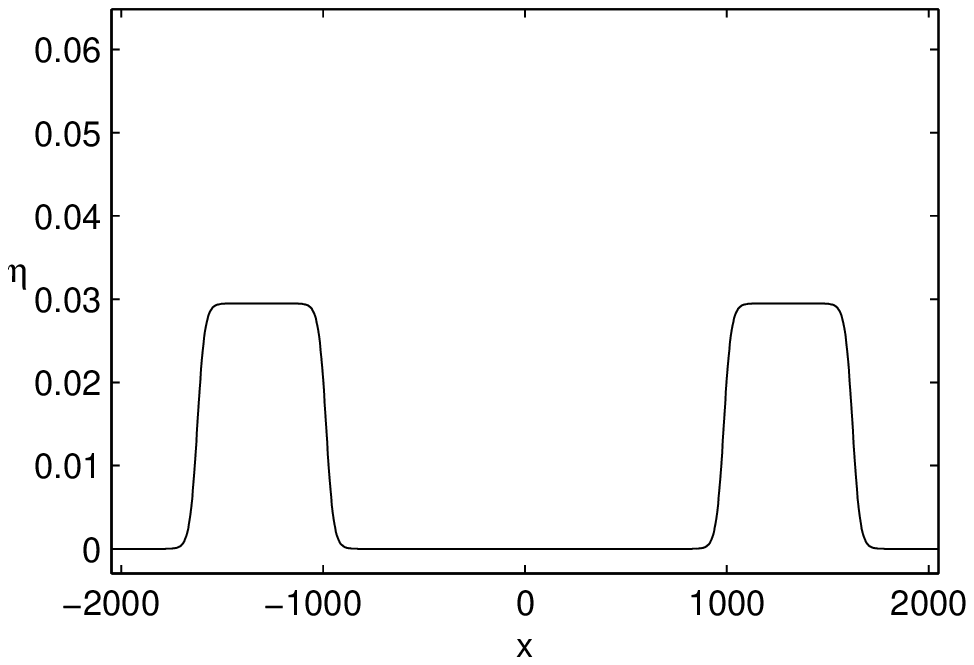}& \includegraphics[width=6cm]{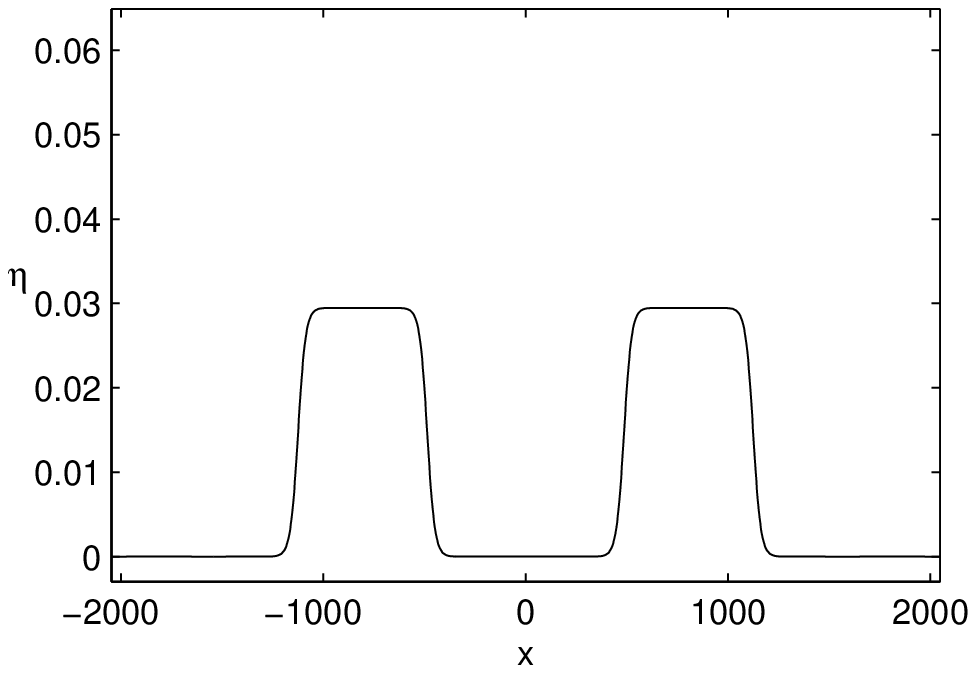} \\
 (c) $t=1100$ & (d) $t=1300$\\
\includegraphics[width=6cm]{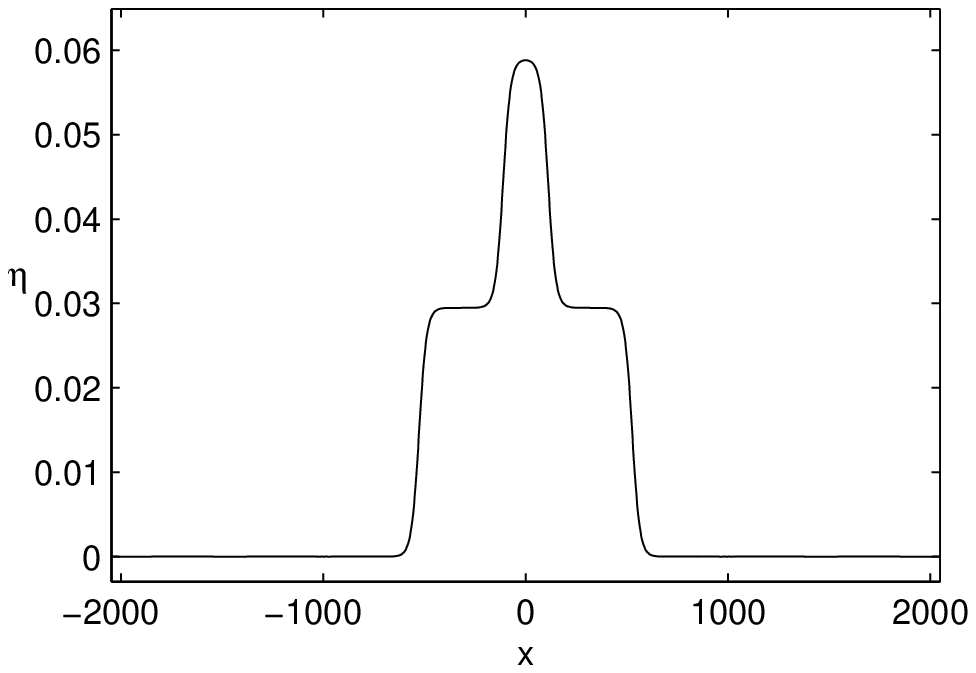} & \includegraphics[width=6cm]{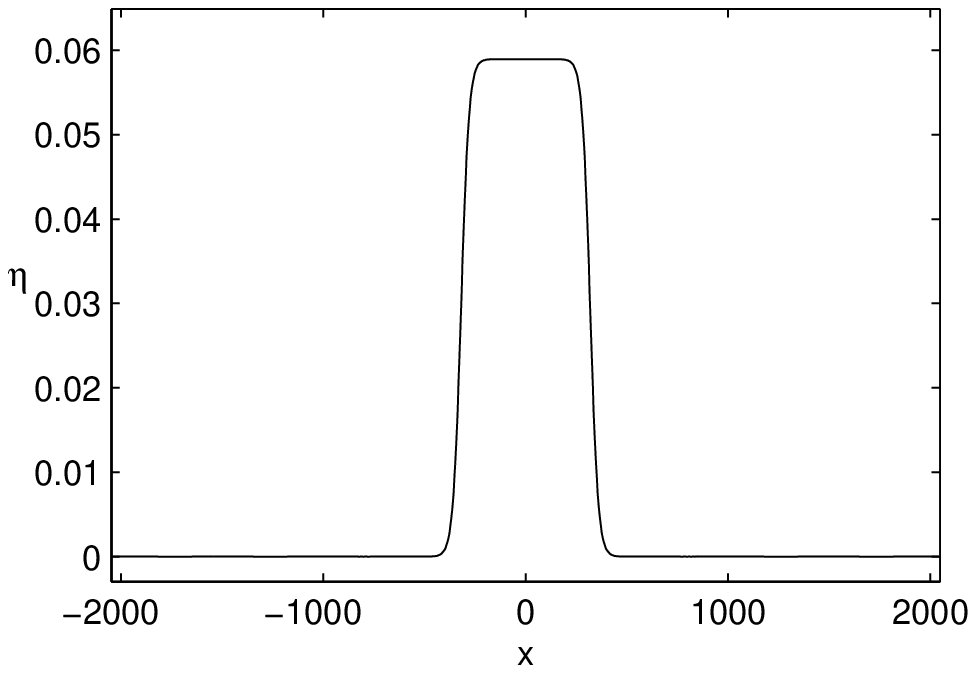}\\
 (e) $t=1700$ & (f) evolution in time\\
\includegraphics[width=6cm]{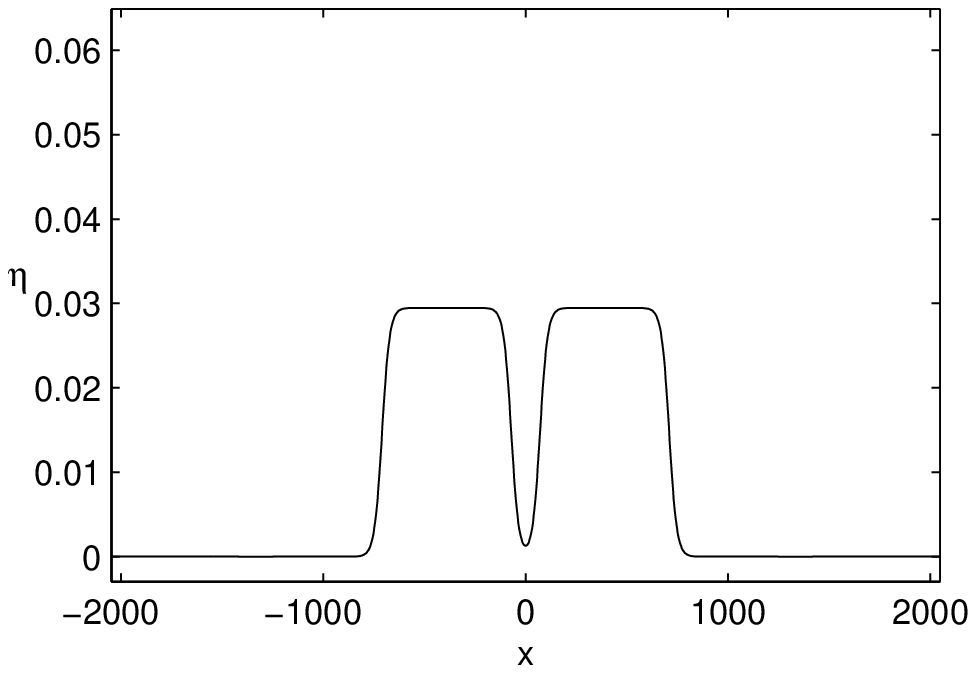} & \includegraphics[width=6cm]{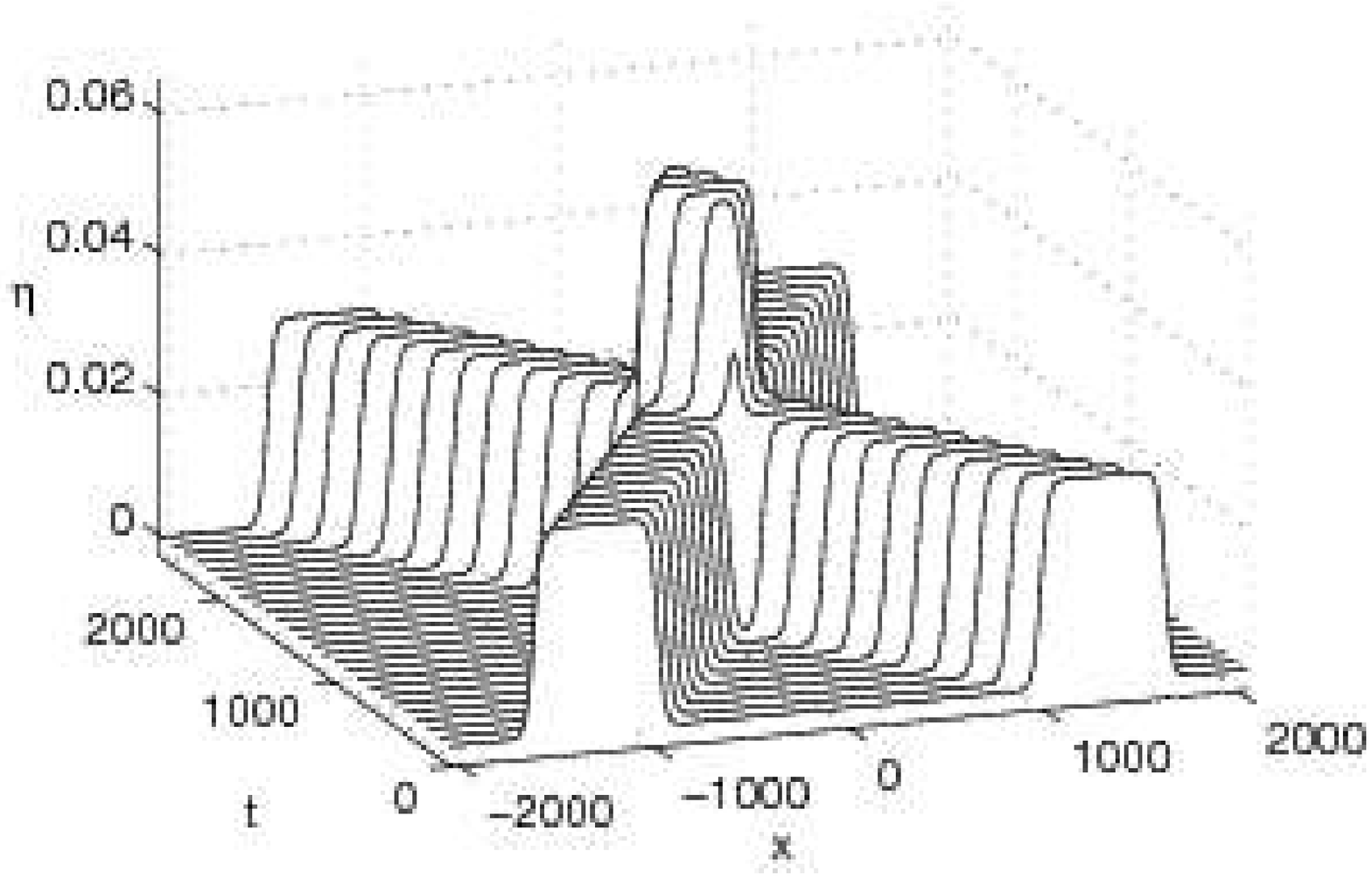}\\
\end{tabular}
\caption{Head-on collision of two approximate `table-top' elevation solitary waves of equal size. This is a solution to the system 
of cubic Boussinesq equations (\ref{ordre5S}), with parameters $H=0.95$, $r=0.8$, $L=4096$, $N=1024$, $S=-1-rH$, 
$V_{\rm max}-V \sim 10^{-17}$.}
\label{cubic_collision_table}
\end{center}
\end{figure}

\begin{figure}
\begin{center}
\begin{tabular}{c c}
 (a) $t=0$ & (b) $t=300$\\
\includegraphics[width=6cm]{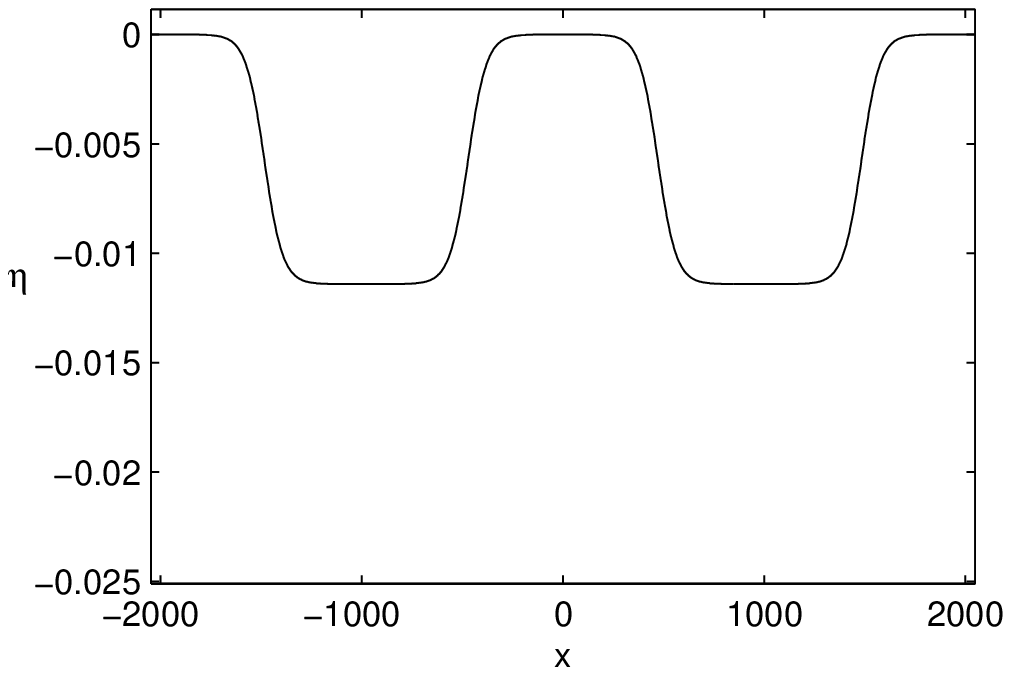}& 
\includegraphics[width=6cm]{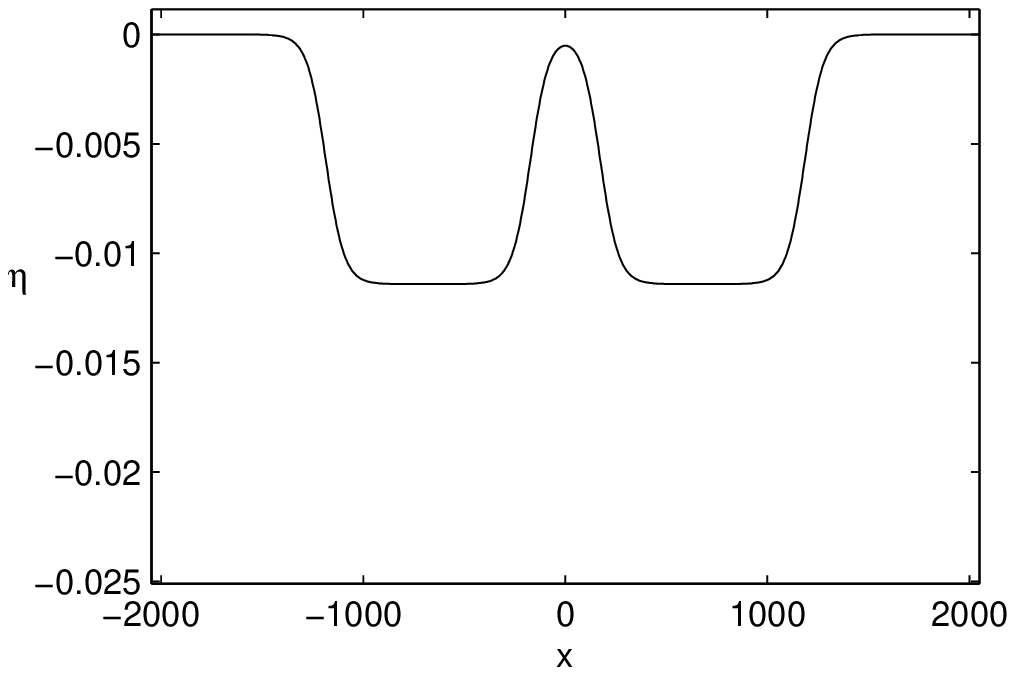} \\
 (c) $t=700$ & (d) $t=1000$\\
\includegraphics[width=6cm]{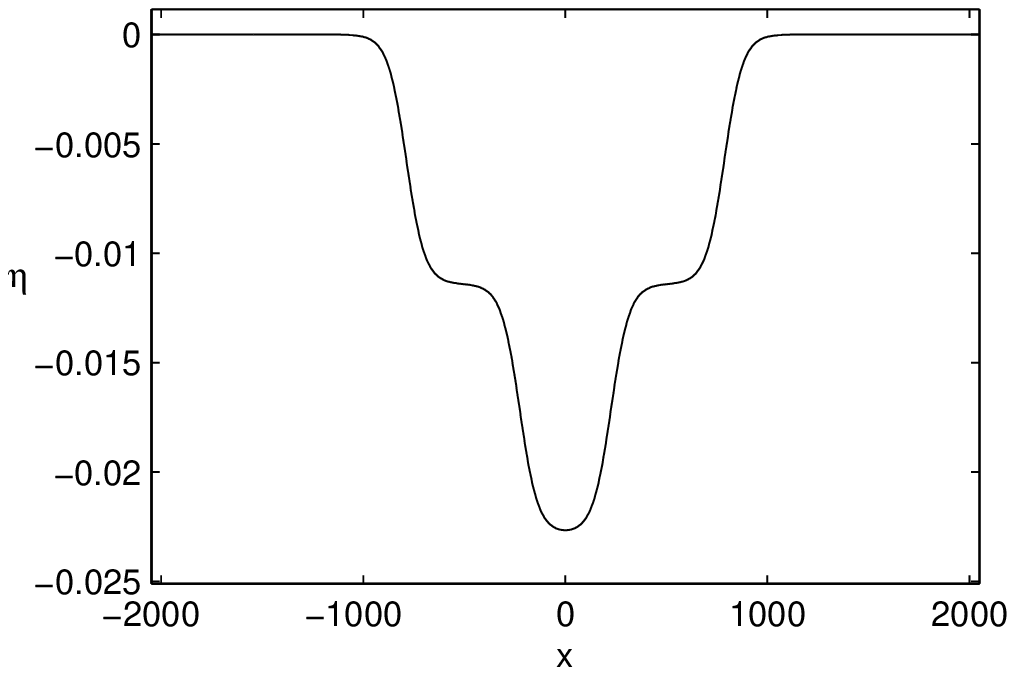} & 
\includegraphics[width=6cm]{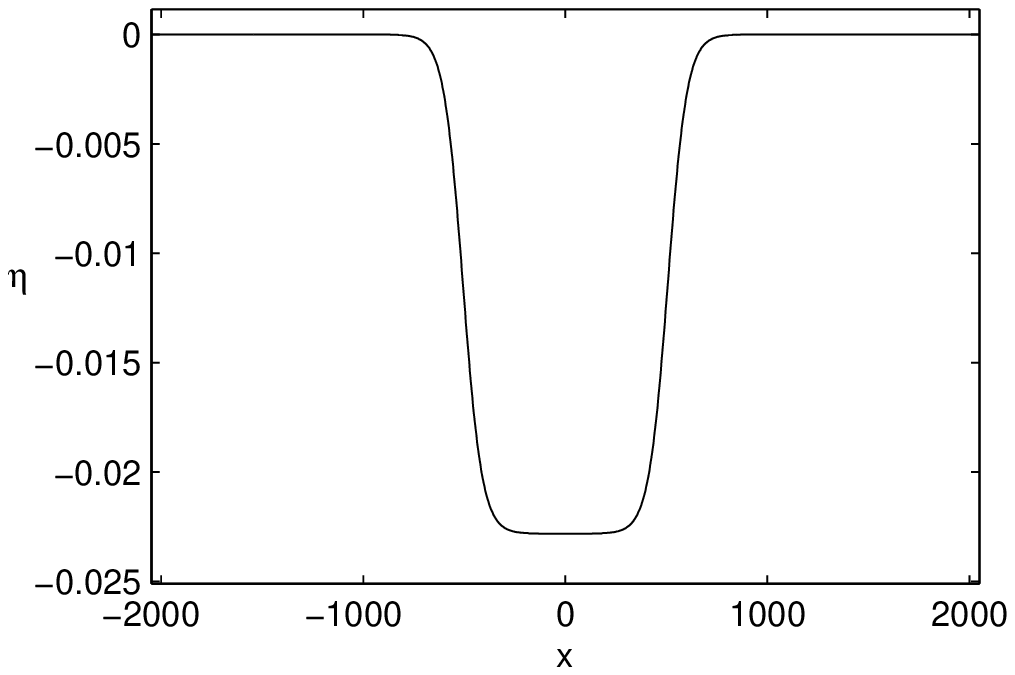}\\
 (e) $t=2200$ & (f) evolution in time\\
\includegraphics[width=6cm]{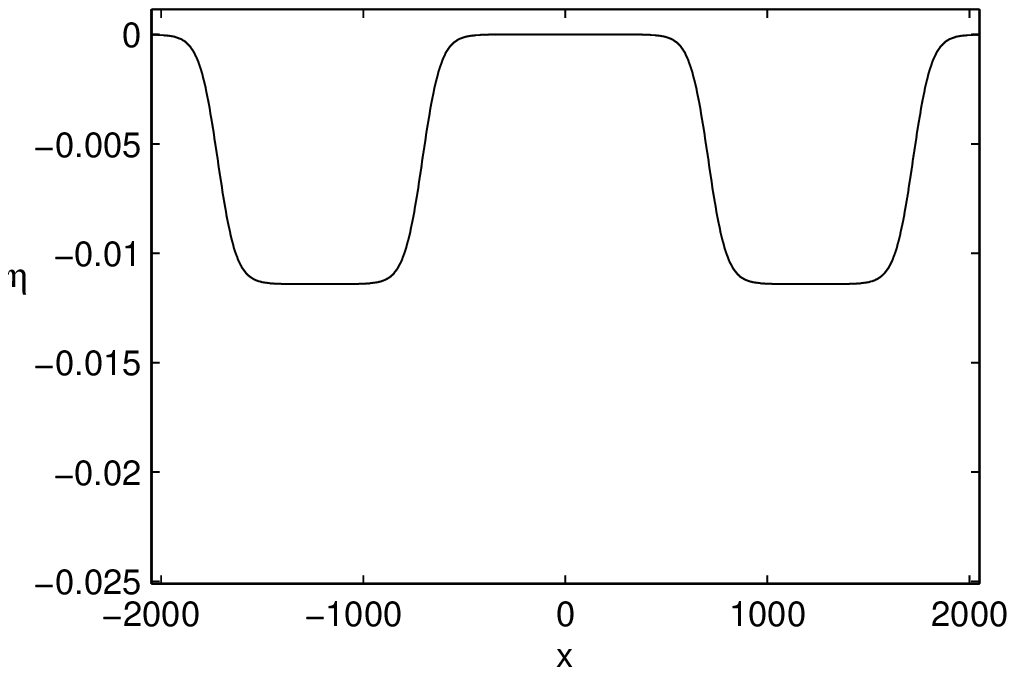} & 
\includegraphics[width=6cm]{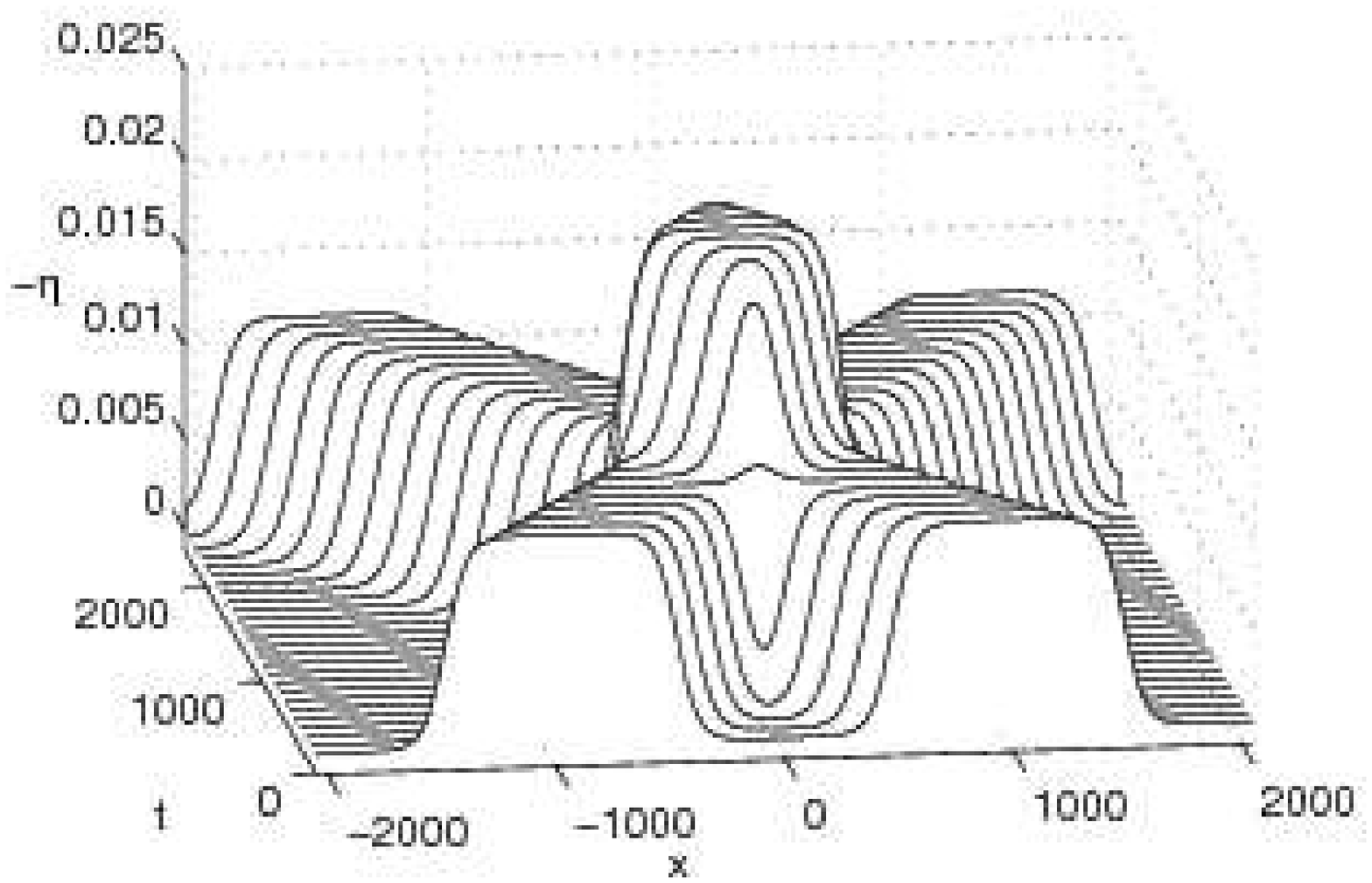}
\end{tabular}
\caption{Head-on collision of two approximate `table-top' depression solitary waves of equal size. This is a solution to the system 
of cubic Boussinesq equations (\ref{ordre5S}), with parameters 
$H=0.9$, $r=0.85$, $L=4096$, $N=1024$, $S=-1-rH$, $V_{\rm max}-V \sim 10^{-14}$.
In plot (f), note that $-\eta(x,t)$ has been plotted for the sake of clarity.}
\label{cubic_collision_table_negative}
\end{center}
\end{figure}

In Figure \ref{cubic_collision_table_negative}, we show the collision of two almost perfect solitary waves of depression of equal 
amplitude moving in opposite directions. The numerical simulations exhibit the same features that have been observed in the
elevation case. 

\begin{figure}
\begin{center}
\begin{tabular}{c c}
 (a) $t=0$ & (b) $t=320$\\
 \includegraphics[width=6cm]{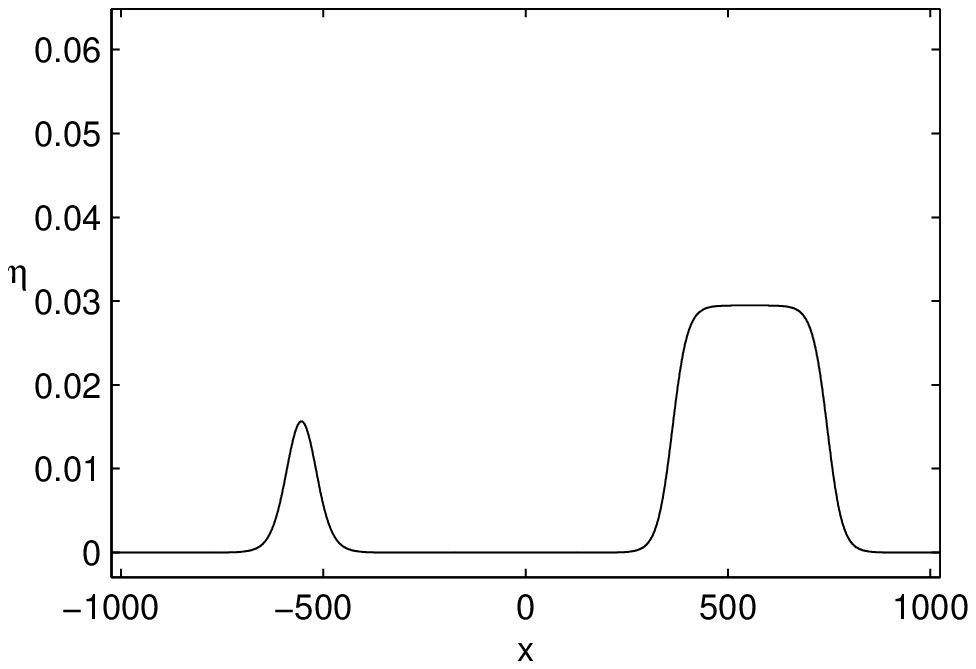}& \includegraphics[width=6cm]{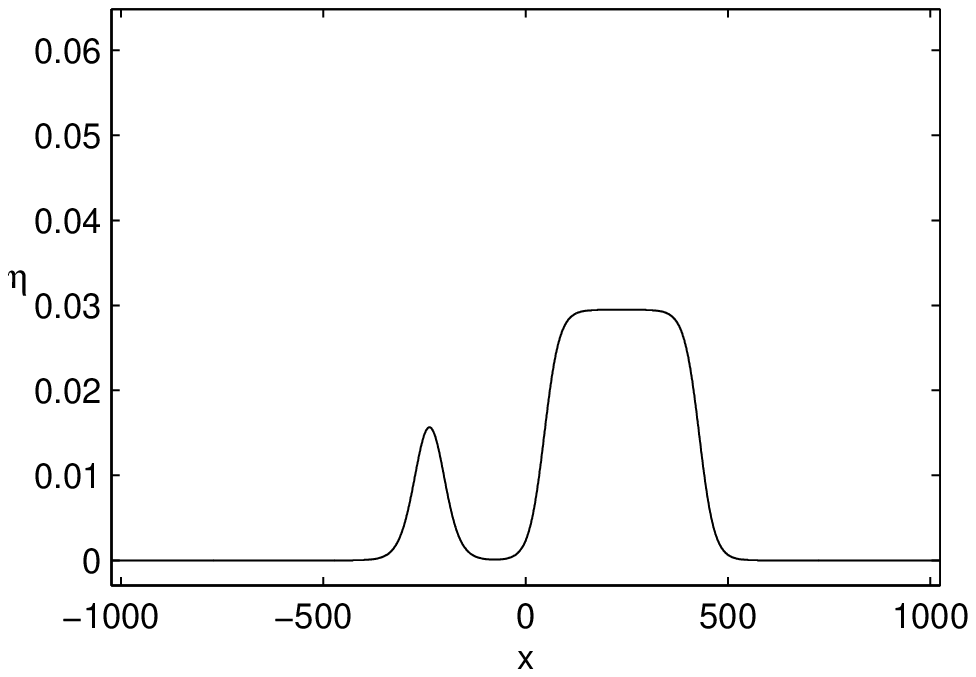} \\
 (c) $t=480$ & (d) $t=600$\\
 \includegraphics[width=6cm]{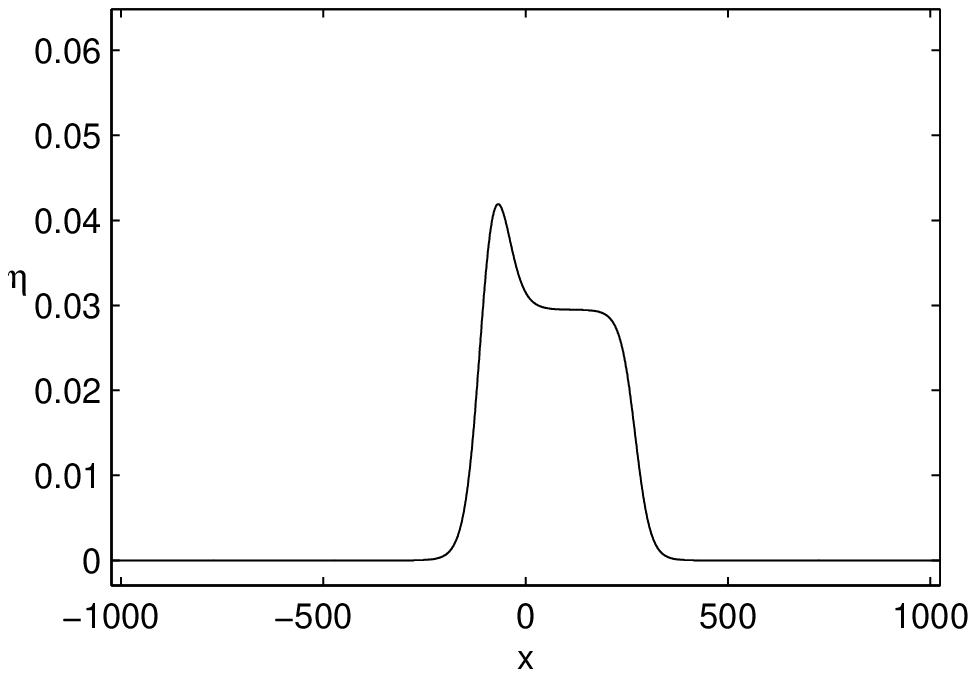} & \includegraphics[width=6cm]{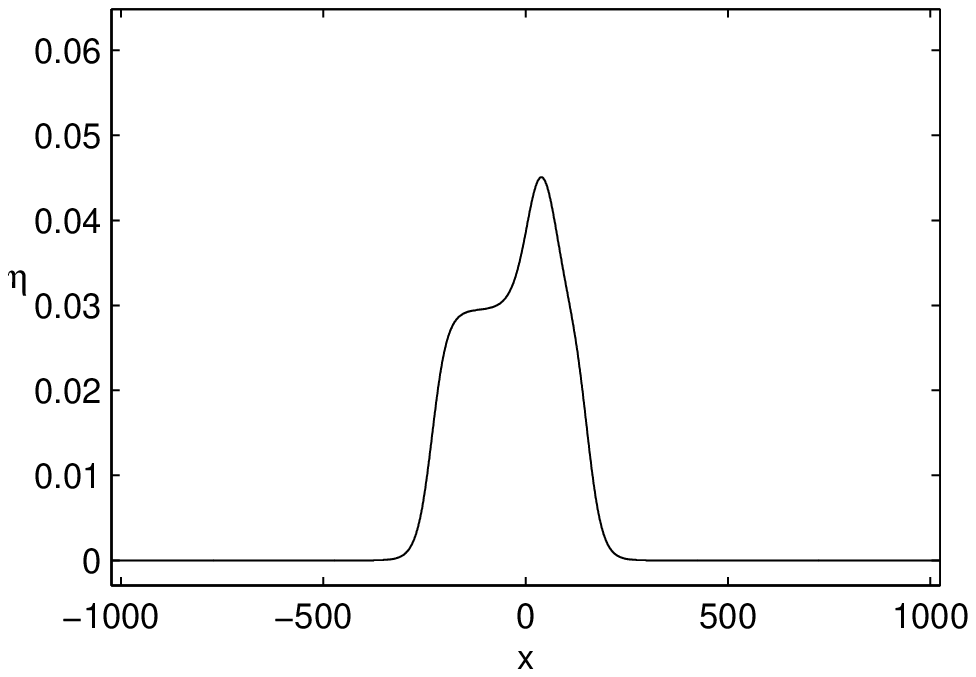}\\
 (e) $t=800$ & (f) evolution in time\\
 \includegraphics[width=6cm]{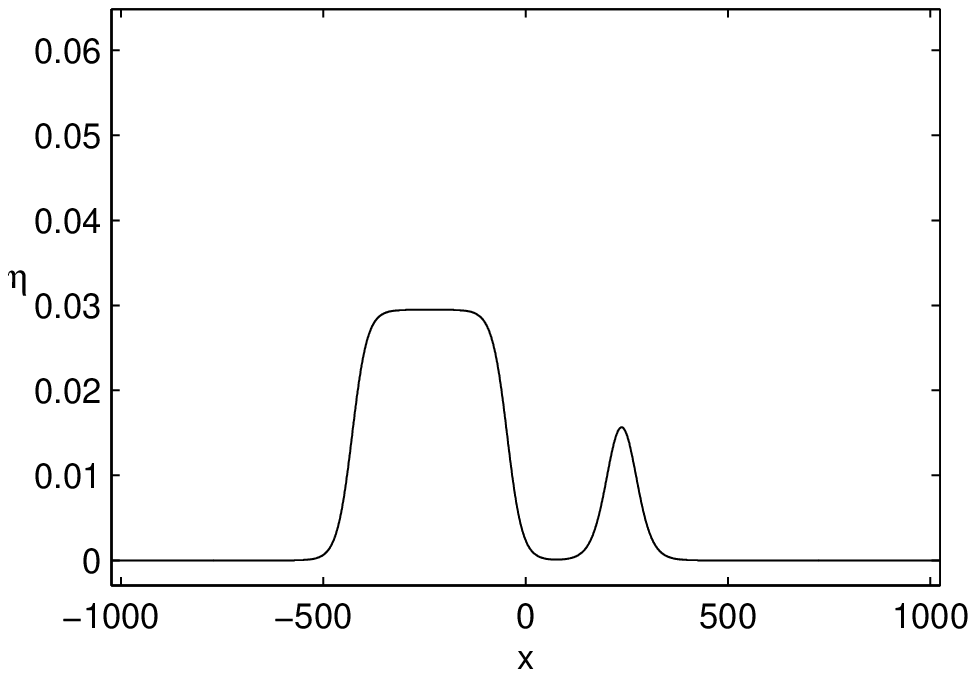} & \includegraphics[width=6cm]{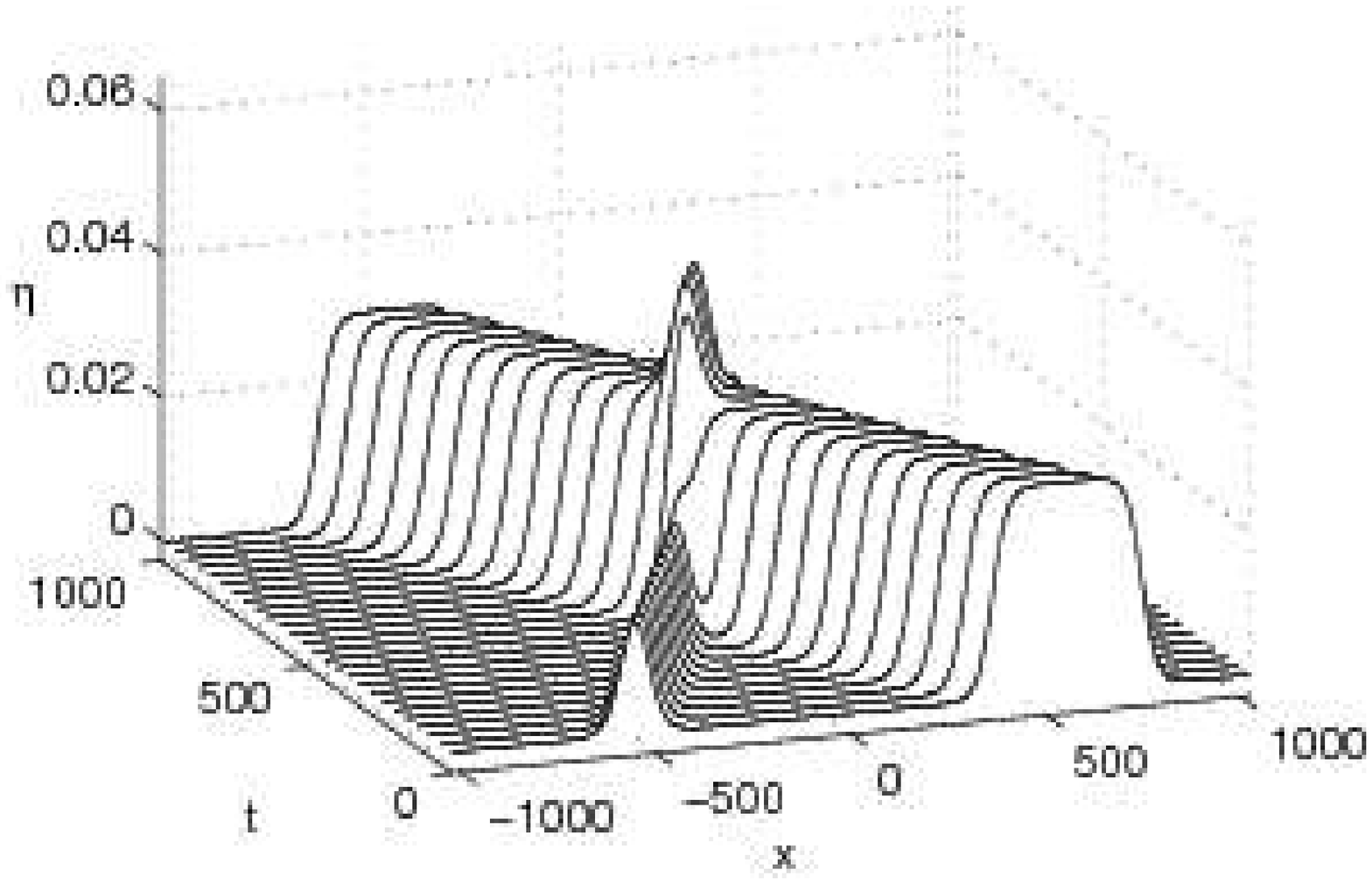}\\
\end{tabular}
\caption{Head-on collision of a solitary wave of elevation and of a `table-top' solitary wave of elevation. This is a solution to the 
system of cubic Boussinesq equations (\ref{ordre5S}), with parameters 
$H=0.95$, $r=0.8$, $L=2048$, $N=1024$, $S=-1-rH$, $V_{\rm max}-V^\ell \sim 10^{-4}$, $V_{\rm max}-V^r \sim 10^{-11}$.}
\label{cubic_collision_table_solitary}
\end{center}
\end{figure}

In Figure \ref{cubic_collision_table_solitary}, we show the collision of an almost perfect `table-top' solitary wave of elevation 
with a solitary wave of elevation moving
in the opposite direction. The numerical simulations exhibit a number of the same features that have been observed in the
symmetric case. The phase lag is asymmetric, with the smaller solitary wave being delayed more significantly than the larger.

Note that in the quadratic as well as in the cubic cases, it is not possible to consider the collision between a solitary wave 
of depression and a solitary wave of elevation. Indeed the sign of $H^2-r$ determines whether the wave is of elevation or of depression.

\section{Conclusion}

In this paper, we derived a system of extended Boussinesq equations in order to describe weakly nonlinear waves at the interface between
two heavy fluids in a `rigid-lid' configuration. To our knowledge we have described for the first time the collision between `table-top' 
solitary waves. The extension to a `free-surface' configuration and to arbitrary wave amplitude is left to future studies. Indeed, since 
the waves we considered are only weakly nonlinear, we do not have to worry about the 
resulting wave reaching the roof or the bottom. However, in a fully nonlinear regime, this could happen. Indeed
the maximum amplitude $A$ for `table-top' solitary waves is given by
$$ \frac{A}{h} = \frac{H-\sqrt r}{1+\sqrt r}. $$
Take the case where $H^2 > r$. It is easy to see that while $A/h$ is always smaller than $H$,
$2A/h$ can exceed $H$, so that the resulting wave will hit the roof. Therefore it will be
interesting to consider the collision of solitary waves of arbitrary amplitudes by using the full Euler equations. On the other
hand, for `table-top' solitary waves of depression, the resulting wave cannot touch the bottom.

\appendix

\section{Additional results on run-ups and phase shifts}

In this appendix, we provide accurate results on run-ups and phase shifts. The terminology `run-up' denotes the fact that
during the collision of two counterpropagating solitary waves the wave amplitude increases beyond the sum of the two
single wave amplitudes. Since run-ups and phase shifts are always very small, they must be computed with high
accuracy. This is why it is important to clean the solitary waves obtained by the approximate expressions (\ref{onde_sol})
or (\ref{onde_plate}). We proceed as follows. We begin with an approximate solution, let it propagate across the domain,
truncate the leading pulse, use it as new initial value by translating it to the left of the domain, let it propagate again and 
distance itself from the trailing
dispersive tail, truncate again, and repeat the whole process over and over until a clean, at least to the eye, solitary
wave is produced. Then we use this new filtered solution as initial guess to study the various collisions. 

For solitary wave solutions to the
system of equations with quadratic nonlinearities (\ref{2eq_clean_AB}), the behavior is the same as the behavior shown for example in
\cite{CGHHS}. In particular we obtain pictures that look very similar to their Figure 2 for the phase shift resulting from
the head-on collision of two solitary waves of equal height, to their Figure 4 for the time evolution of the maximum
amplitude of the solution (it rises sharply to more than twice the elevation of the incident solitary waves, then descends
to below this level after crest detachment, and finally relaxes back to almost its initial level) and to their Figure 12
for the asymmetric head-on collision of two solitary waves of different heights. 

\begin{figure}
\begin{center}
\begin{tabular}{c}
(a) \\
\includegraphics[width=12cm]{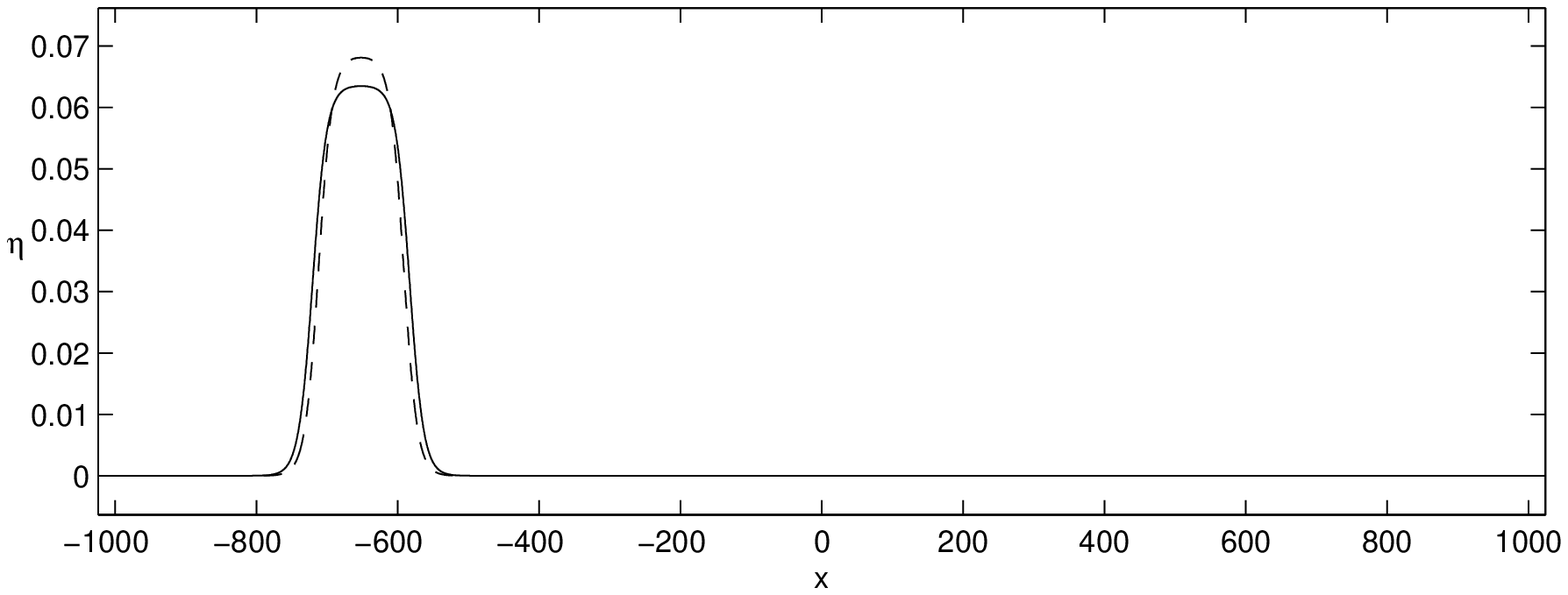}
\end{tabular}
\begin{tabular}{c c}
(b) & (c) \\
\includegraphics[width=6cm]{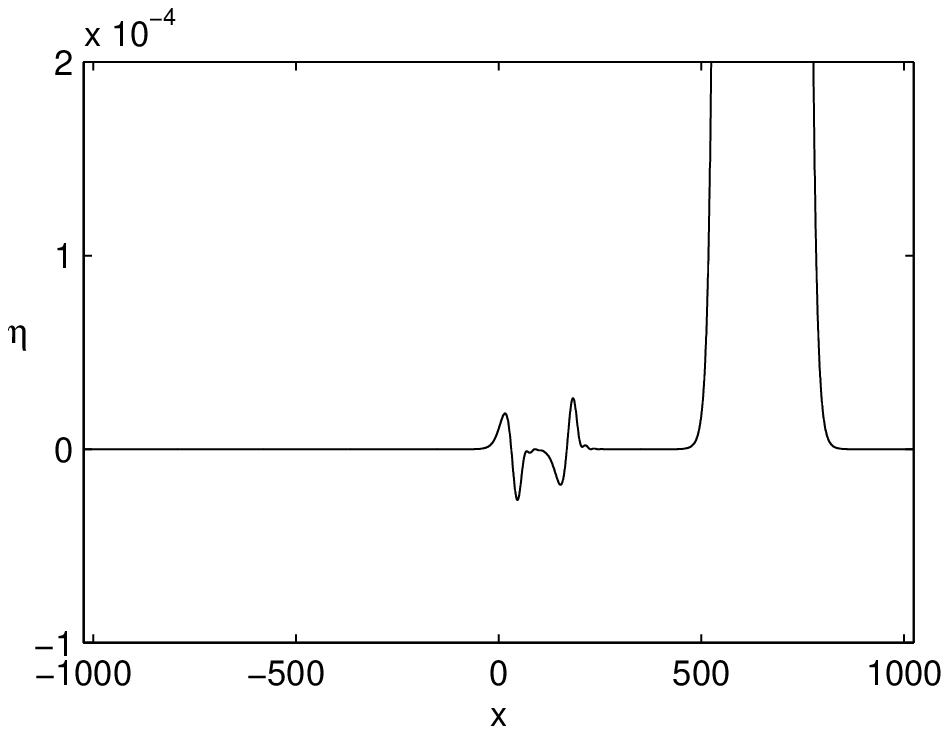} & \includegraphics[width=6cm]{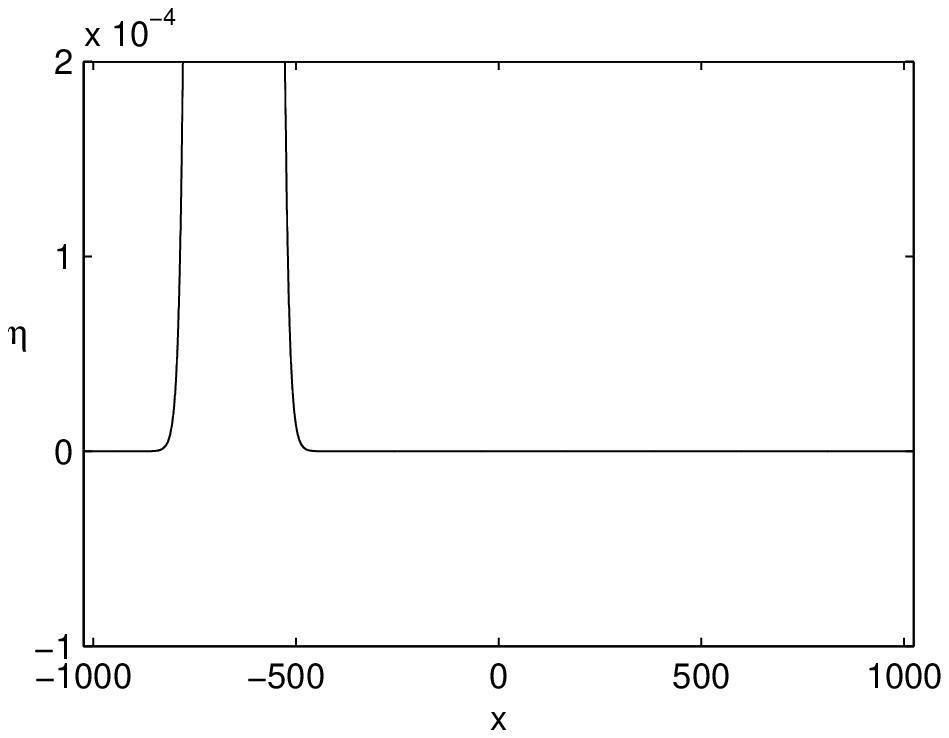} \\
(d) & (e) \\
\includegraphics[width=6cm]{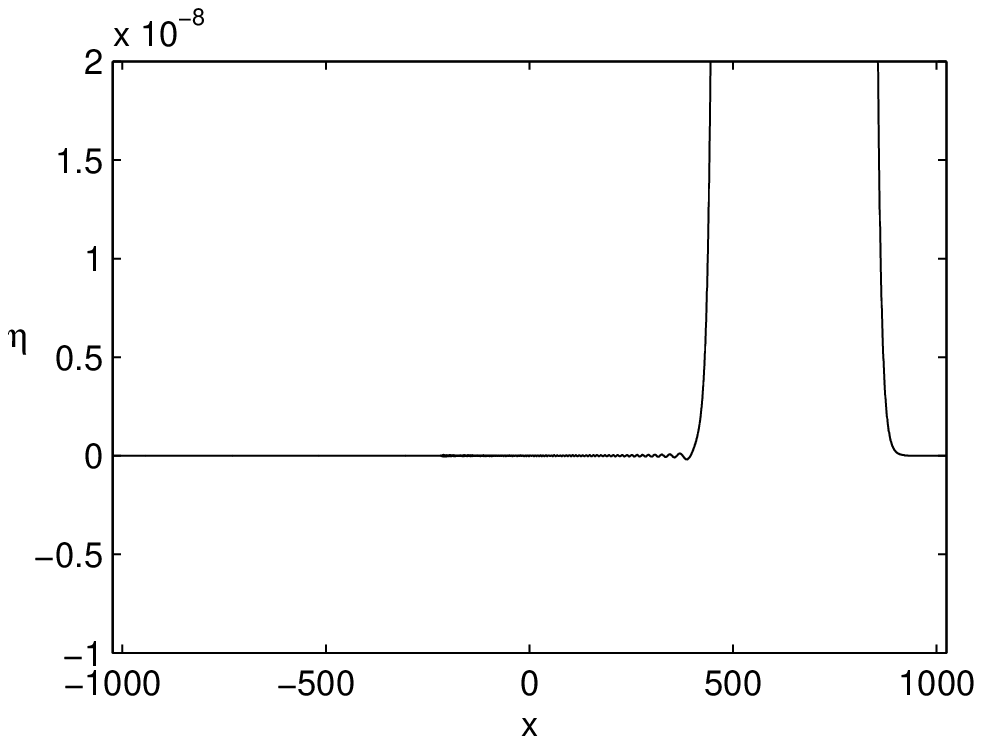}& \includegraphics[width=6cm]{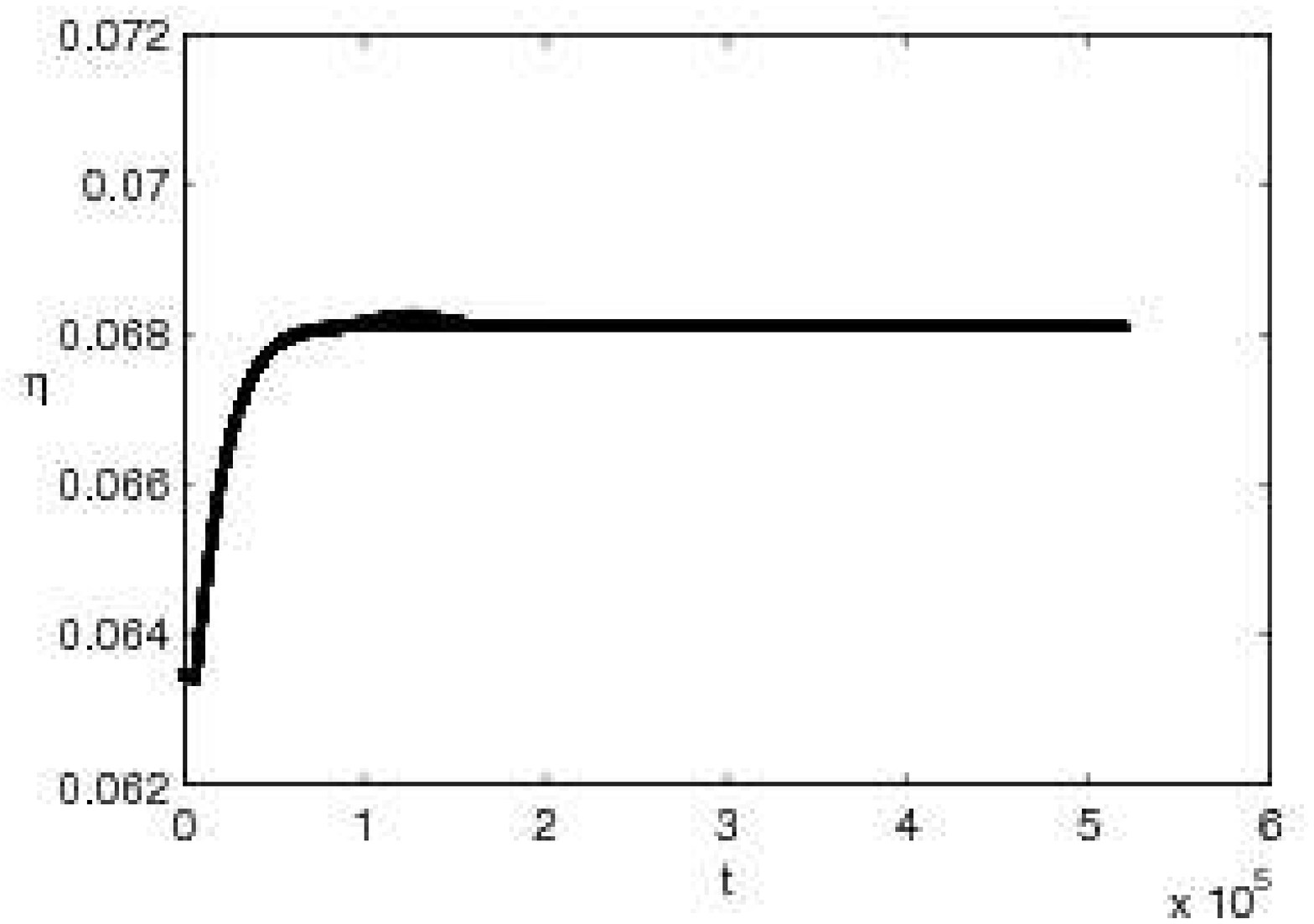} \\
\end{tabular}
\caption{Flat solitary wave produced by iterative cleaning. This is a solution to the system of extended Boussinesq equations
(\ref{ordre5S}). (a) Difference in the profile before (solid line) and after (dashed line) cleaning. (b) Profile of
the approximate solitary wave (\ref{onde_plate}) after one propagation across the domain. (c) Profile (b) after cleaning and 
translation to the left of the domain. (d) Profile after several
cleanings. Notice the change of scale in the vertical axis. (e) Evolution of the maximum amplitude $\eta_{\max}$ as cleaning 
is repeated over and over. The amplitude reaches an asymptotic level. }
\label{accurate_filter}
\end{center}
\end{figure}

\begin{figure}
\begin{center}
\begin{tabular}{c}
(a) \\
\includegraphics[width=12cm]{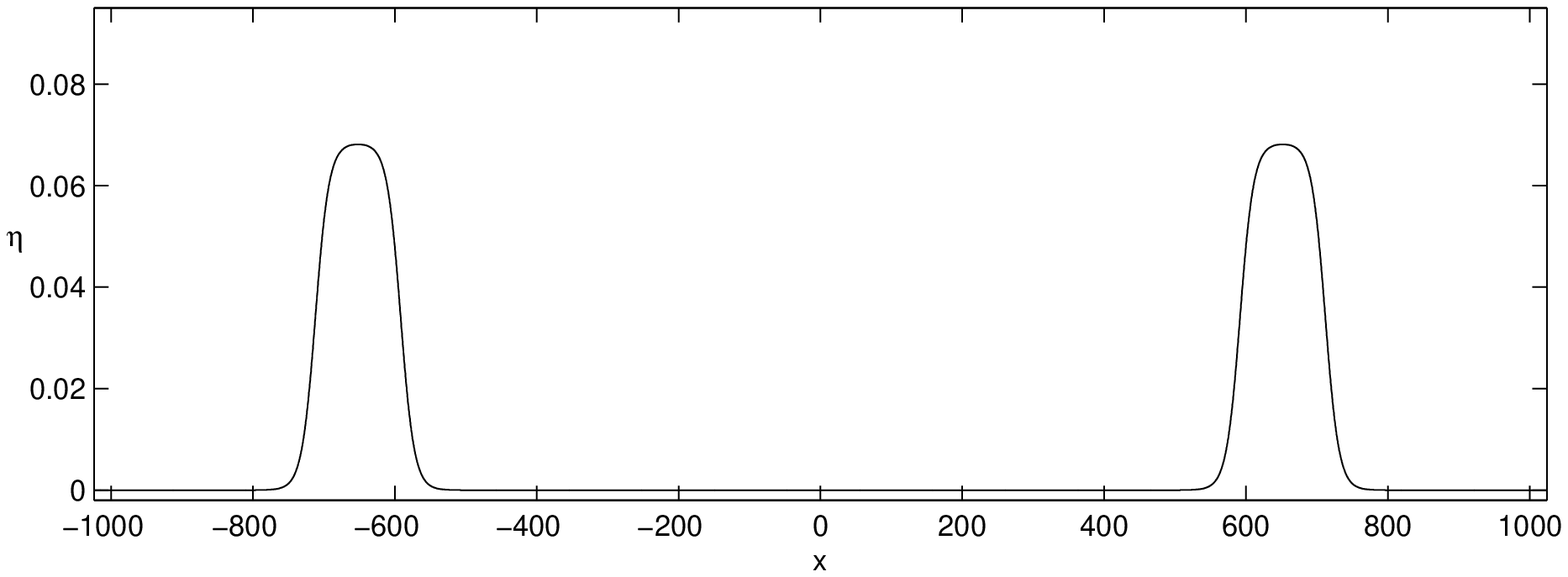}
\end{tabular}
\begin{tabular}{c c}
(b) & (c) \\
\includegraphics[width=6cm]{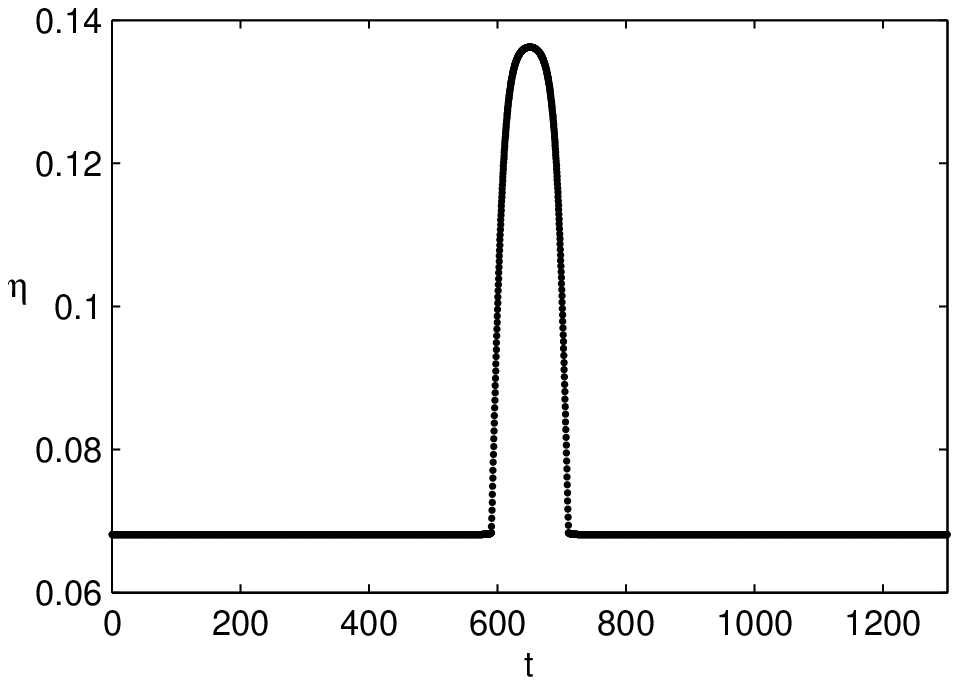}& \includegraphics[width=6cm]{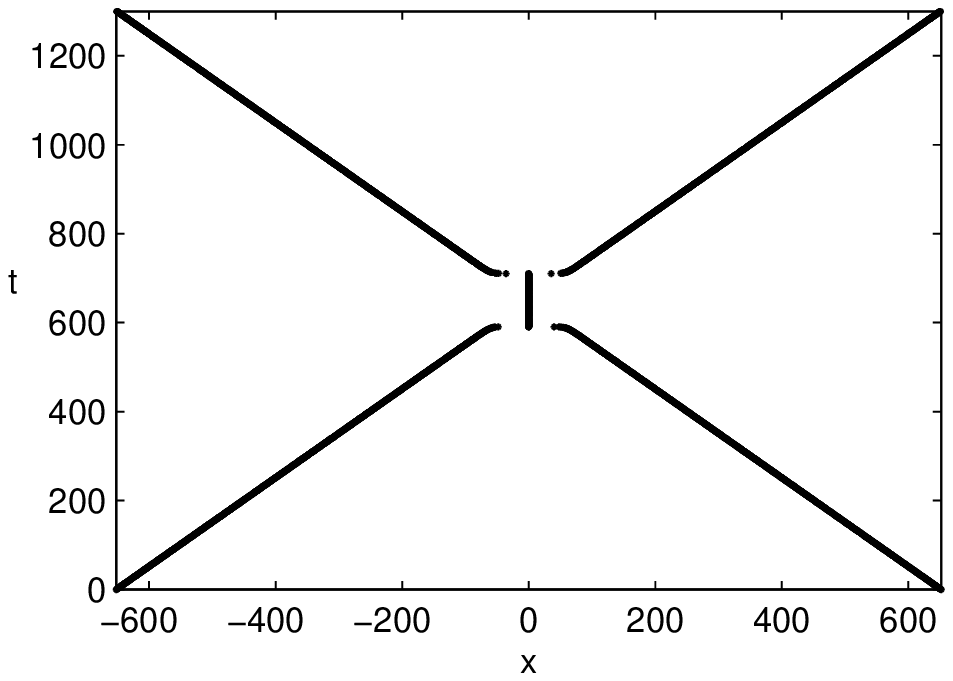} \\
\end{tabular}
\caption{A collision between two clean `table-top' solitary waves of equal height. This is a solution to the system of extended 
Boussinesq equations (\ref{ordre5S}). (a) Initial profiles. (b) Time evolution of the amplitude $\eta_{\rm max}$. (c)
Crest trajectory.}
\label{accurate_2tt}
\end{center}
\end{figure}

\begin{figure}
\begin{center}
\begin{tabular}{c}
(a) \\
\includegraphics[width=12cm]{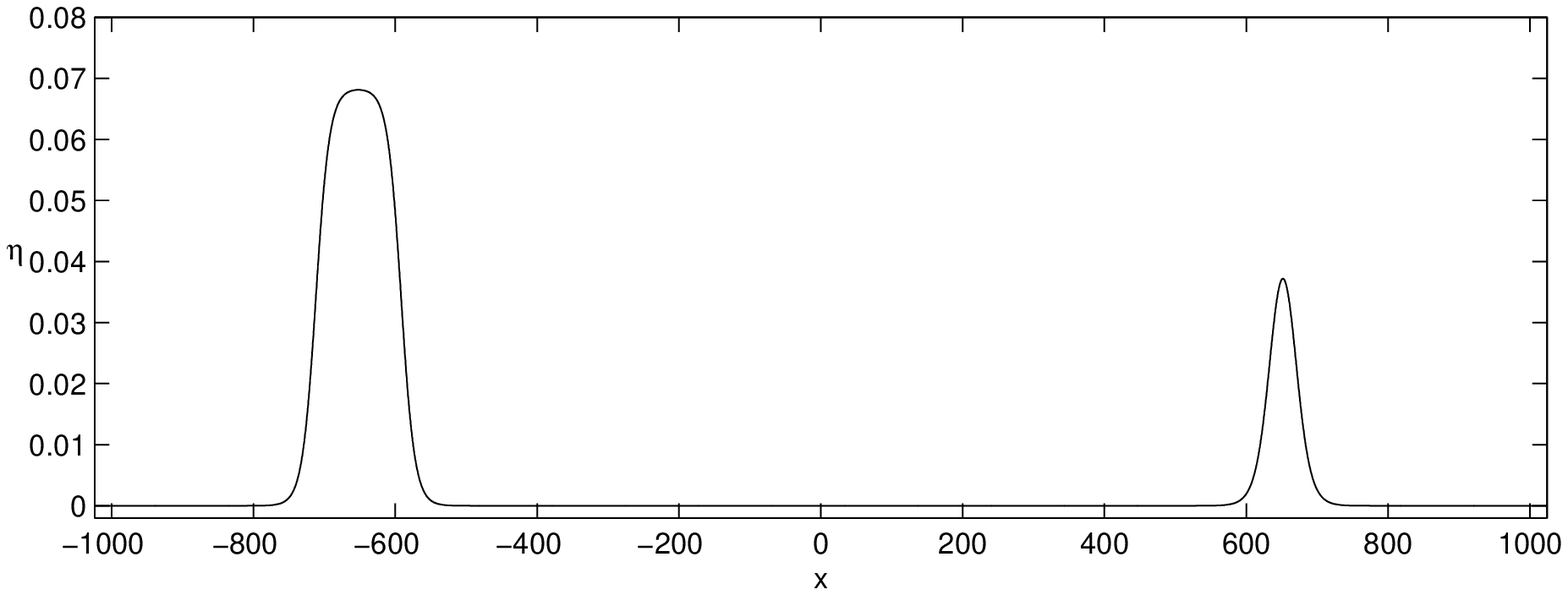}
\end{tabular}
\begin{tabular}{c c}
(b) & (c) \\
\includegraphics[width=6cm]{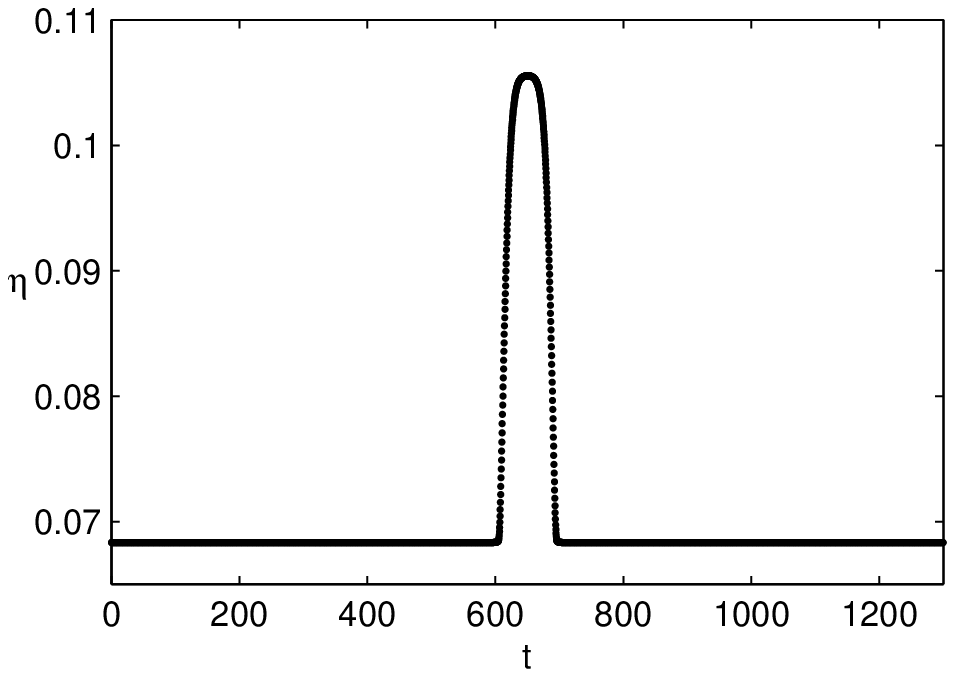}& \includegraphics[width=6cm]{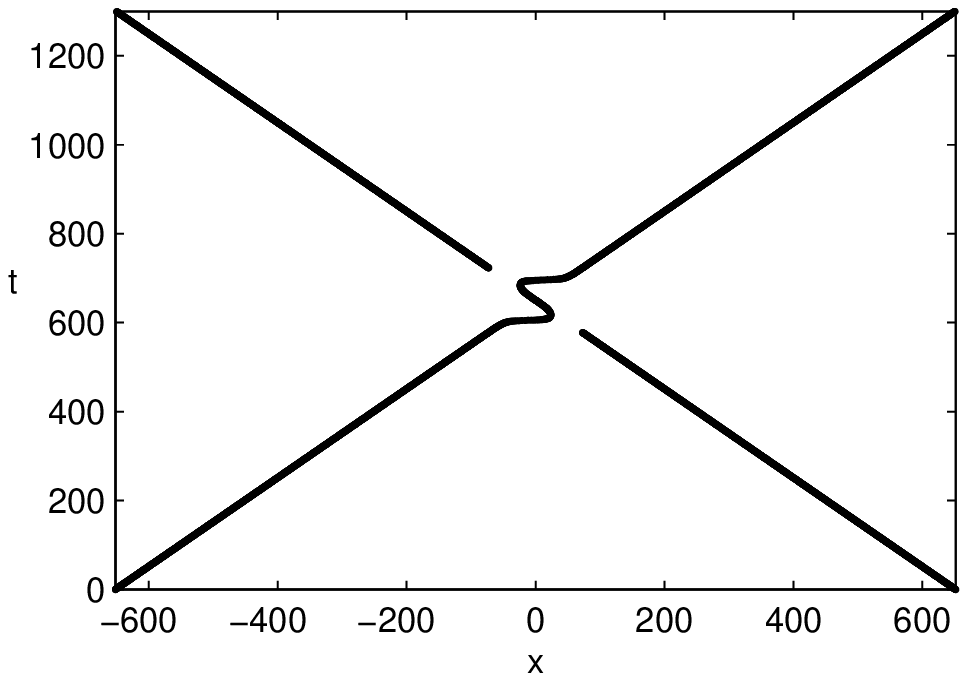} \\
\end{tabular}
\caption{A collision between a clean solitary wave and a clean `table-top' solitary wave. This is a solution to the system of 
extended Boussinesq equations (\ref{ordre5S}). (a) Initial profiles. (b) Time evolution of the amplitude $\eta_{\rm max}$. (c)
Crest trajectory.}
\label{accurate}
\end{center}
\end{figure}

Since the main contribution of the present paper is the inclusion of cubic terms in addition to the quadratic terms, we focus
on results for the extended Boussinesq system (\ref{ordre5_phy}). Figure \ref{accurate_filter} shows the effect of cleaning. In Figure 
\ref{accurate_2tt}, the collision between two clean `table-top' solitary waves (the cleaning has been applied 400 times)
is shown. Their speed is $V=1.00183358$. The amplitude before cleaning was $\eta_{\max}=0.063476$. After iterative cleaning,
it reached $\eta_{\max}=0.06812113$. The run-up during collision is extremely small: indeed $\eta_{\max}=0.13624323$ at
collision, which is slightly larger than $2 \times 0.06812113 = 0.13624226$. The phase shift is also very small.
In Figure 
\ref{accurate}, the collision between the clean `table-top' solitary wave of the Figure \ref{accurate_filter} and a clean solitary wave 
(the cleaning has been applied 230 times) is shown. The maximum amplitude is greater than the sum of the two wave 
amplitudes. The speed of the smaller wave is $V=1.0015$. Its amplitude before cleaning was $\eta_{\max}=0.03647847$. After 
iterative cleaning,
it reached $\eta_{\max}=0.03719492$. The run-up during collision is again extremely small, even if it is larger than in the
previous case: indeed $\eta_{\max}=0.10556057$ at
collision, which is slightly larger than $0.06812113 + 0.03719492 = 0.10531605$. The
phase shift is very small and the crest trajectory shows an interesting path. The overall conclusion is that run-ups and
phase shifts are smaller for `table-top' solitary waves than for `classical' solitary waves.

\section{Intermediate steps in the derivation of the extended Boussinesq system with cubic terms}

Adding $H$ times equation (\ref{eq1u'}) to $r$ times equation (\ref{eq2u'}) yields
\begin{eqnarray}\label{somme12}
(r+H)\eta_t+H(w_x-rw'_x)+\alpha [(Hw+rw')\eta]_x & & \nonumber \\
+\frac{H}{2}\beta\left[(\theta^2-\frac{1}{3}) w_{xxx}
-r(\theta'^2-\frac{1}{3}H^2)w'_{xxx}\right] & & \nonumber \\
+\frac{1}{2}\alpha\beta\left[H(\theta^2-1)(\eta w_{xx})_x+r(\theta'^2-H^2)(\eta w'_{xx})_x\right] & & \nonumber \\
+\frac{5H}{24}\beta^2\left[\left(\theta^2-\frac{1}{5}\right)^2 w_{xxxxx}-r\left(\theta'^2-\frac{1}{5}H^2\right)^2w'_{xxxxx}
\right] & = & 0.
\end{eqnarray}
Let us replace the variables $w$ and $w'$ in (\ref{somme12}) by their expressions (\ref{ww'})-(\ref{w'w}) in terms of $W$ and let
$$ F=\frac{1+H}{r+H}, \quad G=H \frac{ (\theta'^2-\frac{1}{3}H^2)-(\theta^2-\frac{1}{3})}{r+H}. $$
We consider all the terms one by one:

\begin{enumerate}
\item {\it Term $H(w_x-rw'_x)$}
\begin{equation}\label{termeA}
H(w_x-rw'_x)=HW_x
\end{equation}

\item {\it Term $\alpha[(Hw+rw')\eta]_x$}\\ 
From equation (\ref{ww'}), we have
$$ w-rw'=-(r+H)w'-\alpha FW\eta+\frac{1}{2}\beta GW_{xx}, $$
so that
\begin{equation} \label{B1}       
w'=-\frac{1}{r+H}W-\alpha\frac{1}{r+H}FW\eta+\frac{1}{2}\beta \frac{1}{r+H}G W_{xx}.             
\end{equation}
Similarly from equation (\ref{w'w}), we obtain
\begin{equation}\label{B2}
         w = \frac{H}{r+H}W-\alpha\frac{r}{r+H}FW\eta+\frac{1}{2}\beta\frac{r}{r+H}GW_{xx}. 
\end{equation}
Combining (\ref{B1}) and (\ref{B2}) yields
 $$Hw+rw'=\frac{H^2-r}{r+H}W-\alpha\frac{r(1+H)}{r+H}FW\eta+\frac{1}{2}\beta\frac{r(1+H)}{r+H}GW_{xx}.$$
Therefore
\begin{eqnarray*}\label{termeB}
\alpha\Big[(Hw+rw')\eta\Big]_x & = & \alpha\frac{H^2-r}{r+H}(W\eta)_x-\alpha^2\frac{r(1+H)^2}{(r+H)^2}(W\eta^2)_x \\
& & +\frac{1}{2}\alpha\beta\frac{rH(1+H)\left((\theta'^2-\frac{1}{3}H^2)-(\theta^2-\frac{1}{3})\right)}{(r+H)^2}(W_{xx}\eta)_x.
\end{eqnarray*}

\item {\it Term in $w_{xxx}$ and $w'_{xxx}$}\\
Combining (\ref{B1}) and (\ref{B2}) yields
$$ \frac{H}{2}\beta\left[(\theta^2-\frac{1}{3})w_{xxx}-r(\theta'^2-\frac{1}{3}H^2)w'_{xxx}\right] $$
\begin{eqnarray}\label{termeC}
&=&\frac{1}{2}\beta H\frac{H(\theta^2-\frac{1}{3})+r(\theta'^2-\frac{1}{3}H^2)}{r+H}W_{xxx}\nonumber\\
&&+\frac{1}{2}\alpha\beta rH(1+H)\frac{\left(\theta'^2-\frac{1}{3}H^2\right)-\left(\theta^2-\frac{1}{3}\right)}{(r+H)^2}(W\eta)_{xxx}\nonumber\\
&&-\frac{1}{4}\beta^2rH^2\frac{\left((\theta'^2-\frac{1}{3}H^2)-(\theta^2-\frac{1}{3})\right)^2}{(r+H)^2}W_{xxxxx}.
\end{eqnarray}

\item {\it Term in $(\eta w_{xx})_x$ and $(\eta w'_{xx})_x$}\\
Using (\ref{W}) yields
\begin{eqnarray}\label{termeD}
\frac{1}{2}\alpha\beta \left[H(\theta^2-1)(\eta w_{xx})_x+r(\theta'^2-H^2)(\eta w'_{xx})_x\right] & & \nonumber\\
 = \frac{1}{2}\alpha\beta\frac{H^2(\theta^2-1)-r(\theta'^2-H^2)}{r+H} (\eta W_{xx})_x & &
\end{eqnarray}

\item {\it Term in $w_{xxxxx}$ and $w'_{xxxxx}$}\\
Using (\ref{W}) yields
\begin{eqnarray}\label{termeE}
&&\frac{5}{24}H\beta^2\left[\left(\theta^2-\frac{1}{5}\right)^2 w_{xxxxx}-r\left(\theta'^2-\frac{1}{5}H^2\right)^2w'_{xxxxx}\right]
\nonumber\\
&=&\frac{5}{24}H\beta^2\frac{H(\theta^2-\frac{1}{5})^2+r(\theta'^2-\frac{1}{5}H^2)^2}{r+H}W_{xxxxx}
\end{eqnarray}
\end{enumerate}

Combining all terms (\ref{termeA})--(\ref{termeE}) yields the first equation of the extended Boussinesq system
\begin{equation}\label{sommedeux}
\fbox{$
   \begin{array}{ll}
\displaystyle{(r+H)\eta_t+HW_x+\alpha\frac{H^2-r}{r+H}(W\eta)_x}\\ \\
\displaystyle{+\frac{1}{2}\beta\frac{H\left(H(\theta^2-\frac{1}{3})+r(\theta'^2-\frac{1}{3}H^2)\right)}{r+H}W_{xxx}}
\displaystyle{-\alpha^2 \frac{r(1+H)^2}{(r+H)^2} (W\eta^2)_x}\\  \\
\displaystyle{+\frac{1}{2}\alpha\beta\frac{rH(1+H)}{(r+H)^2}\left((\theta'^2-\frac{1}{3}H^2)-(\theta^2-\frac{1}{3})\right)(W_{xx}\eta)_x}
\\ \\
\displaystyle{+\frac{1}{2}\alpha\beta\frac{H^2(\theta^2-1)-r(\theta'^2-H^2)}{r+H} (W_{xx}\eta )_x}\\ \\
\displaystyle{+\frac{1}{2}\alpha\beta{rH(1+H)}\frac{(\theta'^2-\frac{1}{3}H^2)-(\theta^2-\frac{1}{3})}{(r+H)^2}(W\eta)_{xxx}}\\ \\
\displaystyle{-\frac{1}{4}\beta^2\frac{rH^2\Big((\theta'^2-\frac{1}{3}H^2)-(\theta^2-\frac{1}{3})\Big)^2}{(r+H)^2}W_{xxxxx}}\\ \\
\displaystyle{+\frac{5}{24}H\beta^2\frac{H(\theta^2-\frac{1}{5})^2+r(\theta'^2-\frac{1}{5}H^2)^2}{r+H}W_{xxxxx}=0}\\
  \end{array}
  $}
\end{equation}
%%%%%%%%%%%%%%%%%%%%%%%%%%%%%%%%%%%%%%%%%%%%%%%%%%%%%%%%%%%%%%%%%%%%%%%%%%%%%%%%%%%5
We proceed the same way for equation (\ref{eq3u'}).

\begin{enumerate}
\item {\it Term in $w_{xxt}$ and $w'_{xxt}$}\\
$$ \frac{1}{2}\beta\left[(\theta^2-1)w-r(\theta'^2-H^2)w'\right]_{xxt} = $$
\begin{eqnarray}\label{termeM}
& &\frac{1}{2}\beta\frac{H(\theta^2-1)+r(\theta'^2-H^2)}{r+H}W_{xxt}
-\frac{1}{2}\alpha\beta\frac{r(1+H)\Big((\theta^2-1)-(\theta'^2-H^2)\Big)}{(r+H)^2}(W\eta)_{xxt}\nonumber\\
& &+\frac{1}{4}\beta^2\frac{rH\Big((\theta^2-1)-(\theta'^2-H^2)\Big)\Big((\theta'^2-\frac{1}{3}H^2)-
(\theta^2-\frac{1}{3})\Big)}{(r+H)^2}W_{xxxxt} 
\end{eqnarray}

\item {\it Term $\alpha(ww_x-rw'w'_x)$}\\
\begin{eqnarray}\label{termeN}
\alpha(ww_x-rw'w'_x)&=&\alpha\frac{H^2-r}{(r+H)^2}WW_x-\alpha^2\frac{r(H+1)^2}{(r+H)^3}(W^2\eta)_x \nonumber\\
                    & &+\frac{1}{2}\alpha\beta\frac{rH(H+1)\Big((\theta'^2-\frac{1}{3}H^2)-(\theta^2-\frac{1}{3})\Big)}
{(r+H)^3}(WW_{xx})_x \nonumber\\
\end{eqnarray}

\item {\it Term} 
\begin{equation}\label{termeP}
\alpha\beta\Big[(\eta w_{xt})_x+rH(\eta w'_{xt})_x\Big] 
= \alpha\beta\frac{H(1-r)}{r+H}(\eta W_{xt})_x 
\end{equation}

\item {\it Term}
\begin{eqnarray}\label{termeQ}
\frac{1}{2}\alpha\beta\Big[(\theta^2-1)ww_{xxx}-r(\theta'^2-H^2)w'w'_{xxx}\Big] 
&=&\frac{1}{2}\alpha\beta\frac{H^2(\theta^2-1) -r(\theta'^2-H^2)}{(r+H)^2} WW_{xxx} \nonumber\\
\end{eqnarray}

\item {\it Term} 
\begin{eqnarray}\label{termeR}
\frac{1}{2}\alpha\beta\Big[(\theta^2+1)w_xw_{xx}-r(\theta'^2+H^2)w'_xw'_{xx}\Big] 
&=&\frac{1}{2}\alpha\beta\frac{H^2(\theta^2+1)-r(\theta'^2+H^2)}{(r+H)^2} W_xW_{xxx}\nonumber\\
\end{eqnarray}

\item {\it Term} 
\begin{eqnarray}\label{termeS}
\frac{1}{2}\beta^2\Big((\theta^2-1)(5\theta^2-1)w_{xxxxt}-r(\theta'^2-H^2)(5\theta'^2-H^2)w'_{xxxxt}\Big) & & \nonumber \\
=\frac{1}{2}\beta^2\frac{H(\theta^2-1)(5\theta^2-1)+r(\theta'^2-H^2)(5\theta'^2-H^2)}{r+H} W_{xxxxt} & & 
\end{eqnarray}

\end{enumerate}

Combining all terms (\ref{termeM})--(\ref{termeS}) yields
\begin{equation}\label{trois}
\fbox {$
   \begin{array}{rll}
&&(1-r)\eta_x+ W_t \displaystyle{+\alpha\frac{H^2-r}{(r+H)^2} WW_x} \\ \\
&&\displaystyle{+\frac{1}{2}\beta\frac{H(\theta^2-1)+r(\theta'^2-H^2)}{r+H}W_{xxt}}
\displaystyle{-\alpha^2\frac{r(1+H)^2}{(r+H)^3}(W^2\eta)_x}\\ \\
&&\displaystyle{+\frac{1}{2}\alpha\beta\frac{rH(H+1)\Big((\theta'^2-\frac{1}{3}H^2)-(\theta^2-\frac{1}{3})\Big)}{(r+H)^3}(WW_{xx})_x}\\ \\
&&\displaystyle{+\alpha\beta\frac{H(1-r)}{r+H}(\eta W_{xt})_x}
\displaystyle{+\frac{1}{2}\alpha\beta\frac{H^2(\theta^2-1) -r(\theta'^2-H^2)}{(r+H)^2} WW_{xxx}}\\ \\
&&\displaystyle{+\frac{1}{2}\alpha\beta\frac{H^2(\theta^2+1)-r(\theta'^2+H^2)}{(r+H)^2} W_xW_{xx}}\\ \\
&&\displaystyle{-\frac{1}{2}\alpha\beta\frac{r(1+H)\Big((\theta^2-1)-(\theta'^2-H^2)\Big)}{(r+H)^2}(W\eta)_{xxt}}\\\ \\
&&\displaystyle{+\frac{1}{4}\beta^2\frac{rH\Big((\theta^2-1)-(\theta'^2-H^2)\Big)\Big((\theta'^2-\frac{1}{3}H^2)-(\theta^2-\frac{1}{3})\Big)}{(r+H)^2}W_{xxxxt}}\\ \\
&&\displaystyle{+\frac{1}{2}\beta^2\frac{H(\theta^2-1)(5\theta^2-1)+r(\theta'^2-H^2)(5\theta'^2-H^2)}{r+H} W_{xxxxt}=0 }
  \end{array}
    $}
\end{equation}

\end{document}